\documentclass[onecolumn]{emulateapj}

\usepackage{geometry}   
\usepackage[latin1]{inputenc}
\usepackage{graphicx}
\usepackage{epstopdf}
\usepackage{amsmath,amsfonts,amssymb}
\usepackage{soul}

\renewcommand{\(}{\left(}
\renewcommand{\)}{\right)}
\newcommand{\vb}{{\bf v}_b}
\newcommand{\xvec}{{\bf x}}

\newcommand{\rhob}{\rho_b}

\newcommand{\Enu}{E_{\nu}}

\newcommand{\mn}{{\tt n}}

\newcommand{\hii}{H {\footnotesize II}~}

\begin{document}

\title{Fully-Coupled Simulation of Cosmic Reionization. I: Numerical Methods and Tests}


\author{Michael L. Norman\altaffilmark{1,3}}
\author{Daniel R. Reynolds\altaffilmark{4}}
\author{Geoffrey C. So\altaffilmark{1}}
\author{Robert P. Harkness\altaffilmark{2,3}}
\author{John H. Wise\altaffilmark{5}}

\affiliation{
\altaffilmark{1}CASS, University of California, San Diego, 9500 Gilman Drive La Jolla, CA 92093-0424\\
\altaffilmark{2}NICS, Oak Ridge National Laboratory, 1 Bethel Valley Rd, Oak Ridge, TN 37831\\
\altaffilmark{3}SDSC, University of California, San Diego, 9500 Gilman Drive La Jolla, CA 92093-0505\\ 
\altaffilmark{4}Southern Methodist University, 6425 Boaz Ln, Dallas, TX 75205\\
\altaffilmark{5}Center for Relativistic Astrophysics, Georgia Institute of Technology, 837 State St, Atlanta, GA 30332 \\
}

\begin{abstract}
We describe an extension of the {\em Enzo} code to enable fully-coupled radiation hydrodynamical simulation 
of inhomogeneous reionization in large $\sim (100 Mpc)^3$ cosmological volumes with thousands to millions of point sources. We solve
all dynamical, radiative transfer, thermal, and ionization processes self-consistently on the same mesh, 
as opposed to a postprocessing approach which coarse-grains the radiative transfer. We do,
however, employ a simple subgrid model for star formation which we calibrate to observations. 
The numerical method presented is a modification of an earlier method presented in 
\cite{ReynoldsHayesPaschosNorman2009} differing principally in the operator splitting algorithm we we use to advance the system of equations. Radiation transport is done in the grey flux-limited diffusion (FLD) approximation, which is solved by
implicit time integration split off from the gas energy and ionization equations, which are solved
separately. This results in 
a faster and more robust scheme for cosmological applications compared to the earlier method. 
The FLD equation is solved
using the {\em hypre} optimally scalable geometric multigrid solver from LLNL. 
By treating the ionizing radiation as a grid field as opposed to rays, our method
is scalable with respect to the number of ionizing sources, limited only by the parallel scaling
properties of the radiation solver. We test the speed and accuracy of our approach on a 
number of standard verification and validation tests.  We show by direct comparison with {\em Enzo}'s adaptive ray tracing method {\em Moray} that the well-known 
inability of FLD to cast a shadow behind opaque clouds has a minor effect on the evolution of ionized volume and mass fractions
in a reionization simulation validation test. 
We illustrate an application of our method to the
problem of inhomogeneous reionization in a 80 Mpc comoving box resolved with
$3200^3$ Eulerian grid cells and dark matter particles. 
\end{abstract}

\maketitle

\section{Introduction}
\label{sec:introduction}

The epoch of reionization (EoR) is a current frontier of cosmological research both observationally and theoretically. 
Observations constrain the transition from a largely neutral intergalactic medium (IGM) of primordial gas to a largely ionized one 
(singly ionized H and He) to the redshift interval $z \sim 11-6$, which is a span of roughly 500 Myr. 
The completion of H reionization by $z \approx 6$ is firmly established through quasar absorption
line studies to luminous, high redshift quasars which exhibit Ly $\alpha$ Gunn-Peterson absorption troughs \citep{FanCarilliKeating2006}. 
The precise onset of H reionization (presumably tied to the formation of the first luminous ionizing sources) is presently unknown observationally, however CMB measurements of the Thomson optical depth 
to the surface of last scattering by the WMAP and Planck satellites indicates that the IGM was substantially ionized by $z\sim 10$ 
\citep{Spergel2003,Komatsu2009,JarosikEtAl2011,Planck2013}.
Since the optical depth measurement is redshift integrated  and averaged over the sky, the CMB observations provide no information about how reionization proceeded or the nature of the radiation sources that caused it.

 It is generally believed that reionization begins with the formation of Population III stars at $z \sim 20-30$ \citep{ABN02,Yoshida03,BL04,Sokasian04}, but that soon the ionizing photon budget becomes dominated by young, star forming galaxies (see e.g., \cite{Wise12,Xu13}), and to a lesser extent by the first quasars \citep{MadauEtAl1999, BoltonHaehnelt2007, HaardtMadau2012, BeckerBolton2013}. Observations of galaxies in the redshift interval $6 \leq z \leq 10$ using the Hubble Space Telecope support the galaxy reionizer hypothesis, with the caveat that the faint end of the luminosity function which contributes substantially to the number of ionizing photons has not yet been measured \citep{Robertson10,Bouwens12}. 

Given the paucity of observational information about the {\em process} of cosmic reionization, researchers have resorted to theory and
numerical simulation to fill in the blanks. As reviewed by Trac \& Gnedin (2011), progress in this area has been dramatic, driven by a synergistic interplay between semi-analytic approaches and numerical simulations. The combination of these two approaches have converged on a qualitative picture of how H reionization proceeds assuming the primary ionizing sources are young, star-forming galaxies. The physics of the reionization process is determined by the physics of the source and sinks of ionizing radiation in an expanding universe. Adopting the $\Lambda$CDM model of structure formation, galaxies form hierarchically through the merger of dark matter halos. The structure and evolution of the dark matter density field is now well understood through ultra-high resolution numerical N-body simulations \citep{Millenium,Bolshoi} and through analytic models based on these simulations \citep{CooraySheth2002}. By making certain assumptions about how ionizing light traces mass and the dynamics of \hii regions, a basic picture of the reionization process has emerged \citep {Furlanetto04,Furlanetto06,Iliev06,Zahn07} that is confirmed by detailed numerical simulations; e.g., \cite{Zahn11}.

The basic picture is that galaxies form in the peaks of the dark matter density field and drive expanding \hii regions into their surroundings by
virtue of the UV radiation emitted from young, massive stars. These \hii regions are initially isolated, but begin to merge into larger, Mpc-scale \hii regions due to the clustering of the galaxy distribution (expansion phase). Driven by a steadily increasing global star formation rate and recombination time (due to cosmic expansion) this process goes on until \hii regions completely fill the volume (overlap phase). In this picture, rare peaks in the density field ionize first while regions of lower density ionize later from local sources that themselves formed later. In this picture, referred to as ``inside-out reionization", void regions are the last to ionize because they have few local sources of ionization and remain neutral until an I-front from a denser region has swept over it. 

To date numerical simulations of reionization have fallen into two basic classes \citep{TracGnedin2011}: small-scale simulations that resolve the sources and sinks of ionizing radiation; and large-scale simulations that account for the diversity and clustering of sources. While ideally one would like to do both in a single simulation, this has not been feasible until now owing to numerical limitations. Historically, small-scale simulations came first. These simulations self-consistently modeled galaxy formation, radiative transfer, and photoionization/recombination within a hydrodynamic cosmological code. Comoving volumes were typically $\leq (10 Mpc)^3$ with spatial resolution of a few comoving kpc--sufficient to resolve high redshift dwarf galaxies and the baryonic cosmic web  \citep{Gnedin00a,RazoumovEtAl2002,RicottiEtAl2002,Petkova11,Finlator11}. Large-scale simulations followed, however these were not self-consistent radiation hydrodynamic cosmological simulations. Rather, density fields were simulated with a cosmological N-body or hydrodynamics code, and then ionization was computed in a post-processing step using a standalone radiative transfer code, typically a Monte Carlo or ray-tracing code
\citep{Ciardi01,SokasianEtAl2001,Sokasian03,Iliev06,Iliev2012,Iliev2014,Zahn07,TracCen2007,TracCenLoeb2008,ShinTracCen2008,Finlator09}.
Recently, a new class of reionization simulations is emerging which combines fully-coupled
radiation hydrodynamics in large volumes \citep{So2014,Gnedin14b}.

Regardless whether reionization is simulated in a fully-coupled or  post-processing approach, the 3D radiative transfer equation must be solved in some fashion to some level of approximation. Fortunately, reionization is a continuum radiative transfer problem dominated by photon removal, and therefore the full 6-dimensional phase space of the specific intensity does not need to be computed directly taking into account scattering and energy redistribution. The basic methods employed are ray tracing, using long or short characteristics,  \citep{AbelEtAl1999,NakamotoEtAl2001,SokasianEtAl2001,RazoumovEtAl2002,AbelWandelt2002,
RazoumovCardall2005,MellemaEtAl2006,RijkhorstEtAl2006,TracCen2007,Zahn07,
PawlikSchaye2008,PawlikSchaye2011,WiseAbel11},
Monte Carlo methods \citep{Ciardi01,MaselliEtAl2003}, and moment methods \citep{NormanAbelPaschos98,GnedinAbel2001,PaschosEtAl2007,AubertTeyssier2008,Petkova09,
Finlator09}. Ray tracing methods have the advantage of resolving the angular structure of the radiation field to the desired accuracy at the expense of carrying many angles per spatial cell. 
Monte Carlo methods are attractive because they directly sample the photon distribution function in energy and angle, but are very costly because a large number of photons must be tracked to control noise in the average properies of the radiation field. Moment methods are interesting because they represent the radiation field using a small number of angular moments of the specific intensity, and thus can be quite efficient computationally at the expense of a high degree of angular fidelity. A number of these methods have been compared against one another as a part of the ``cosmological radiative transfer codes comparison project", and have been shown to perform quite well on a suite of static and dynamic ionization tests \citep{IlievEtAl2006,IlievEtAl2009}.

The most well known and widely used moment method for astrophysical calculations is flux limited diffusion (FLD) \citep{MihalasMihalas1984}, which closes the hierarchy of moment equations at the zeroth order, resulting in a diffusion equation for the radiation energy density. A variety of flux limiters have been developed that ensure correct radiation propagation speeds in optically thick and thin limits \citep{LevermorePomraning1981}. The application of moment methods beyond FLD to cosmological reionization simulations was first suggested by Norman, Paschos \& Abel (1998). They proposed closing the hierarchy of moment equations at the first order through a variable tensor Eddington factor (VTEF) closure. The VTEF method was first described and implemented in the non-cosmological ZEUS-2D code in 1992 \citep{StoneMihalasNorman92}, and later in the ZEUS-MP code \citep{HayesNorman2003,HayesEtAl2006}. A technical difficulty of this approach for reionization simulations is the evaluation of the Eddington tensor (ET) at every point in space, which formally requires a solution of the 3D radiative transfer equation--something one wants to avoid. Several approximate ET evaluation methods have been introduced which avoid solving the 3D RT equation. The first is the OTVET method of \cite{GnedinAbel2001}. In this method the ET
is evaluated at every point in space assuming the medium is optically thin. Since the radiation intensity decreases inversely as the square of the distance from point sources, just as gravity does, fast gravity solvers can be applied for a large collection of point sources. The second is the method of \cite{AubertTeyssier2008} which applies
the M1 closure model developed by \cite{Levermore1984} to the problem of cosmological reionization. The M1 model is a local, analytic closure relation between the radiation pressure and energy density which relies on the assumption that the radiation angular distribution is axysymmetric around the flux vector. It is designed to be accurate in the optically thick limit while retaining some accuracy in the optically thin regime. The OTVET method has been adapted to both Eulerian adaptive mesh codes \citep{Gnedin14a} and Lagrangian SPH codes \citep{Petkova09,Finlator09}. The M1 model has recently been implemented in the RAMSES adaptive mesh refinement code \citep{RosdahlEtAl2013}.

The need to simulate large cosmological volumes coupled with the cost or limited scalability of available radiative transport methods led to a trend which continues to this day of using different numerical resolutions to model the N-body dynamics and the radiative transfer/ionization calculations. For example Iliev et al. (2006a) used a particle-mesh code to
simulate N-body dynamics in a volume 100 Mpc/h on a side with a force resolution of 31 kpc/h, while performing the RT calculation on grids with comoving resolutions of 246 and 492 kpc/h. Similarly Trac \& Cen (2007,2008) and Shin, Trac \& Cen (2008) achieved a force resolution of 8.7 kpc/h in particle-mesh N-body simulations in 50 and 100 Mpc/h boxes, but performed the RT calculation with a mesh resolution of 278 kpc/h. 

More recent simulations are typically far better resolved than that, using either P$^3$M (e.g. \cite{Iliev2012} or tree methods (e.g. \cite{PawlikSchaye2011, Petkova11}), which typically give 1-2 kpc/h or better force resolution. These references illustrate the trade-off between volume and resolution. The Iliev simulations are performed in boxes varying in size from 37 Mpc/h to 114 Mpc/h. Radiative transfer is performed on a mesh of 256$^3$ cells, regardless of the box size, resulting in a resolution mismatch ranging from 144 to 445 kpc/h. The other two references perform the radiative transfer calculation at the same resolution as the the underlying SPH simulation, however they have only been applied to small volumes.  Although \cite{PawlikSchaye2011} and \cite{Petkova11} both use SPH for the cosmological hydrodynamics, they use totally different methods for the radiative transfer. The former have devised a novel photon packet scheme that resembles a short characteristics method implemented on the unstructured mesh defined by the SPH particle positions, while the latter have implemented both FLD and OTVET within the SPH formalism. 

An interesting variant of the post-processing approach is the work of Trac, Cen \& Loeb (2008) who carried out a hydrodynamic cosmological simulation of reionization in a 100 Mpc/h box with $1536^3$ grid cells and particles, taking the emitting source population and subgrid clumping from a much higher resolution N-body simulation. This is an advance over previous work in that large scale baryonic flows are included self-consistently. However small scale radiation hydrodynamic effects, such as the retardation of I-fronts by minihalos \citep{Shapiro04} or the photoevaporation of gas from halos are not modeled. 

In this paper we present a numerical method for simulating cosmological reionization in large cosmological volumes in which 
all relevant processes (dark matter dynamics, hydrodynamics, chemical ionization and recombination, radiation transport, local
star formation and feedback) are computed self-consistently on the same high resolution computational grid. We refer to this as {\em resolution matched}, to distinguish our simulations from the fine/coarse dual resolution scheme used in previous large-scale simulations. The numerical method presented is a modification of an earlier method presented in \cite{ReynoldsHayesPaschosNorman2009} differing principally in the operator splitting algorithm we we use to advance the system of equations. Radiation transport is done in the grey flux-limited diffusion (FLD) approximation, which is solved by
implicit time integration split off from the gas energy and ionization equations, which are solved
separately. This results in 
a faster and more robust scheme for cosmological applications compared to the earlier method.

The key numerical requirement for performing simulations that {\em both} resolve the sources and sinks of ionizing radiation {\em and} correctly model the abundance and clustering of sources is {\em algorithmic scalability}. Parallel scalability is also important, but of secondary importance to algorithmic scalability. Algorithmic scalability refers to how the time to solution scales with the number of unknowns $N$. Direct force evalulation gravitational N-body problems scale as $N^2$. While this is the most accurate approach, it is impractical for $N \sim 10^{10}$ which characterizes modern cosmological N-body simulations. Reionization simulations pose a similar scaling problem. If $N$ is the number of fluid elements (particles, cells) and $S$ is the number of ionizing sources, then the work scales as $N \times S$. At fixed resolution $S$ scales as $N$, since both are proportional to the volume simulated. If ray tracing is the method used for radiative transfer, and $R$ is the number of rays propagated per source, then the work scales as $N^2 R$. The factor $R$ is typically of order 100, but may be compensated for by the fact that $S/N \ll 1$. Therefore work scales as $N^2$ with the commonly used ray tracing approach, and this approach is not tenable for very large $N$. This is the underlying reason why previous large box simulations perform the radiative transfer calculation on a coarser grid than the dark matter calculation. For example, in the work of Trac \& Cen (2007), the disparity in scales is 32.

What is desired is an algorithm that is ideally $\mathcal O(N)$, but lacking that, no worse that $\mathcal O(N\log N)$.  SPH-based methods \citep{PawlikSchaye2008, PawlikSchaye2011, Petkova11} inherit the scalability of the underlying SPH simulation. Others alleviate the scaling problem with adaptive rays, ray merging, short characteristics ray tracing or other techniques \citep{RazoumovCardall2005, Mellema2006, TracCen2007, ShinTracCen2008, WiseAbel11}. We have achieved $\mathcal O(N\log N)$ scaling by numerically representing the radiation field as a {\em grid field}, and employing optimally scalable geometric multigrid methods for the solution of the radiation field equation. In this work the radiation field is treated in the flux-limited diffusion (FLD) approximation, and discretized on the same grid as used for the dark matter and hydrodynamics.  The method we describe below is currently implemented on uniform Cartesian grids within version 2 {\bf of} the community code  {\em Enzo} \citep{BryanEtAl2014}; an adaptive mesh version of this is under development and will be reported on in a forthcoming paper (Reynolds et al., {\em in prep}). 

In Sec. 2 we describe the mathematical formulation of the problem. In Sec. 3 we present the numerical method of solution, focusing on the solution of the coupled radiation diffusion, chemical ionization, and gas energy equations within the {\em Enzo} code framework. As {\em Enzo} has been described elsewhere, only a brief summary of its methods are included.  Section 4 contains results from verification tests (Sec. 4.1), validation tests (Sec. 4.2), parallel scaling tests (Sec. 4.3), and execution speed tests (Sec. 4.4).  We then illustrate the applicability of our method to cosmic reionization in Sec. 5, confining ourselves to a qualitative description of the results; a quantitative analysis of the results is presented \citep{So2014}. We present a summary and conclusions in Sec. 6.

\section{Mathematical Formulation}
\label{sec:math_formulation}


We solve the coupled equations of multispecies gas dynamics, dark matter dynamics, self-gravity, 
primordial gas chemistry, radiative transfer, and gas cooling/heating in a comoving volume
of the expanding universe. In this paper we assume the governing equations are discretized on a cubic uniform Cartesian mesh  in comoving 
coordinates assuming periodic boundary conditions. In Reynolds et
al. (in prep.) we generalize our method to adaptive meshes. 
The background spacetime is assumed to be a FRW model with $\Lambda$CDM cosmological parameters \citep{WMAP7}. In this work we consider only the 5 ionic states of H and He and $e^-$; i.e., the commonly used ``6-species" model of primordial gas \citep{AbelEtAl1997, Anninos97}. Molecular hydrogen chemistry is ignored as we are primarily concerned with the later stages of H reionization driven by star formation in atomic line cooling galaxies. Star formation is modeled phenomenologically through a subgrid model described in the next section. Newly formed stars are sources of UV radiation, and the radiation is transported in the grey flux-limited diffusion approximation. Star formation in spatially distributed galaxies thus sources an inhomogeneous and evolving ionizing radiation field, which is used to calculate the local ionization and thermal state of the gas. This in turn controls the local cooling rate of the gas, and by virtue of the subgrid star formation model, the local star formation rate. We thus have a closed system of equations that we can evolve forward in time subject to the choice of initial conditions. In all but the verification test problems, cosmological initial conditions are generated using standard methods.  

The choice of flux-limited diffusion (FLD) is motivated by its
simplicity and its ability to smoothly transition between optically
thin and thick regimes. Its properties as well as its limitations are
well understood, and efficient numerical methods exist for parallel
computation (e.g., Hayes et al. 2006). A second motivation is that we
are interested in modeling reionization in large cosmological volumes
and field-based solvers scale independently of the number of sources, unlike ray tracing
methods. In the early stages of reionization, when \hii regions are
largely isolated, FLD provides accurate I-front speeds, as shown by
our verification tests in Sec. 4.1. At late times, during and after
overlap, the gas is bathed in a diffuse radiation field arising from
numerous point sources for which the angular structure of the
radiation field is unimportant. It is during the early percolation
phase when several \hii regions merge that FLD is inaccurate with
regard to the angular distribution of the radiation field. This leads
to some inaccuracies of the shapes of the I-fronts compared to a
solution obtained using ray tracing (see Sec. 4.3). However we
consider these shape differences of secondary importance since we are
interested in globally averaged ionization properties. 

A well known
limitation of FLD is that opaque blobs do not cast shadows if they are illuminated from one side (e.g.,
Hayes \& Norman 2003). Instead, the radiation flows around the
backside of the irradiated blob. By contrast, a ray tracing method
will cast a sharp shadow (Iliev et al. 2009, Wise \& Abel 2011). What
matters for global reionization simulations however is how long the
opaque blobs remain self-shielded; i.e., their photoevaporation
times. We have compared the photoevaporation times for identically
resolved blobs using FLD and the ray tracing method of Wise \& Abel
(2011), and find them comparable despite the inability of FLD to cast
a shadow (sec Sec.~\ref{subsubsec:test7}). 

Finally we comment on the use of a {\em grey} treatment of the radiation field. 
Grey FLD is an improvement over monochromatic radiative transfer as it provides a formalism for calculating the contributions of higher energy photons above the ionization threshold to the frequency-integrated photoionization rate and photoheating rate. It is not as accurate as multifrequency/multigroup radiative transfer in that it does not model spectral hardening of the radiation field and preionization ahead of the I-front.

We consider the coupled system of partial differential equations
\citep{ReynoldsHayesPaschosNorman2009},
\begin{align}
  \label{eq:gravity}
  \nabla^2 \phi &= \frac{4\pi g}{a}(\rhob + \rho_{dm} - \langle \rho \rangle), \\
  \label{eq:cons_mass}
  \partial_t \rhob + \frac1a \vb \cdot \nabla
    \rhob &= -\frac1a \rhob \nabla\cdot\vb - \dot{\rho}_{SF}, \\
  \label{eq:cons_momentum}
  \partial_t \vb + \frac1a\(\vb\cdot\nabla\)\vb &=
    -\frac{\dot{a}}{a}\vb - \frac{1}{a\rhob}\nabla p - \frac1a
    \nabla\phi, \\
  \label{eq:cons_energy}
  \partial_t e + \frac1a\vb\cdot\nabla e &=
    - \frac{2\dot{a}}{a}e
    - \frac{1}{a\rhob}\nabla\cdot\left(p\vb\right) 
    - \frac1a\vb\cdot\nabla\phi + G - \Lambda  + \dot{e}_{SF} \\
  \label{eq:chemical_ionization}
  \partial_t \mn_i + \frac{1}{a}\nabla\cdot\(\mn_i\vb\) &=
    \alpha_{i,j} \mn_e \mn_j - \mn_i \Gamma_{i}^{ph}, \qquad
    i=1,\ldots,N_s \\
  \label{eq:cons_radiation}
  \partial_t E + \frac1a \nabla\cdot\(E \vb\) &= 
    \nabla\cdot\(D\nabla E\) - \frac{\dot{a}}{a}E - c \kappa E + \eta.
\end{align}
The comoving form of Poisson's equation \eqref{eq:gravity} is used to
determine the modified gravitational potential, $\phi$, 
where $g$ is the gravitational constant, $\rhob$ is the comoving
baryonic density, $\rho_{dm}$ is the dark matter density, and 
$\langle \rho \rangle$ is the cosmic mean density.  The collisionless
dark matter density $\rho_{dm}$ is evolved using the Particle-Mesh
method, as described in
\cite{HockneyEastwood1988,NormanBryan1999,BryanEtAl2014}.  The
conservation equations \eqref{eq:cons_mass}-\eqref{eq:cons_energy}
correspond to the compressible Euler equations in a comoving
coordinate system \cite{BryanEtAl1995}.  These relate the density to
the proper peculiar baryonic velocity $\vb\equiv a(t)\dot{\xvec}$, the
proper pressure $p$, and the total gas energy per unit mass $e$.   
The equations \eqref{eq:chemical_ionization} model
ionizaton processes between the chemical species HI, HII, HeI, HeII,
HeIII and the electron density.  Here, $\mn_i$ denotes the
$i^{th}$ comoving elemental species number density, $\mn_e$ is the
electron number density, $\mn_j$ corresponds to ions that react
with the species $i$, and $\alpha_{i,j}$ are the reaction rate
coefficients defining these interactions
\citep{AbelEtAl1997,HuiGnedin1997}.  The equation
\eqref{eq:cons_radiation} describes the flux-limited diffusion (FLD)  
approximation of radiation transport in a cosmological medium
\citep{HayesNorman2003,Paschos2005}, where $E$ is the comoving grey
radiation energy density.  Within this equation, the function
$D$ is the {\em flux limiter} that depends on face-centered values of
$E$, $\nabla E$ and the opacity $\kappa$ \citep{Morel2000},
\begin{align}
  D &= \min\left\{c \(9\kappa^2 + R^2\)^{-1/2}, D_{max}\right\},\quad\mbox{and}\quad
  R = \max\left\{\frac{|\partial_{x} E|}{E},R_{min}\right\}.
\end{align}
Here the spatial derivative within $R$ is computed using
non-dimensional units at the computational face adjoining two
neighboring finite-volume cells, $D_{max}=0.006\,c\,L_{unit}$ and
$R_{min}=10^{-20}/L_{unit}$ with $L_{unit}$ the length
non-dimensionalization factor for the simulation, and the 
face-centered radiation energy density and opacity are computed using
the arithmetic and harmonic means, respectively,
\[
   E = \frac{E_1 + E_2}{2}, \qquad
   \kappa = \frac{2\kappa_1 \kappa_2}{\kappa_1 + \kappa_2}.
\]
Among the many available limiter formulations we have tested
(\citep{HayesNorman2003,Morel2000,ReynoldsHayesPaschosNorman2009}), this 
version performs best at producing causal radiation propagation
speeds in the low-opacity limit typical of the late stages of reionization simulations.

Cosmic expansion for a smooth homogeneous background is modeled by the
function $a(t)\equiv(1+z)^{-1}$ 
, where the redshift $z$ is a function
of time. $a(t)$ is obtained from a solution of the Friedmann equation
for the adopted cosmological parameters. All comoving densities $\rho_i$ relate to the proper
densities through $\rho_i \equiv \rho_{i,\text{proper}}a(t)^3$. All
spatial derivatives are taken with respect to the comoving position
$\xvec\equiv{\bf r}/a(t)$.  We use a standard ideal gas equation of
state to close the system,
\begin{equation}
\label{eq:eos}
  e = \frac{p}{2 \rhob/3} + \frac12|\vb|^2.
\end{equation}

\subsection{Model Coupling}
\label{subsec:equation_coupling}

The equations \eqref{eq:gravity}-\eqref{eq:cons_radiation} are coupled
through a variety of physical processes.   In defining our grey
radiation energy density $E$, we allow specification of an assumed
spectral energy distribution (SED), $\chi_E(\nu)$.  Here, we write
the frequency-dependent radiation density using the
decomposition $\Enu(\xvec,t,\nu)=\tilde{E}(\xvec,t)\,\chi_E(\nu)$.  This
relates to the grey radiation energy density $E$ through the equation
\begin{equation}
\label{eq:grey_definition}
  E(\xvec,t) = \int_{\nu_1}^{\infty} \Enu(\xvec,t,\nu)\,\mathrm d\nu =
  \tilde{E}(\xvec,t) \int_{\nu_1}^{\infty} \chi_E(\nu)\,\mathrm d\nu,
\end{equation}
where $\tilde{E}$ is an intermediate quantity that is never computed.
We note that this relationship is valid only if the indefinite
integral of $\chi_E(\nu)$ exists, as is the case for quasar and
stellar type spectra.  Implemented in {\em Enzo} are a variety of user-selectable
SEDs including black body, monochromatic, and powerlaw (some of these
are used for the verification tests; see Sec. 4.2). 
In our application to cosmic reionization, we utilize the SED for low
metallicity Pop II stars from \cite{RicottiEtAl2002}.

With this in place, we define the radiation-dependent photoheating
and photoionization rates \citep{Osterbrock1989},
\begin{align}
  \label{eq:photoheating}
  G &= \frac{c E}{\rho_b} \sum_i^{N_s} \mn_i \left[\int_{\nu_i}^{\infty}
    \sigma_i(\nu) \chi_E(\nu)\left(1-\frac{\nu_i}{\nu}\right)\,\mathrm
    d\nu\right]   \bigg /
  \left[\int_{\nu_1}^{\infty} \chi_E(\nu)\,\mathrm d\nu\right], \\
  \label{eq:photoionization}
  \Gamma_i^{ph} &= \frac{c E}{h} \left[\int_{\nu_i}^{\infty}
    \frac{\sigma_i(\nu) \chi_E(\nu)}{\nu}\,\mathrm d\nu \right]  \bigg /
  \left[\int_{\nu_1}^{\infty} \chi_E(\nu)\,\mathrm d\nu\right].
\end{align}
Here, $\sigma_i(\nu)$ is the ionization cross section for the species 
$\mn_i$, $h$ is Planck's constant, and $\nu_i$ is the frequency
ionization threshold for species $\mn_i$ ($h\nu_{HI}$ = 13.6 eV, 
$h\nu_{HeI}$ = 24.6 eV, $h\nu_{HeII}$ = 54.4 eV).

In addition, gas cooling due to chemical processes occurs through the
rate $\Lambda$ that depends on both the chemical number densities 
and current gas temperature \citep{AbelEtAl1997,Anninos97},
\begin{equation}
\label{eq:temperature}
  T = \frac{2\, p\,\mu\, m_p}{3\, \rhob\, k_b},
\end{equation}
where $m_p$ corresponds to the mass of a proton, $\mu$ corresponds to
the local molecular weight, and $k_b$ is Boltzmann's constant.
In addition, the reaction rates $\alpha_{i,j}$ are highly temperature
dependent \citep{AbelEtAl1997,HuiGnedin1997}. 
The opacity $\kappa$ depends on the local ionization states $\mn_i$
and the assumed SED $\chi_E$, 
\begin{align}
  \label{eq:opacity}
  \kappa &= \sum_{i=1}^{N_s} \mn_i \left[ \int_{\nu_i}^{\infty}
    \sigma_i(\nu) \chi_E(\nu)\,\mathrm d\nu \right]  \bigg /
  \left[\int_{\nu_i}^{\infty} \chi_E(\nu)\,\mathrm d\nu\right].
\end{align}
The emissivity $\eta$ is based on a star-formation ``recipe'' described below.

\section{Numerical Method}
\label{sec:numerical}


\subsection{The Enzo Code}

Our radiation hydrodynamical cosmology is built on top of the publicly available hydrodynamic cosmology code {\em Enzo} ({\tt enzo-project.org}), whose numerical methods have been documented elsewhere \citep{OSheaEtAl2004,NormanEtAl2007,BryanEtAl2014}. Here we provide a brief summary. 
The basic {\em Enzo} code couples an N-body particle-mesh (PM) solver,
which is used to follow the evolution of collisionless dark matter, with an
Eulerian adaptive mesh refinement (AMR) method for ideal gas dynamics. 
Dark matter is assumed to behave as a collisionless phase fluid, obeying the Vlasov-Poisson equation. We use the second order-accurate Cloud-In-Cell (CIC) formulation, together with leapfrog time integration, which is formally second order-accurate in time. 
{\em Enzo} hydrodynamics utilizes the piecewise parabolic method (PPM) \citep{ColellaWoodward1984}
to evolve the mass density field for each chemical species of interest assuming a common velocity field (i.e., multispecies hydrodynamics.) PPM is formally second order-accurate in space and time. 
The gravitational potential is computed by solving the Poisson
equation on the uniform Cartesian grid using 3D FFTs. When AMR is
employed (which is not the case in this work), the subgrid
gravitational potential is computed using a local multigrid solve of
the Poisson equation with boundary conditions supplied from the parent
grid.  

The non-equilibrium chemical and cooling properties of primordial (metal-free) gas
are determined using optional 6--, 9--, and 12--species models; in this work we restrict ourselves
to the 6-species model involving 
$H$, $H^+$, $He$, $He^+$, $He^{++}$, and $e^-$.  This reaction network
results in a stiff set of rate equations which are solved with the
first-order semi-implicit method described in \cite{Anninos97}, or a
new second-order semi-analytic method described below.  
{\em Enzo} also calculates radiative heating and cooling following atomic line excitation, recombination,
collisional excitation, free-free transitions, molecular line cooling, and Compton
scattering of the cosmic microwave background as well as different models for a metagalactic
ultraviolet background that heats the gas via photoionization and/or photodissociation. 

To this we add our flux-limited diffusion radiation transport solver, which is solved using an optimally scalable geometric multigrid algorithm detailed here. When simulating inhomogeneous reionization, the metagalactic UV radiation field is solved for directly as a function of position and time, rather than input to the code as an externally-evaluated homogeneous background (e.g., \cite{HaardtMadau2012}). 

\subsection{Star Formation and Feedback}
\label{subsec:starform}

Because star formation occurs on scales not resolved by our uniform mesh simulation, 
we rely on a subgrid model which we calibrate to observations of star formation in high
redshift galaxies. The subgrid model is a variant of the Cen \& Ostriker (1992)
prescription with two important modifications as described in Smith et al. (2011). In the original Cen \& Ostriker recipe, a computational cell forms a collisionless ``star particle" if a number of criterial are met: the baryon density exceeds a certain numerical threshold; the gas velocity divergence is negative, indicating collapse; the local cooling time is less than the dynamical time; and the cell mass exceeds the Jeans mass. In our implementation, the last criterion is removed because it is always met in large scale, fixed-grid simulations, and the overdensity threshold is taken to be $\rho_b/(\rho_{c,0}(1+z)^3) > 100$, where $\rho_{c,0}$ is the critical density at z=0. If the three remaining criteria are met, then a star particle representing a large collection of stars is formed in that timestep and grid cell with a total mass

\begin{equation}
m_* = f_* m_{cell} \frac{\Delta t}{t_{dyn}},
\end{equation}
where $f_*$ is an efficiency parameter we adjust to match observations of the cosmic star formation rate density (SFRD) \citep{Bouwens11}, $m_{cell}$ is the cell baryon mass, $t_{dyn}$ is the dynamical time of the combined baryon and dark matter fluid, and $\Delta t$ is the hydrodynamical timestep. An equivalent amount of mass is removed from the grid cell to maintain mass conservation. 

Although the star particle is formed instantaneously (i.e., within one timestep), the conversion of removed gas into stars is assumed to proceed over a longer timescale, namely $t_{dyn}$, which more accurately reflects the gradual process of star formation. In time $\Delta t$, the amount 
of mass from a star particle converted into newly formed stars is given by

\begin{equation}
\Delta m_{SF} = m_* \frac{\Delta t}{t_{dyn}} \frac{t-t_*}{t_{dyn}} e^{-(t-t_*)/t_{dyn}},
\end{equation}
where $t$ is the current time and $t_*$ is the formation time of the star particle. To make the 
connection with Eq. 4, we have $\dot{\rho}_{SF} =\Delta m_{SF}/(V_{cell}\Delta t)$, 
where $V_{cell}$ is the volume of the grid cell. 

Stellar feedback consists of the injection of thermal energy, gas, metals, and radiation
to the grid, all in proportion to $\Delta m_{SF}$. The thermal energy $\Delta e_{SF}$, gas
mass $\Delta m_g$, and metals $\Delta m_Z$ returned to the grid are given by

\begin{equation}
  \Delta e_{SF} = \Delta m_{SF} c^2 \epsilon_{SN}, \qquad
  \Delta m_g = \Delta m_{SF} f_{m*}, \qquad
  \Delta m_Z = \Delta m_{SF} f_{Z*},
\end{equation}

where $c$ is the speed of light, $\epsilon_{SN}$ is the supernova energy efficiency parameter, and $f_{m*}=0.25, f_{Z*}=0.02$ is the fraction of the stellar mass returned to the grid as gas and metals, respectively. Rather than add
the energy, gas, and metals to the cell containing the star particle, as was done in
the original Cen \& Ostriker (1992) paper, we distribute it evenly among the cell and its
26 nearest neighbors to prevent overcooling. As shown by Smith et al. (2011), this 
results in a star formation recipe which can be tuned to reproduce the observed SFRD. This is critical for us, as we use the observed high redshift SFRD to calibrate our reionization simulations. 

To calculate the radiation feedback, we define an emissivity field $\eta(x)$ on the grid which accumulates
the instantaneous emissivities $\eta_i(t)$ of all the star particles within each cell. To calculate the contribution of each star particle $i$ at time $t$ we assume an equation of the same form for supernova energy feedback, but with a different energy conversion efficiency factor $\epsilon_{UV}$. Therefore

\begin{equation}
\label{eq:emissivity}
  \eta= \sum_\mathrm{i}\epsilon_\mathrm{uv}\frac{\Delta m_\mathrm{SF} c^2}{V_\mathrm{cell}\Delta t}
\end{equation}

Emissivity $\eta$ is in units of erg/s/cm$^3$.   The UV efficiency factor $\epsilon_\mathrm{uv}$ is taken from \cite{RicottiEtAl2002} as 4$\pi\times 1.1 \times 10^{-5}$, where the factor $4\pi$ comes from the conversion from mean intensity to radiation energy density.

\subsection{Operator Split Solution Procedure}

We implement the model \eqref{eq:gravity}-\eqref{eq:cons_radiation} in
the open-source community cosmology code, {\em Enzo} \citep{BryanEtAl2014}.  This
simulation framework utilizes a method-of-lines approach, in which
space and time are discretized separately.  To this end, we use a
finite-volume spatial discretization of the modeling equations.  For
this study, all of our simulations were run in {\em unigrid} mode, so
that the cosmological volume is discretized using a regular grid.
Although {\em Enzo} was built to enable block-structured adaptive mesh
refinement (AMR) using a standard Berger-Colella formalism
\citep{BergerColella89}, that mode does not currently allow as
extreme parallel scalability as the unigrid version.  Due to our
desire to simulate very large cosmological volumes for understanding
reionization processes, this scalability was paramount.

We discretize in time using an operator split time-stepping approach,
wherein separate components are treated with solvers that have been
tuned for their specific physics. To this end, we break apart the
equations into four distinct components.  The first component
corresponds to the self-gravity equation \eqref{eq:gravity}, 
\begin{equation}
\label{eq:self_gravity}
  \nabla^2 \phi = \frac{4\pi g}{a}(\rhob + \rho_{dm} - \langle \rho
  \rangle), 
\end{equation}
that solves for the instantaneous gravitational potential $\phi$,
which contributes to sources in the momentum and energy conservation
equations.  We perform this solve using our own 3D Fast Fourier Transform
solver built on the publicly available FFTE library.
These solves take as sources
the gridded baryon density and dark matter density fields $\rho_b$
and $\rho_{dm}$. The former is defined as a grid based Eulerian field. 
The latter is computed from the dark matter particle positions $\xvec_i^n$
using the CIC mass assignment algorithm \citep{HockneyEastwood1988}. 

The second component in our splitting approach corresponds to the
cosmological Euler equations, along with passive advection of other
comoving density fields,
\begin{align}
  \notag
  \partial_t \rhob + \frac1a \vb \cdot \nabla
    \rhob &= -\frac1a \rhob \nabla\cdot\vb, \\
  \notag
  \partial_t \vb + \frac1a\(\vb\cdot\nabla\)\vb &=
    -\frac{\dot{a}}{a}\vb - \frac{1}{a\rhob}\nabla p - \frac1a
    \nabla\phi, \\
 \label{eq:hydro}
  \partial_t e + \frac1a\vb\cdot\nabla e &=
    - \frac{2\dot{a}}{a}e
    - \frac{1}{a\rhob}\nabla\cdot\left(p\vb\right) 
    - \frac1a\vb\cdot\nabla\phi, \\
  \notag
  \partial_t E + \frac1a \nabla\cdot\(E \vb\) &= 0, \\
  \notag
  \partial_t \mn_i + \frac{1}{a}\nabla\cdot\(\mn_i\vb\) &=
    0, \qquad i=1,\ldots,N_s.
\end{align}
We point out that the above energy equation does not include
photo-heating, chemical cooling, or supernova feedback processes,
which are included in subsequent components.  These equations are
solved explicitly using the {\em Piecewise Parabolic Method}
\citep{ColellaWoodward1984}, to properly track hydrodynamic shocks,
while obtaining second-order accuracy away from shock
discontinuities. 

The third solver component corresponds to the grey radiation energy
equation \eqref{eq:cons_radiation},
\begin{equation}
\label{eq:fld}
  \partial_t E = \nabla\cdot\(D\nabla E\) - \frac{\dot{a}}{a}E - 
  c \kappa E + \eta. 
\end{equation}
Our solver for this component is based on the algorithm described
in \cite{ReynoldsHayesPaschosNorman2009} with a modified timestepping algorithm than what is described there. Specifically,
since the time scale for radiation transport is much faster than for
hydrodynamic motion, we use an implicit $\theta$-method for time
discretization, allowing both backwards Euler and trapezoidal
implicit quadrature formulas.  Moreover, we evaluate the limiter $D$
using the previous-time solution, $E^n$ when calculating the
time-evolved solution, $E^{n+1}$.  Under these approximations, our
implicit FLD approximation for the radiative transport results in a
linear system of equations over the computational domain, as opposed
to a nonlinear system of equations, as used in our previous work
\citep{NormanEtAl2007,ReynoldsHayesPaschosNorman2009,NormanReynoldsSo2009}.
This linear system is posed in residual-correction form, in which we
solve for the change in the radiation field, $\delta E = E^{n+1}-E^n$,
over the course of a time step.  To solve this linear system, we
employ a multigrid-preconditioned conjugate gradient solver from the
{\em hypre} library \citep{hypre-site}, that allows optimal $\mathcal O(n\log n)$ 
parallel scalability to the extents of modern supercomputer 
architectures. Specific parameters used in this solve are found in
Table \ref{table:HYPRE_params}.
\begin{table}
\caption{Parameters used in the {\em hypre} linear solver}
\label{table:HYPRE_params}
\centerline{
\begin{tabular}{p{5cm}p{6cm}}
\hline\noalign{\smallskip}
Parameter & Value  \\
\hline
Outer Solver & PCG \\
CG iterations & 50 \\
CG tolerance & $10^{-8}$ \\
Inner Preconditioner & PFMG \\
MG iterations & 12 \\
MG relaxation type & nonsymmetric Red/Black Gauss-Seidel \\
MG pre-relaxation sweeps & 1 \\
MG post-relaxation sweeps & 1 \\
\hline
\end{tabular}}
\end{table}

The fourth physical component within our operator-split formulation
corresponds to photoionization, photoheating, chemical ionization and
gas cooling processes,
\begin{align}
  \label{eq:heat_chem}
  \partial_t e &= G - \Lambda, \\
  \notag
  \partial_t \mn_i &= \alpha_{i,j} \mn_e \mn_j - \mn_i
  \Gamma_{i}^{ph}, \qquad i=1,\ldots,N_s.
\end{align}
Since these processes occur on time scales commensurate with the
radiation transport, and much faster than hydrodynamic motion, they
are also solved implicitly in time, using adaptive-step,
time-subcycled solves of these spatially-local processes.  
We have two different algorithms for solving these equations.  The
first, more loosely coupled, solver uses a single Jacobi
iteration of a linearly-implicit backwards Euler discretization for
each species in each cell.  Although this solver does not attempt to
accurately resolve the nonlinearity in these equations, nor does it
iterate between the different species in each cell to achieve a fully
self-consistent solution, its adaptive time stepping strategy enables
this single iteration to achieve results that are typically accurate
to within 10\% relative error,
and results in highly efficient calculations.  

Our second solver for the system \eqref{eq:heat_chem} 
approximates the equations using an implicit quasi-steady-state
formulation, in which the source terms for the energy equation assume
a fixed ionization state $(\mn_i^{n-1} + \mn_i^n)/2$, and the
chemistry equations assume a fixed energy $(e^{n-1}+e^n)/2$ when
evolving the time step $t^{n-1}\to t^n$.  Under this
quasi-steady-state approximation, we solve the resulting set of
differential equations analytically, to obtain the new values $e^n$
and $\mn_i^n$.  However, since these updated solutions implicitly
contribute to the source terms for one another, we wrap these
analytical solvers within a nonlinear Gauss-Seidel iteration to
achieve full nonlinear convergence.  As a result of this much tighter
coupling between the gas energy and chemical ionization, this solver
is more expensive per time step, but may result in a more accurate and
stable solution than the more loosely-split algorithm.

The fifth solver component computes star formation and feedback processes,
and evaluates the emissivity field for use in the next step. It corresponds
to integrating the equations
\begin{align}
  \label{eq:SF_mass}
  \partial_t \rhob =  - \dot{\rho}_{SF}, \\
  \label{eq:SF_energy}
  \partial_t e = \dot{e}_{SF}
\end{align}
and evaluating Eq. 18 using the procedures described in Sec. 3.2.

These distinct components are coupled together through the potential
$\phi$ (gravity $\to$ hydrodynamics+DM dynamics), opacity $\kappa$ (chemistry
$\to$ radiation), emissivity $\eta$ (star formation $\to$ radiation),
photoheating $G$ (radiation $\to$ energy), cooling $\Lambda$
(chemistry $\to$ energy), temperature $T$ (energy $\to$ chemistry),
and photoionization $\Gamma_i^{ph}$ (radiation $\to$ chemistry).  Each
of these couplings is handled using one of two mechanisms, direct
manipulation of the solution components ($\Lambda, \kappa, T$), or
filling new fields over the domain containing each term that are
passed between modules ($\nabla\phi, \eta, G, \Gamma_i^{ph}$).

\subsection{Radiation Subcycling}
Since both the radiation \eqref{eq:fld} and chemistry/energy
\eqref{eq:heat_chem} subsystems evolve at similar time scales that are
typically much faster than the hydrodynamic time scale, consistency
between these processes is maintained through an adaptive
time-stepping strategy, wherein the radiation system limits the
overall time step selection strategy, using a conservative time step
to ensure consistency between the physical processes.  This
additionally ensures that each radiation solve only requires
relatively minor corrections as time evolves, resulting in a highly
efficient CG/MG iteration.  The time step estimation algorithm is the
same as in \cite{ReynoldsHayesPaschosNorman2009}, but in the current
work we use the time step tolerance $\tau_{tol} = 10^{-4}$, which
ensures a relative change-per-step in the radiation field of 0.01\%,
when measured in a vector RMS norm.

For increased robustness, we have enabled subcycling within the
radiation solver.  While this technically allows the radiation solver
to subcycle faster than the coupled processes, we only employ this
functionality in time steps where the CG/MG solver fails.  This
situation typically only occurs in the initial step after the first
stars are created.  Prior to star formation the dynamical time scale
due to hydrodynamics and gravity is much longer than the time scales
of radiation transport and chemical ionization after star formation.
Since we adapt our time step estimates using the behavior in previous
steps, our estimation strategy does not predict the abrupt change in
physics when the first stars are created, so the step size estimate from
the previous step is too large, causing the CG/MG solver to diverge.
Once this occurs, the radiation subsystem solver decreases its time
step size and then subcycles to catch up with the overall time step of
the remaining physics.

When using the loosely-coupled ionization/heating solver, the sequence
of these processes within a time step $t^{n-1} \to t^n$ are as follows: 
{\tt
\begin{quote}
Set $t_{hydro}=t_{chem}=t_{rad}=t_{dm} = t^{n-1}$.\\
Set $\Delta t = \min\{\Delta t_{hydro}, \Delta t_{expansion}, \Delta
t_{rad}\}$, and $t^n = t^{n-1}+\Delta t$.\\
While ($t_{rad} < t^n$)
\begin{quote}
  Try to evolve the $E(t)$ according to \eqref{eq:fld}.\\
  If failure, set $\Delta t_{rad} = 0.1\Delta t_{rad}$.\\
  Else set $t_{rad} = t_{rad} + \Delta t_{rad}$ and update $\Delta
  t_{rad}$ based on accuracy estimates.
\end{quote}
Post-process $E(t^n)$ to compute $G$ and $\Gamma_i^{ph}$.\\
Compute $\phi$ using \eqref{eq:self_gravity}, and post-process to
generate $\nabla\phi$.\\ 
Evolve the hydrodynamics sub-system \eqref{eq:hydro}, $t_{hydro} \to
t_{hydro} + \Delta t$.\\
While ($t_{chem} < t^n$)
\begin{quote}
  Set $\Delta t_{chem}$ based on accuracy estimates. \\
  Evolve the chemical and gas energy subsystem \eqref{eq:heat_chem},
  $t_{chem} \to t_{chem} + \Delta t_{chem}$.
\end{quote}
Evolve the dark matter particles, $t_{dm} \to t_{dm} + \Delta t$.\\
Compute $\eta$ using equation \eqref{eq:emissivity}.
\end{quote}
}
When using the tightly-coupled ionization/heating
solver, this sequence of processes differs slightly: 
{\tt
\begin{quote}
Set $t_{hydro}=t_{chem}=t_{rad}=t_{dm} = t^{n-1}$.\\
Set $\Delta t = \min\{\Delta t_{hydro}, \Delta t_{expansion}, \Delta t_{rad}\}$.\\
While ($t_{rad} < t^n$)
\begin{quote}
  Try to evolve the radiation field according to \eqref{eq:fld}.\\
  If failure, set $\Delta t_{rad} = 0.1*\Delta t_{rad}$.\\
  Else
  \begin{quote}
    Set $t_{rad} = t_{rad} + \Delta t_{rad}$ and update $\Delta t_{rad}$
    based on accuracy estimates. \\
    Post-process $E(t_{rad})$ to compute $G$ and $\Gamma_i^{ph}$.\\
    While ($t_{chem} < t_{rad}$)
    \begin{quote}
      Set $\Delta t_{chem}$ based on accuracy estimates. \\
      Evolve the chemical/energy subsystem \eqref{eq:heat_chem},
      $t_{chem} \to t_{chem} + \Delta t_{chem}$.
    \end{quote}
  \end{quote}
\end{quote}
Compute $\phi$ using \eqref{eq:self_gravity}, and post-process to
generate $\nabla\phi$.\\ 
Evolve the hydrodynamics sub-system \eqref{eq:hydro}, $t_{hydro} \to
t_{hydro} + \Delta t$.\\
Evolve the dark matter particles, $t_{dm} \to t_{dm} + \Delta t$.\\
Compute $\eta$ using equation \eqref{eq:emissivity}.
\end{quote}
}


\section{Tests}
\label{sec:tests}

In this section we present three kinds of tests: (1) {\em verification tests}--tests with analytic solutions--that allow us to compare the accuracy of our operator-split method with the unsplit method described in \cite{ReynoldsHayesPaschosNorman2009}; (2) {\em validation tests} chosen for their relevance to the target application; and (3) {\em execution speed tests} where we quantify the relative speed of a large, fully-coupled radiation hydrodynamic simulation with a hydrodynamic simulation with a standard optically thin treatment of photoionization and photoheating.

\subsection{Verification Tests}
\label{subsec:verification}

The radiation, hydrodynamics and chemistry solvers in {\em Enzo} have been
verified in previous work \citep{ReynoldsHayesPaschosNorman2009}, so we
will not focus on the performance of each individual solver here.
However, what is new in this work is our updated coupling strategy
between the radiation transport and chemistry, that unlike the fully
coupled implicit solver in
\cite{NormanEtAl2007,ReynoldsHayesPaschosNorman2009,NormanReynoldsSo2009},
now splits these solvers apart, with coupling instead based on our
adaptive time-stepping strategy.  

To this end, we focus our verification tests in this paper on two
tests with analytical solutions that exercise only the radiation
transport and chemical ionization/recombination components of {\em Enzo}.
These tests were previously described in
\cite{ReynoldsHayesPaschosNorman2009} (sections 4.5 and 4.6); we
summarize them again here.

\subsubsection{Isothermal ionization of a static neutral hydrogen region}
\label{subsec:test1}

This verification test problem, matching Test 1 in
\cite{IlievEtAl2006}, focuses on the expansion of an ionized 
hydrogen (HII) region in a uniform gas surrounding a radiation
source.  The problem is simplified through assumption of a static gas
field, and a fixed temperature.  Under these assumptions, the
emitted radiation should rapidly ionize the nearby hydrogen, and then
this ionized region should propagate spherically outward until it
reaches a terminal radius at which ionizations balance with
recombinations, called the Str{\" o}mgren radius.  The radius of
this ionization front, $r(t)$, may be analytically computed as
\begin{equation}
  \label{Iliev1_solution}
  r(t) = r_s \left(1-e^{-t/t_{rec}}\right)^{1/3}, \quad\mbox{where}\quad
  r_s = \left(\frac{3\,\dot{N}_{\gamma}}{4\pi\,\alpha_B\,n_H^2}\right)^{1/3}.
\end{equation}
Here, $r_s$ is the Str{\" o}mgren radius, $t_{rec} =
(\alpha_B\,n_H)^{-1}$ is the recombination time, $\dot{N}_{\gamma}$ is
the photon emission rate, $n_H$ is the hydrogen number
density of the gas, and $\alpha_B$ is the case B hydrogen
recombination rate.

In our tests, we use parameters $\dot{N}_{\gamma} = 5\times10^{48}$
photons s$^{-1}$, $n_H = 10^{-3}$ cm$^{-3}$, $\alpha_B =
2.59\times10^{-12}$ cm$^2$s$^{-1}$, domain $[0,6.6\, \mbox{kpc}]^3$,
temperature $T=10^4$ K, and time interval $[0,5\, \mbox{Myr}]$.  The
ionization source is assumed to be monochromatic, at the HI ionization
frequency $h\nu = 13.6$ eV, and is located at the location $(0,0,0)$.
For initial conditions, we use $E = 10^{-45}$ erg cm$^{-3}$ and
ionization fraction HII/H = 0.0012.  We employ reflecting boundary
conditions for the radiation field at the $x=0$, $y=0$, and $z=0$
faces, and outflow boundary conditions at the other three faces.

\begin{figure}[t]
\centerline{\hfill
  \includegraphics[scale=0.3, trim=1.0cm 1.0cm 1.0cm 0.5cm]{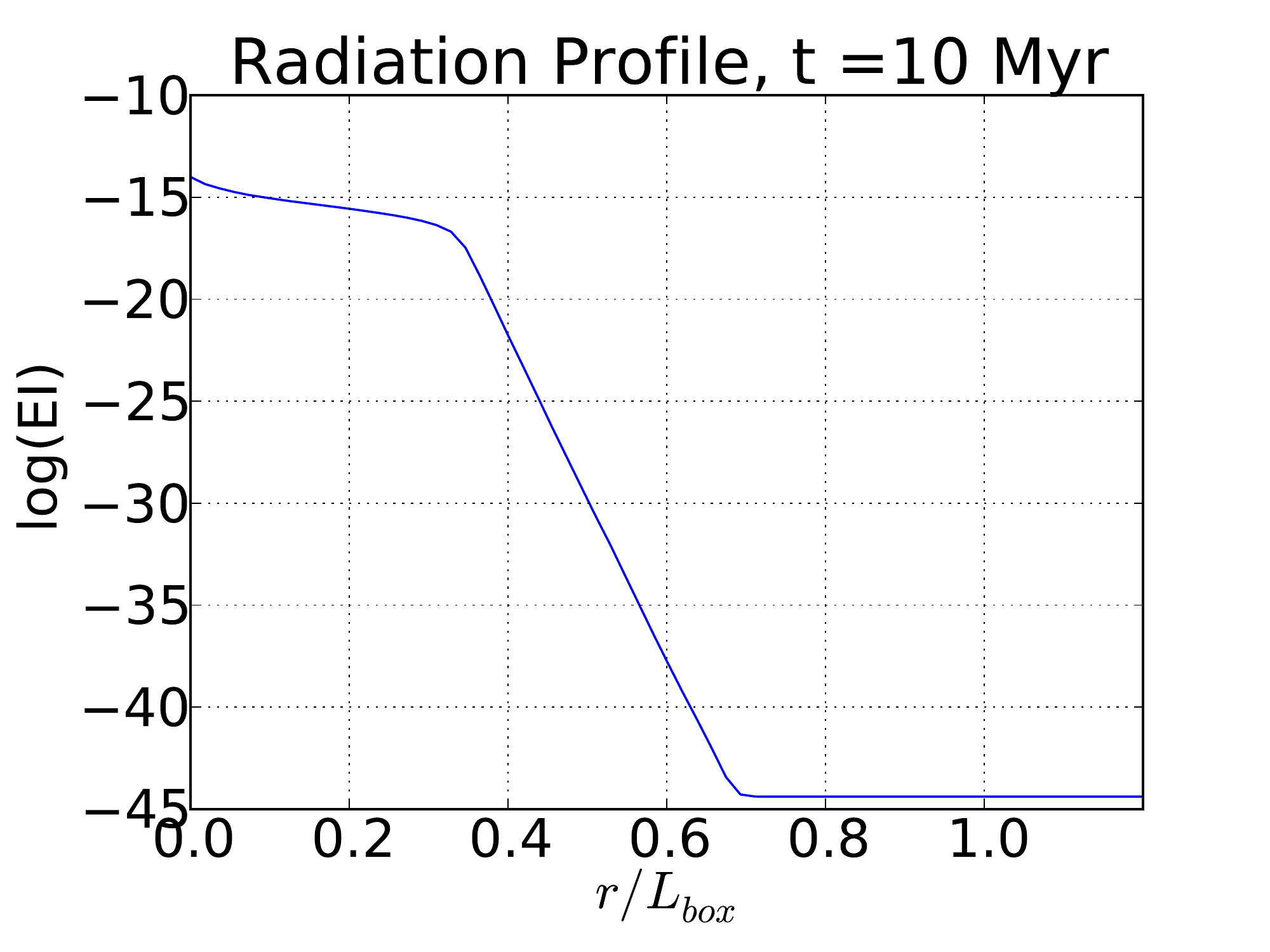}
  \includegraphics[scale=0.3, trim=1.0cm 1.0cm 1.0cm 0.5cm]{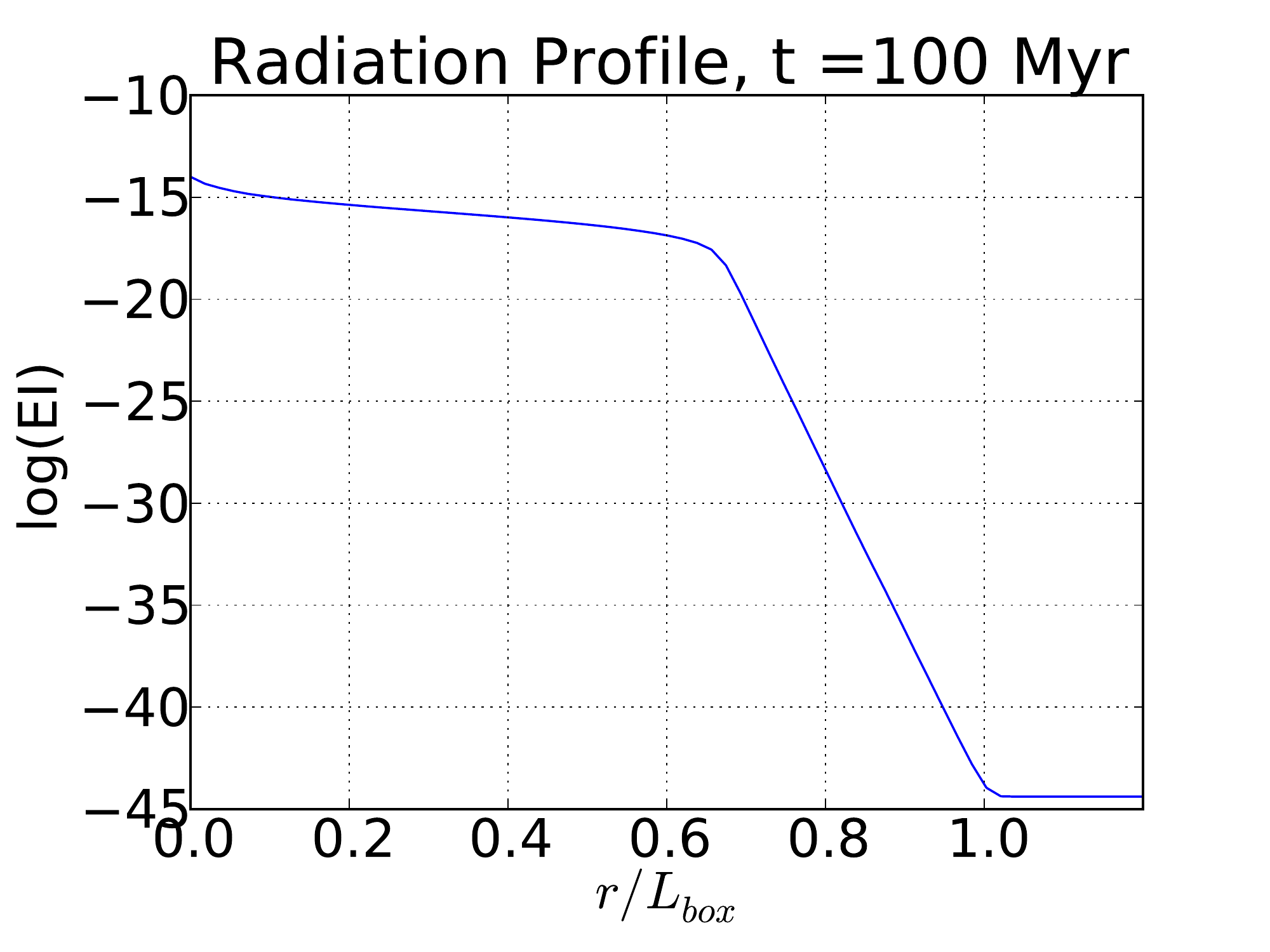}
  \includegraphics[scale=0.3, trim=1.0cm 1.0cm 1.0cm 0.5cm]{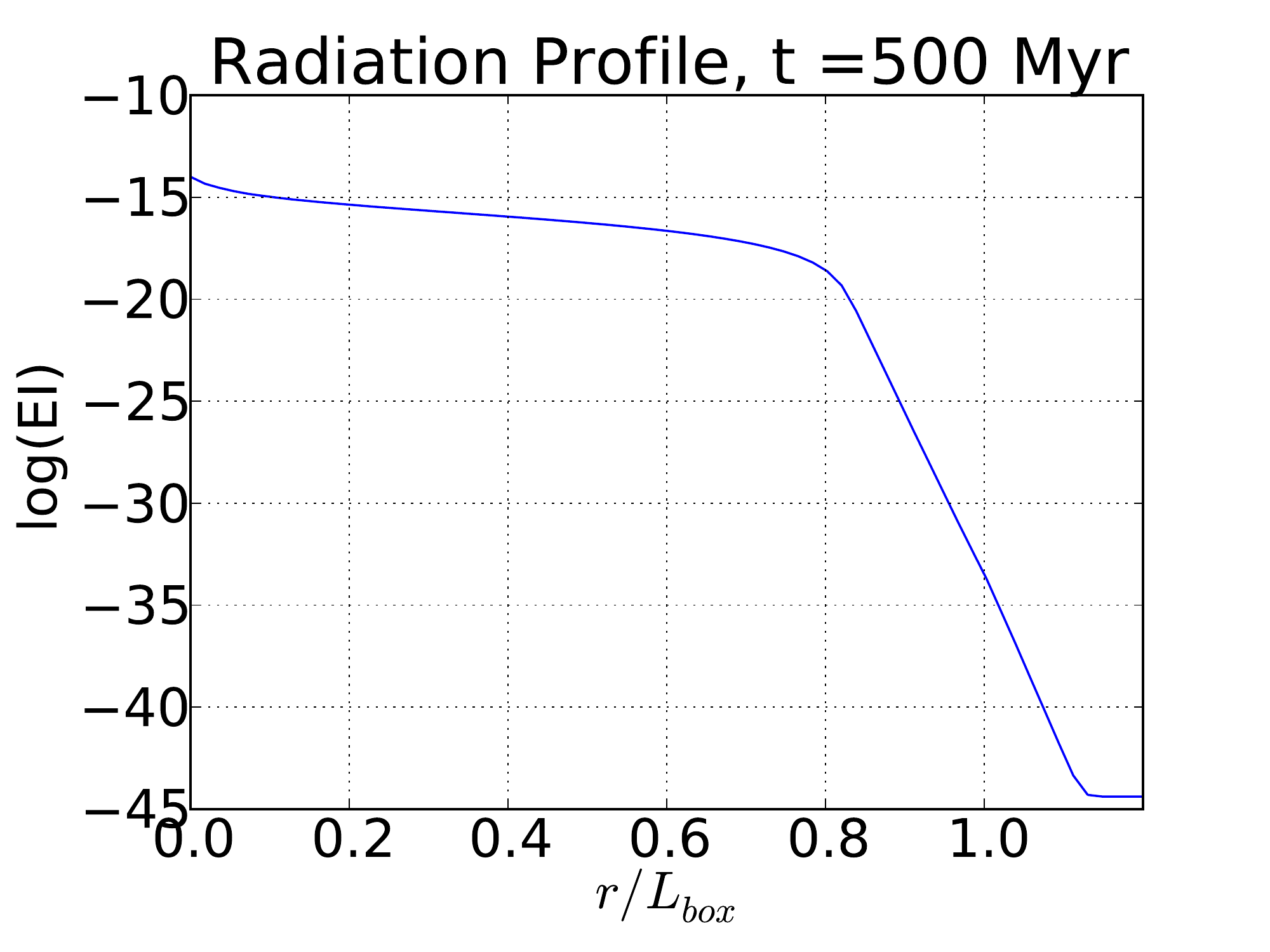}
  \hfill}
\centerline{\hfill
  \includegraphics[scale=0.3, trim=1.0cm 0.5cm 1.0cm 0.5cm]{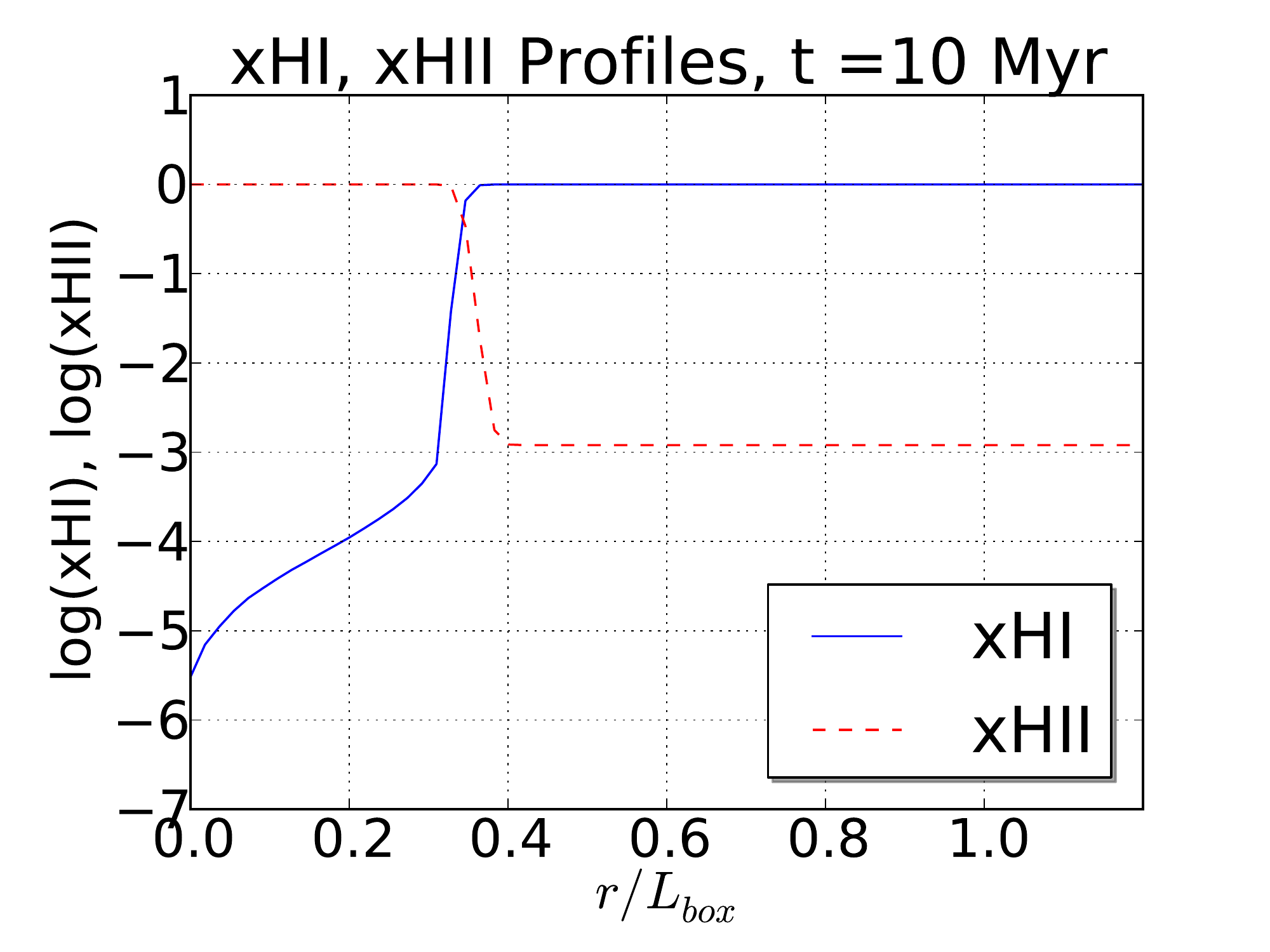}
  \includegraphics[scale=0.3, trim=1.0cm 0.5cm 1.0cm 0.5cm]{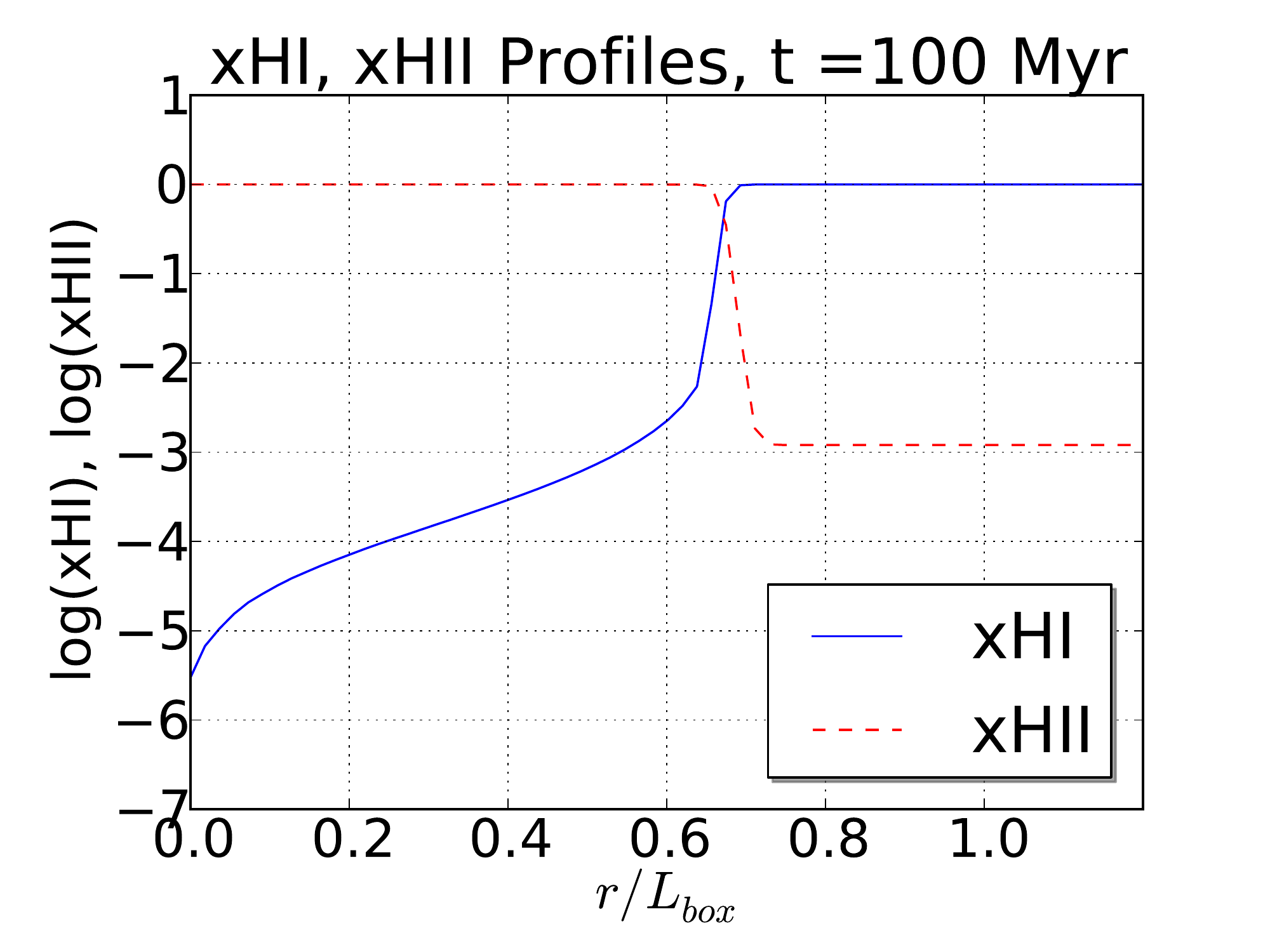}
  \includegraphics[scale=0.3, trim=1.0cm 0.5cm 1.0cm 0.5cm]{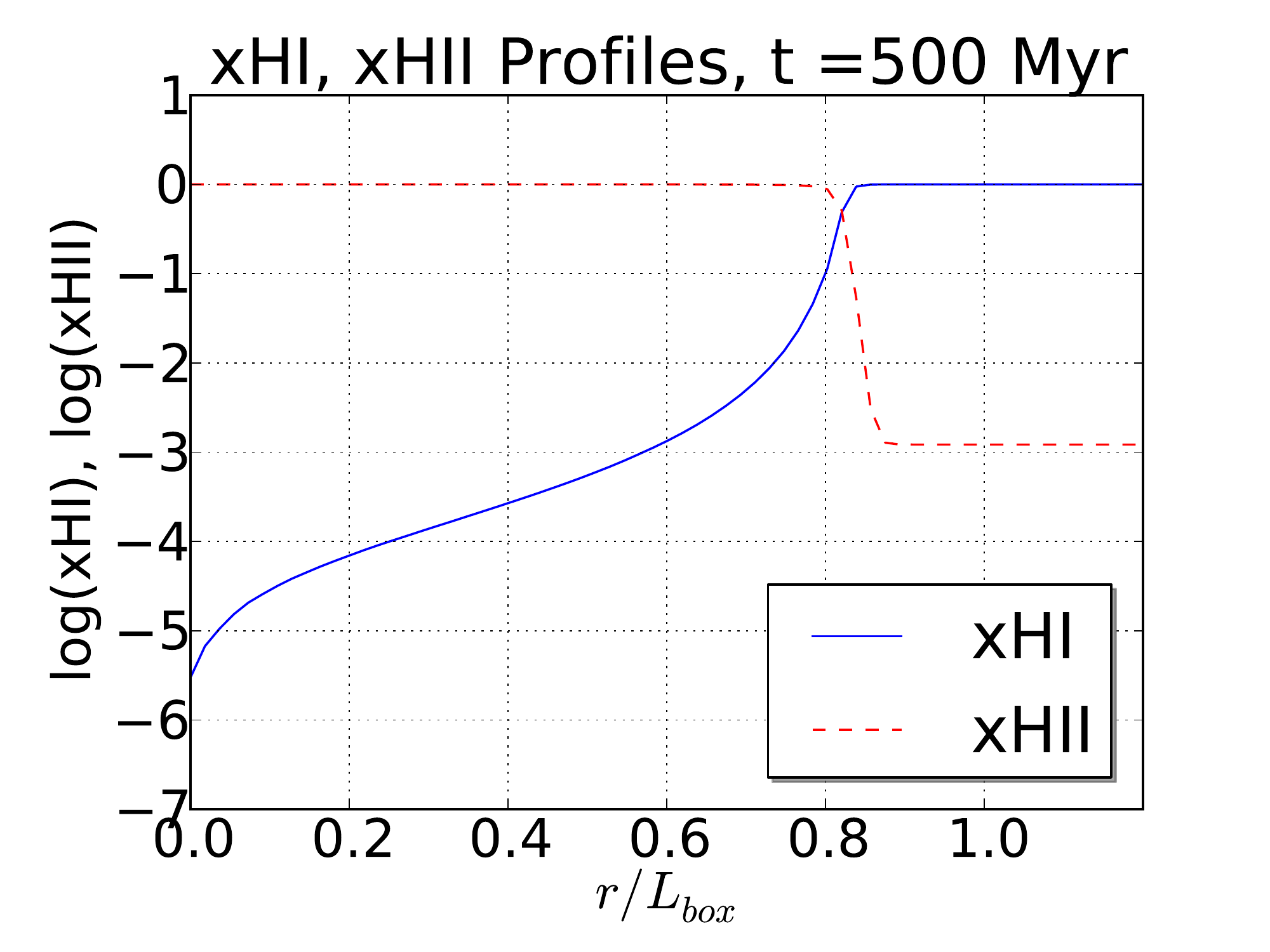}
  \hfill}
  \caption{Spherically-averaged radial profiles of radiation energy
    density and ionization fractions for the isothermal ionization
    test in section \ref{subsec:test1} using a $128^3$ mesh and
    time step tolerance $\tau_{tol} =10^{-4}$.  Plots are shown at 10, 100
    and 500 Myr (left to right), with the radiation energy density on
    the top row and ionization fractions on the bottom row.}
  \label{fig:i1_results}
\end{figure}

\begin{figure}[t]
\centerline{\hfill
  \includegraphics[scale=0.45, trim=1.0cm 0.5cm 1.0cm 0.5cm]{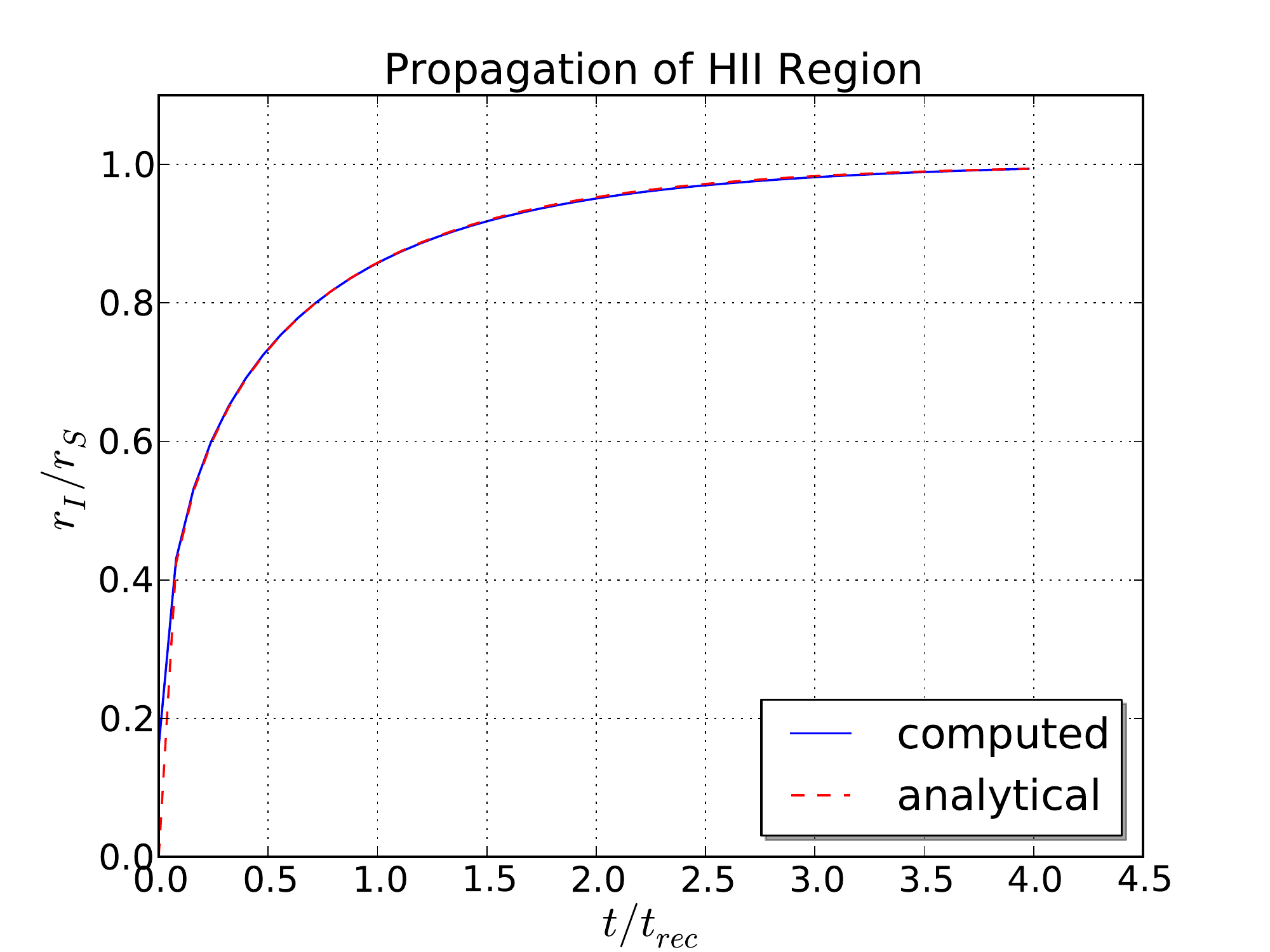}
  \includegraphics[scale=0.45, trim=1.0cm 0.5cm 1.0cm 0.5cm]{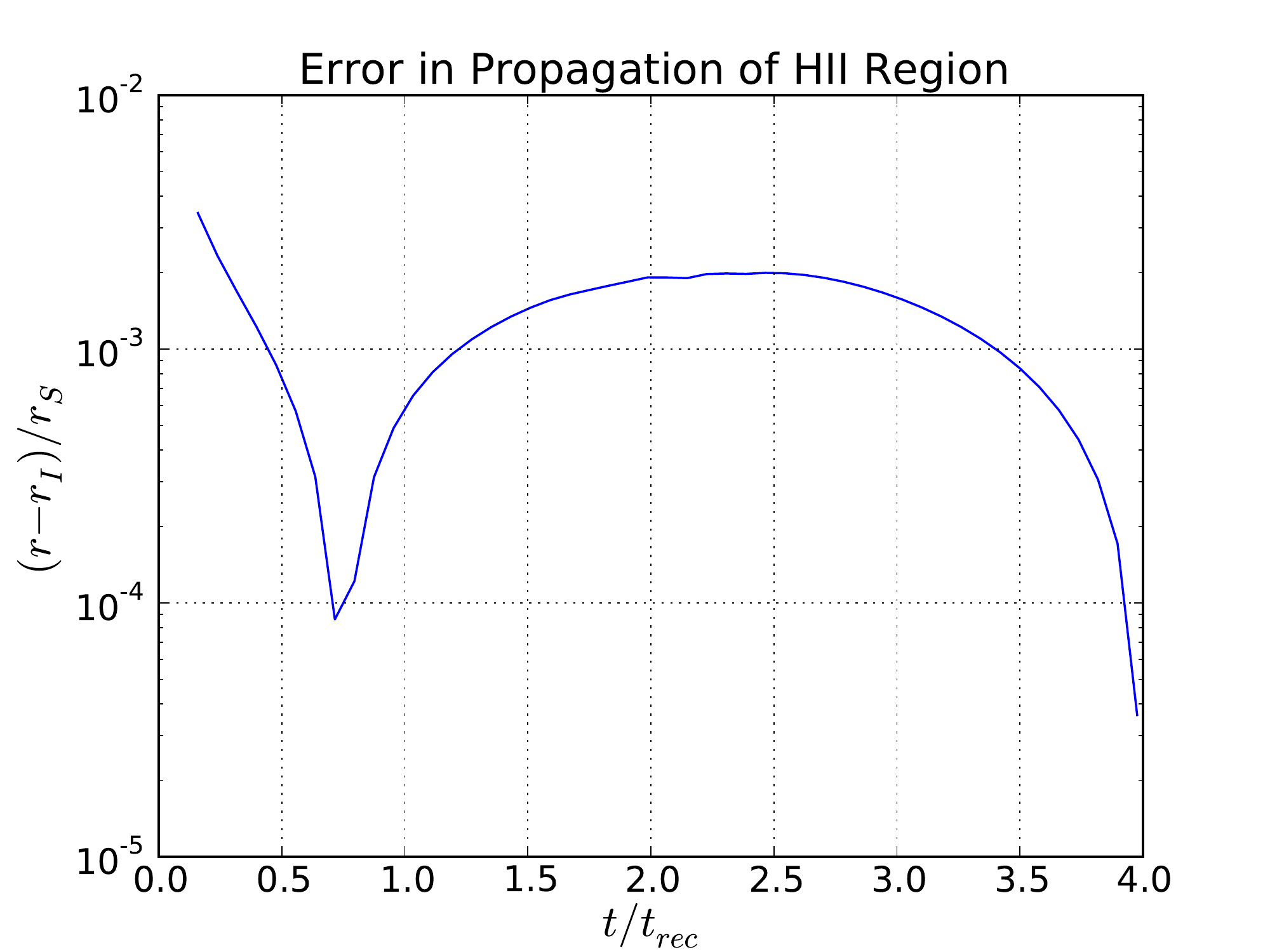}
  \hfill}
  \caption{Comparison between computed and analytical I front position
    for the isothermal ionization test in section \ref{subsec:test1}
    using a $128^3$ mesh and time step tolerance $\tau_{tol}
    =10^{-4}$. Solution on left, error on right.} 
  \label{fig:i1_radius}
\end{figure}

\begin{figure}[t]
\centerline{\hfill
  \includegraphics[scale=0.45, trim=1.0cm 0.0cm 1.0cm 0.5cm]{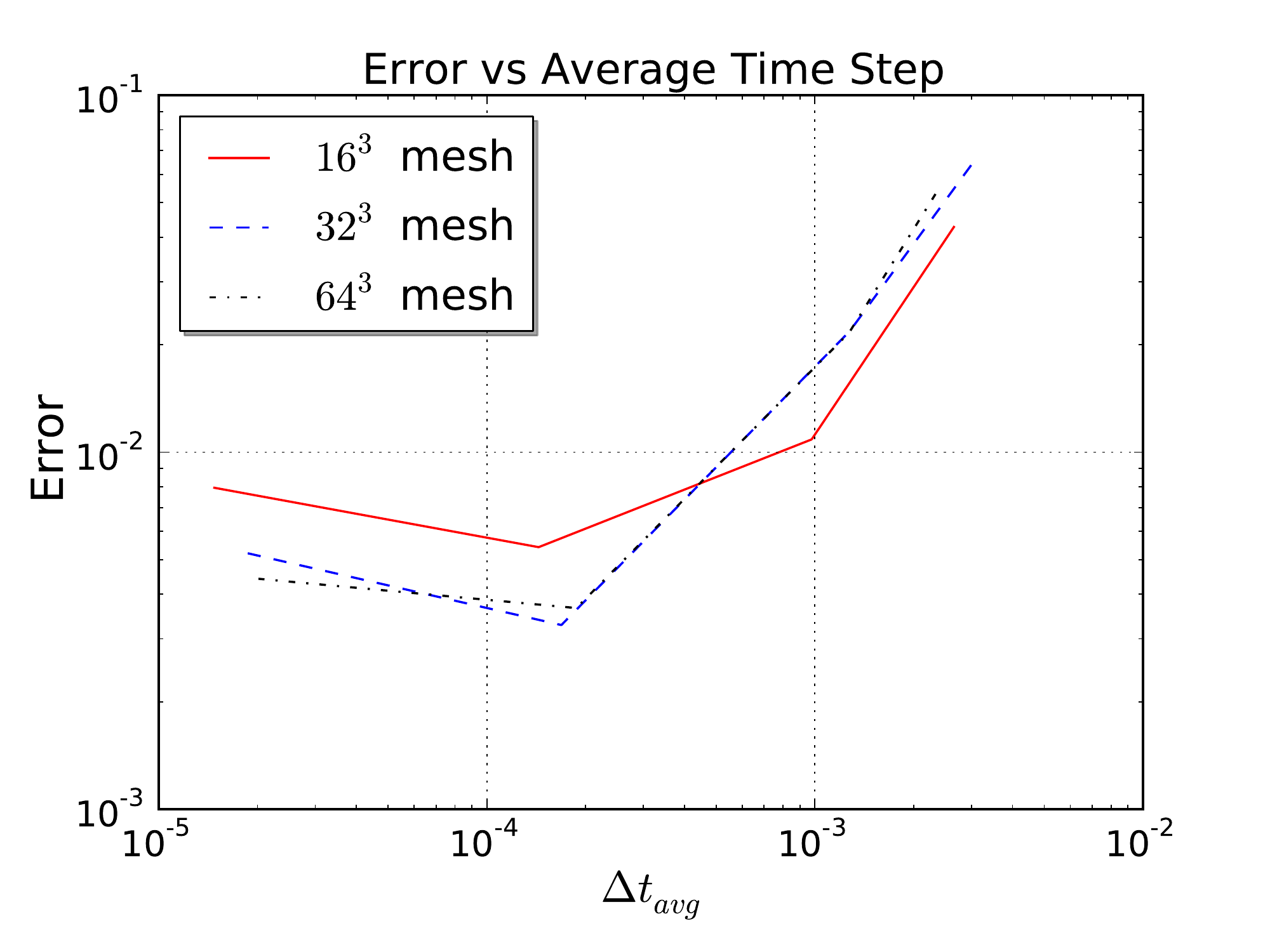}
  \includegraphics[scale=0.45, trim=1.0cm 0.0cm 1.0cm 0.5cm]{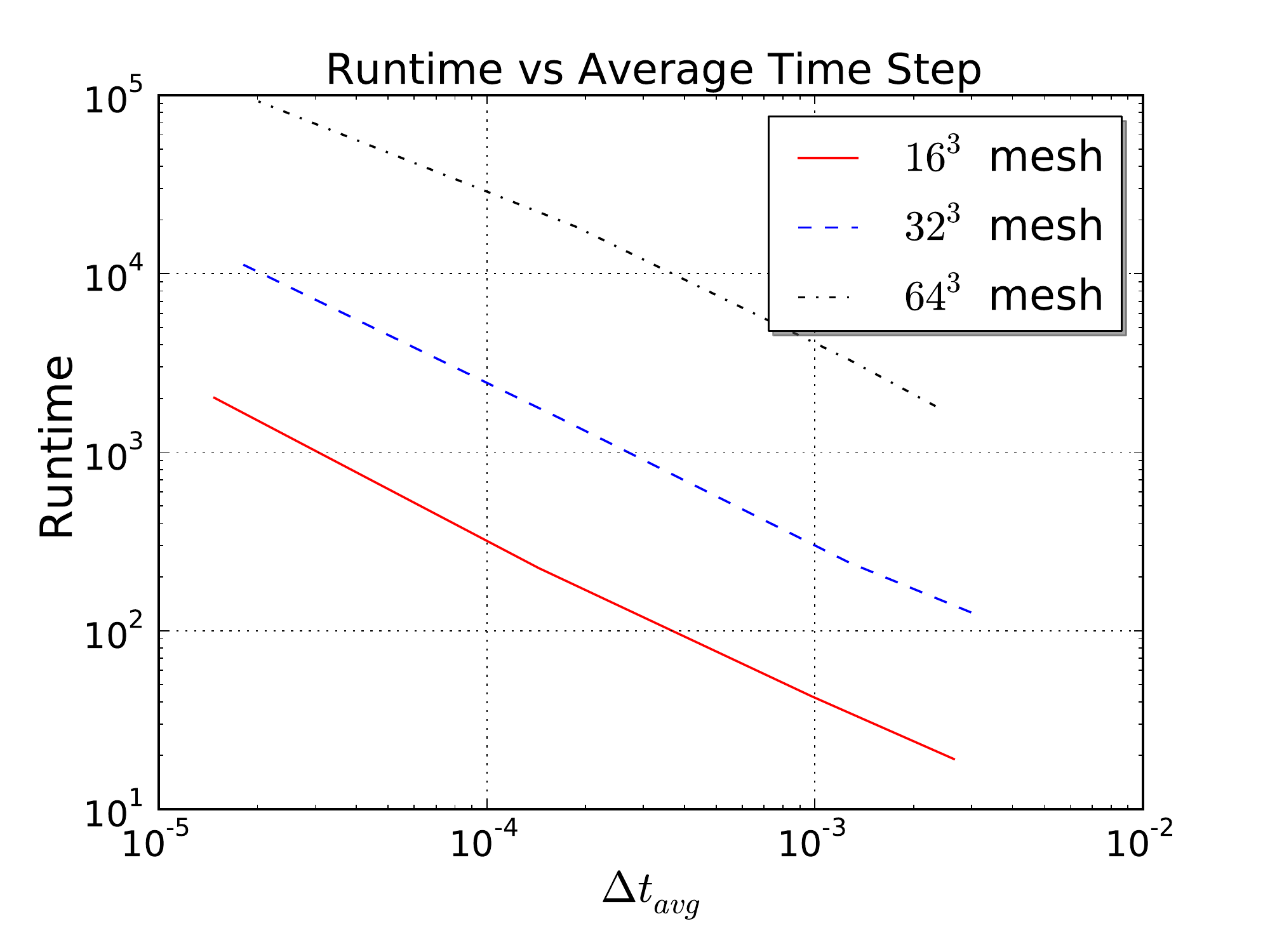}
  \hfill}
  \caption{We ran tests using mesh sizes of
    $16^3$, $32^3$ and $64^4$, and time step tolerances of
    $10^{-2}$, $10^{-3}$, $10^{-4}$ and $10^{-5}$, and plot the I
    front position error \eqref{eq:i1_error} as a function of the
    average time step size.  As expected, the runtime scales linearly
    with the inverse $\Delta t_{avg}$, and the error scales linearly
    with $\Delta t_{avg}$, at least until other sources of error
    dominate the calculation.}
  \label{fig:i1_stats}
\end{figure}

We plot spherically-averaged radial profiles of the radiation energy
density and the ionization fractions at 10 Myr, 100 Myr and 500 Myr
from a simulation using a 128$^3$ spatial grid and time step
tolerance $\tau_{tol} = 10^{-4}$ in Figure \ref{fig:i1_results},
showing the expected propagation of the radiation front and resulting
I-front in time.
Plots of the computed and analytical I front position and resulting
error for this run are provided in Figure \ref{fig:i1_radius}.
To further investigate the accuracy of our new splitting approach
between the radiation and chemistry solvers, we then performed these
same tests at a variety of mesh sizes and time step tolerances 
$\tau_{tol}$.  For mesh sizes of $16^3$, $32^3$ and $64^3$, and for
tolerances $10^{-2}$, $10^{-3}$, $10^{-4}$ and $10^{-5}$, we
compute the error in the I front position as
\begin{equation}
\label{eq:i1_error}
   error \ = \ \left\| \frac{r_{computed} - r_{true}}{r_s} \right\|_{RMS}
   \ = \ \left(\frac{1}{N_t} \sum_{i=1}^{N_t} \left(\frac{r_{computed,i} -
       r_{true,i}}{r_s}\right)^2 \right)^{1/2}.
\end{equation}
In Figure \ref{fig:i1_stats}, we plot the solution error as a function
of the average time step size, as well as the total runtime as a
function of the average time step size.  
As can be seen in these plots, as the tolerance decreases, the
temporal solution accuracy increases linearly until we reach a minimum
accuracy that results from other components in Enzo (spatial
discretization accuracy, accuracy within Enzo's chemistry solver,
etc.).  Moreover, it is evident that as we decrease the time step
tolerances, the required runtime increases linearly.  These results
imply that there is a ``sweet spot'' in our approach, wherein a
tolerance of $\tau_{tol}=10^{-4}$ achieves the solution with optimal
accuracy before we begin to waste additional effort without achieving
accuracy improvements. While this specific value is likely
problem-dependent, we use it as a starting point in subsequent
simulations.

\begin{figure}[t]
\centerline{\hfill
  \includegraphics[scale=0.3]{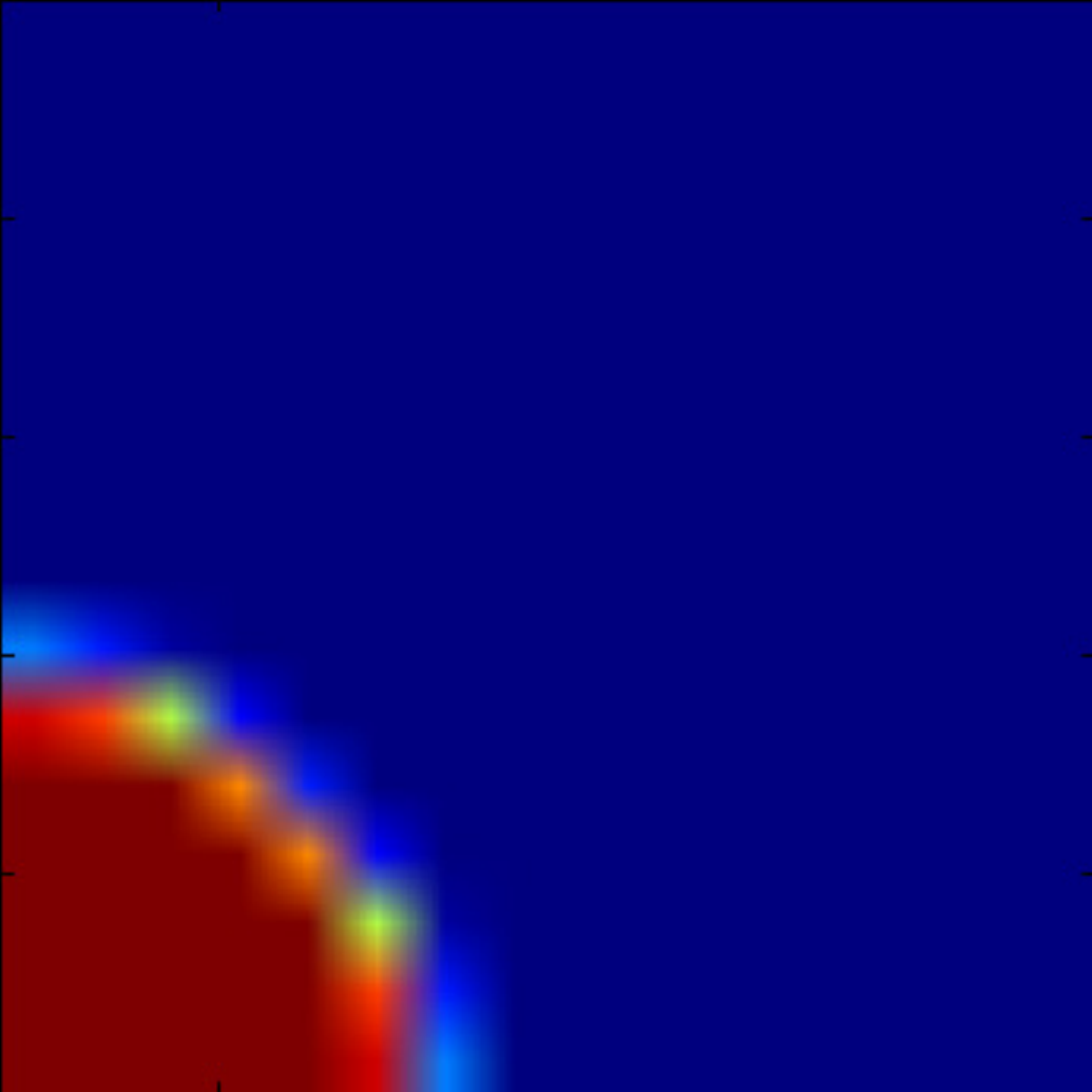}
  \hspace{0.1cm}
  \includegraphics[scale=0.3]{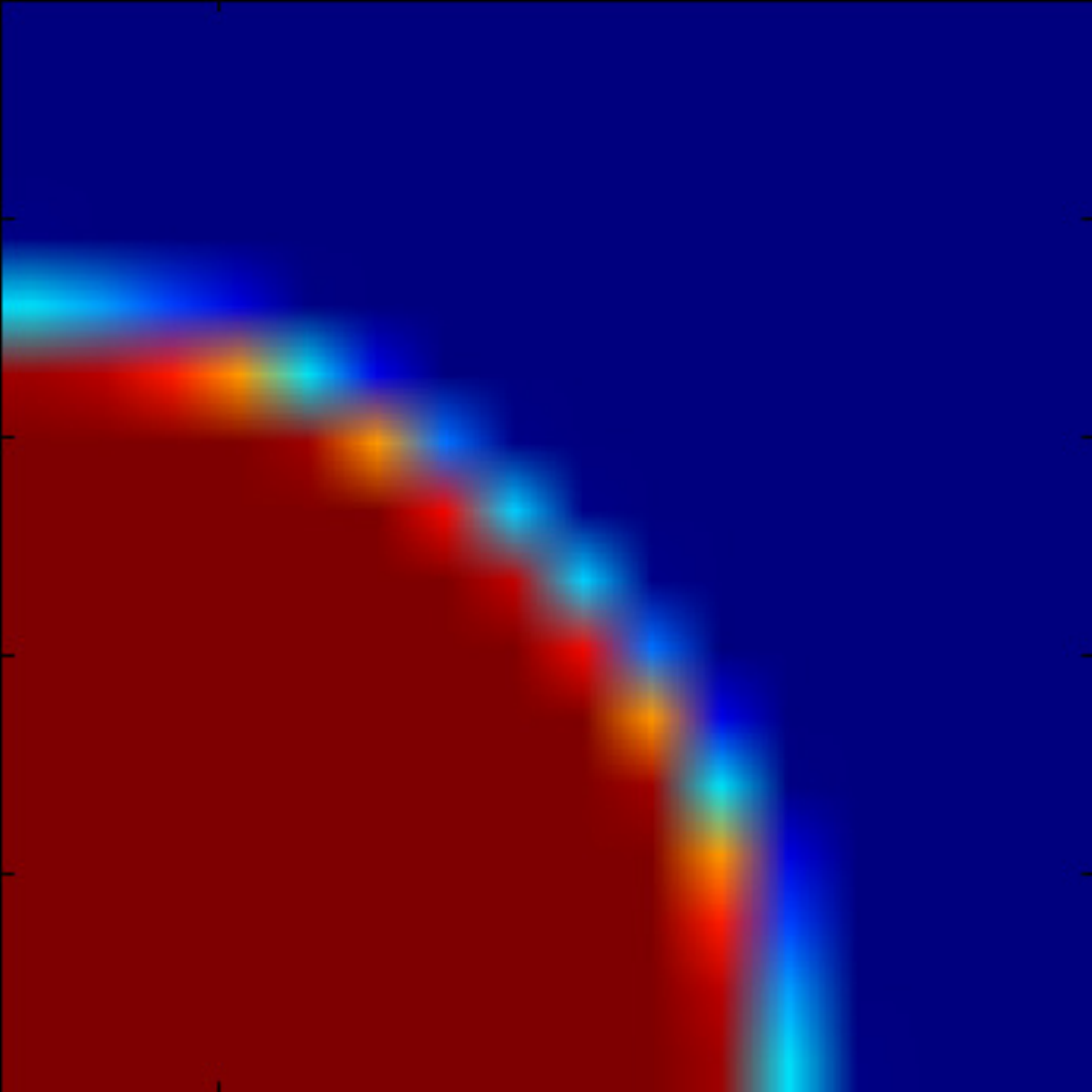}
  \hspace{0.1cm}
  \includegraphics[scale=0.3]{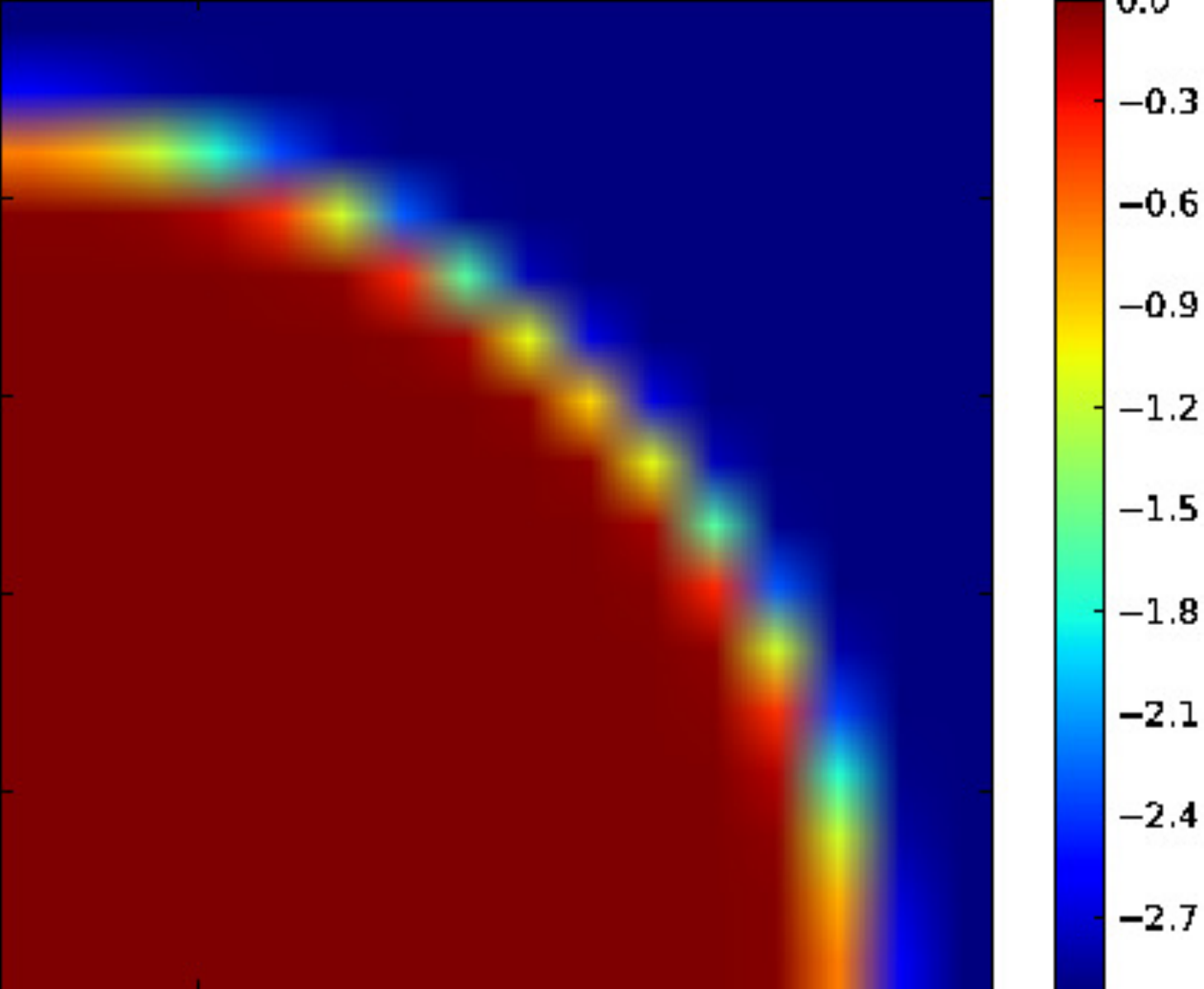}
  \hfill}
\vspace{0.2cm}
\centerline{\hfill
  \includegraphics[scale=0.3]{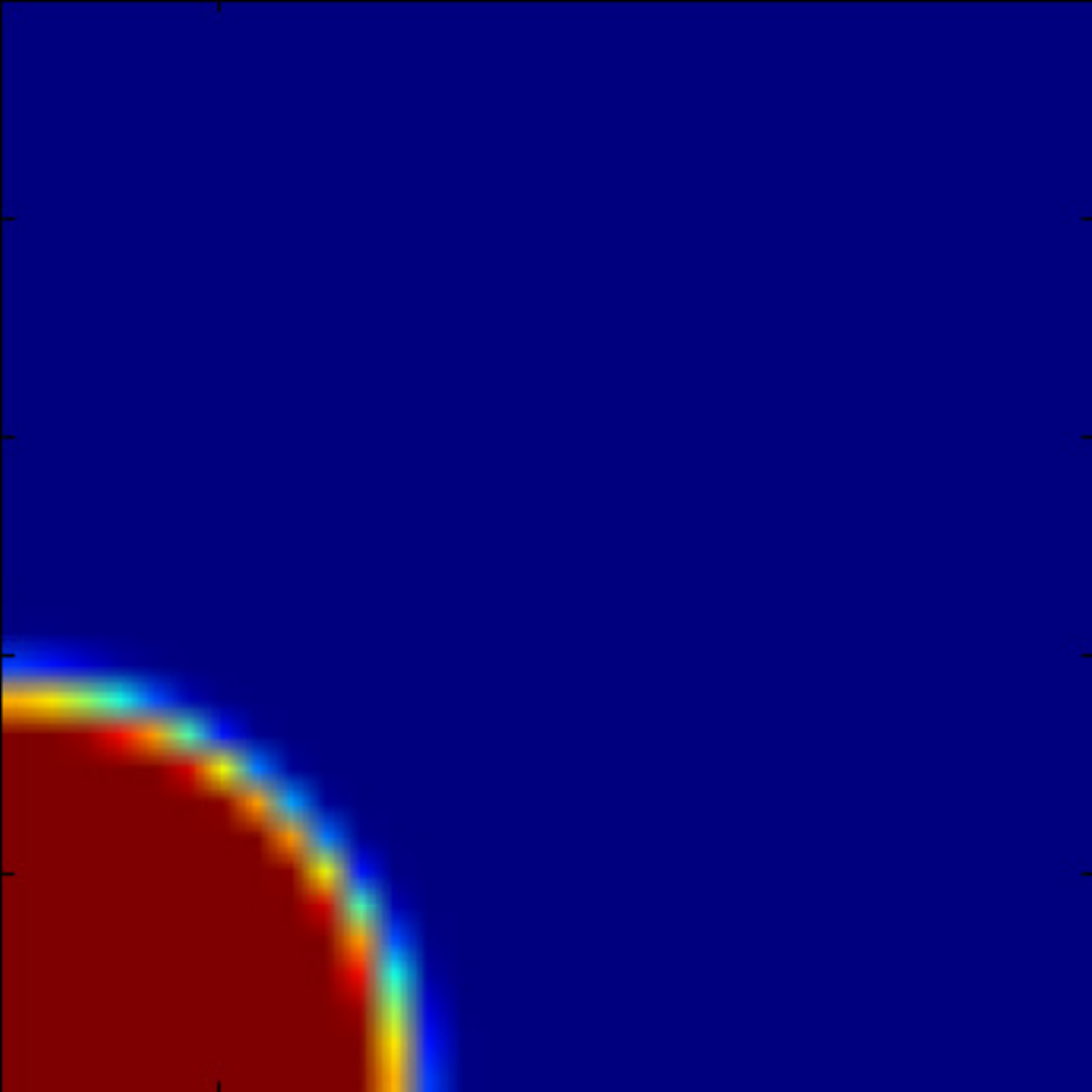}
  \hspace{0.1cm}
  \includegraphics[scale=0.3]{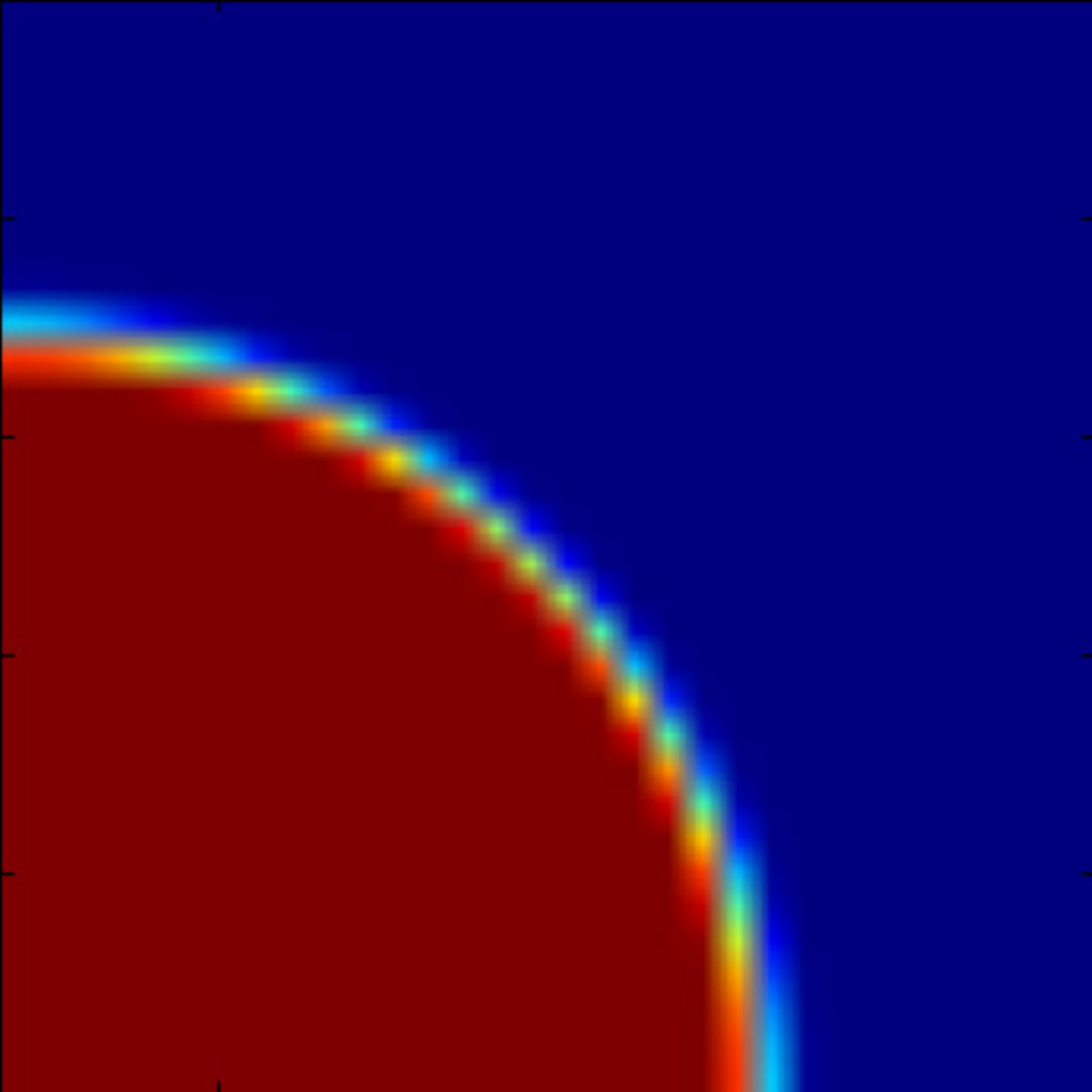}
  \hspace{0.1cm}
  \includegraphics[scale=0.3]{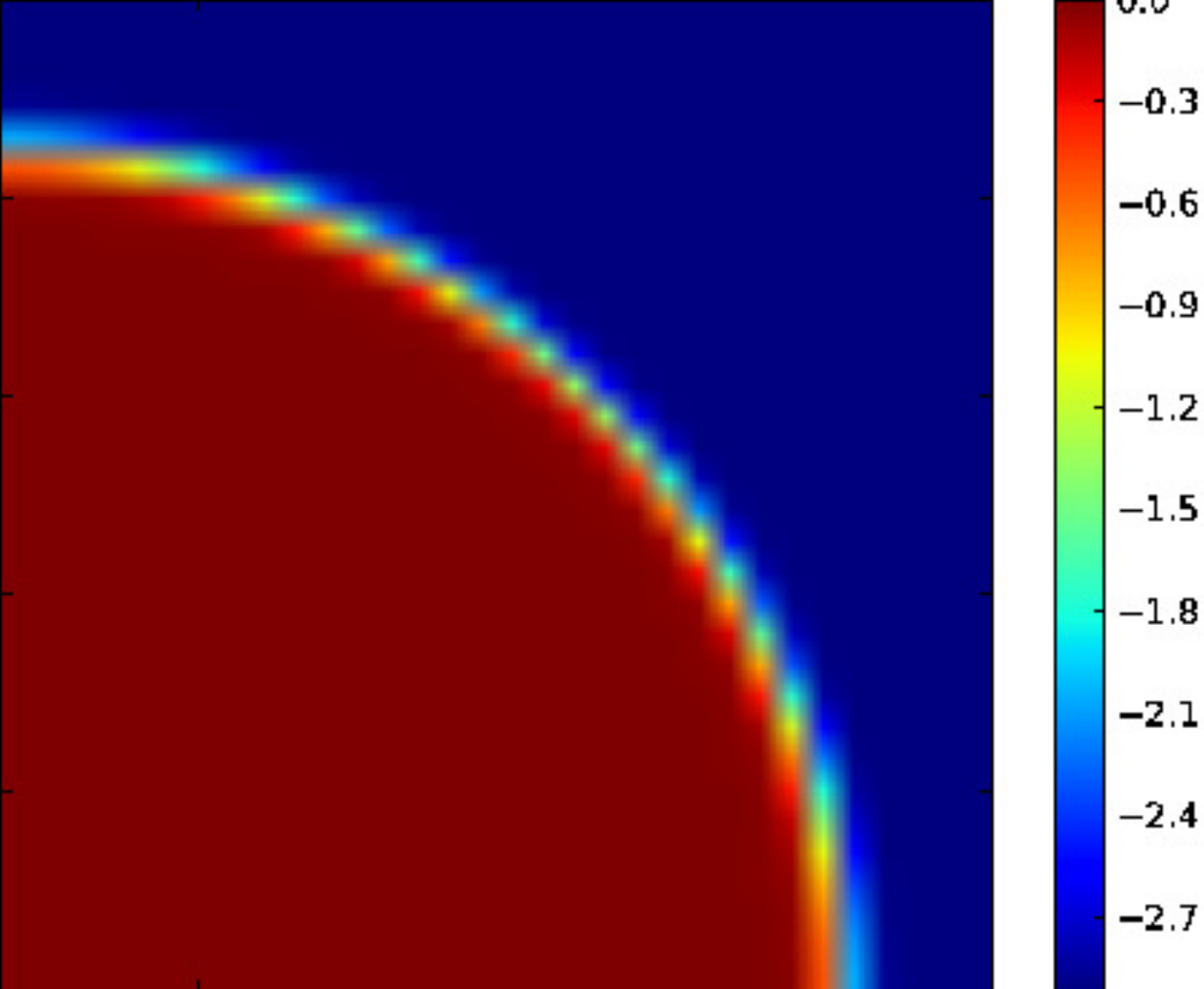}
  \hfill}
\vspace{0.2cm}
\centerline{\hfill
  \includegraphics[scale=0.3]{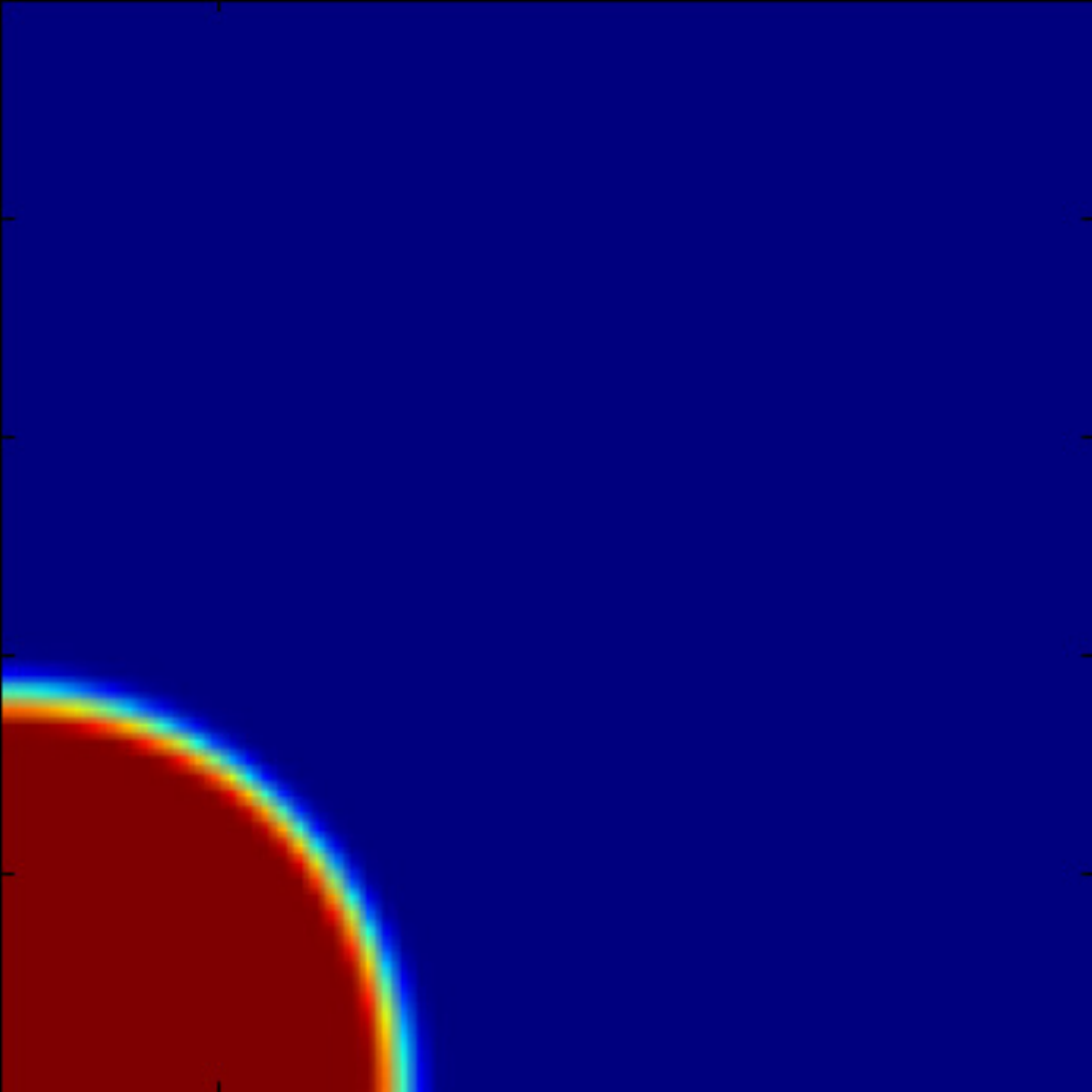}
  \hspace{0.1cm}
  \includegraphics[scale=0.3]{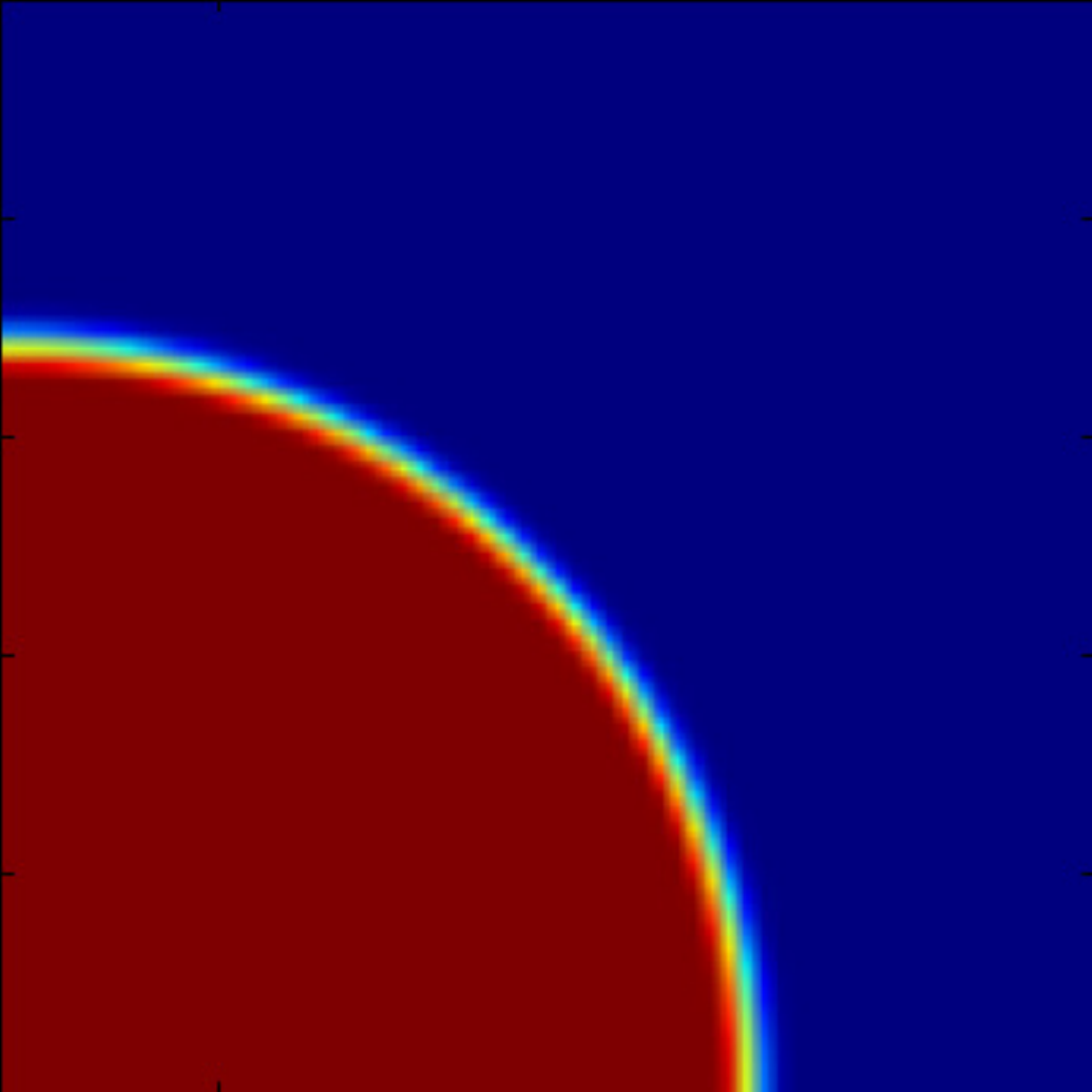}
  \hspace{0.1cm}
  \includegraphics[scale=0.3]{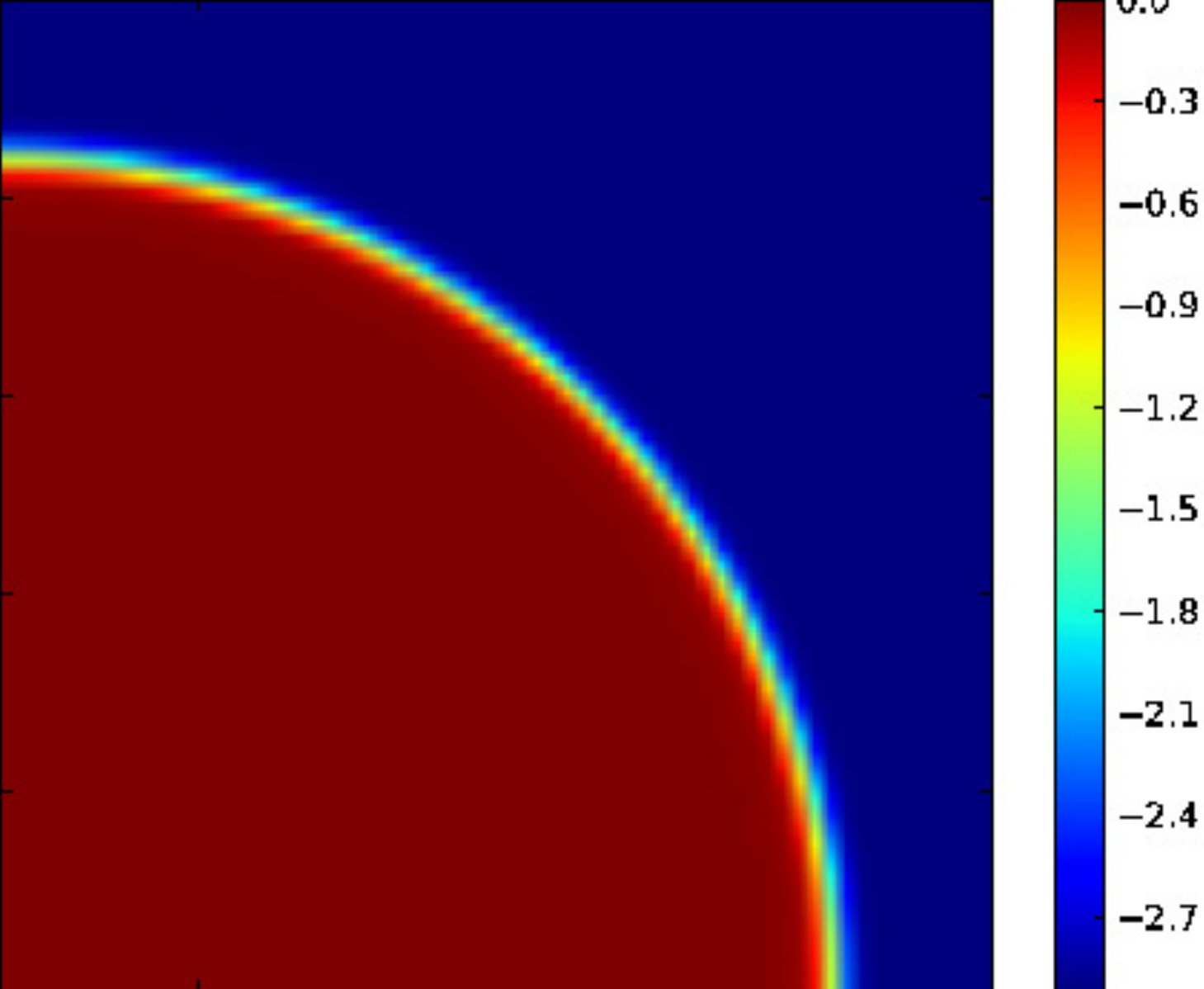}
  \hfill}
\vspace{0.2cm}
\centerline{\hfill
  \includegraphics[scale=0.3]{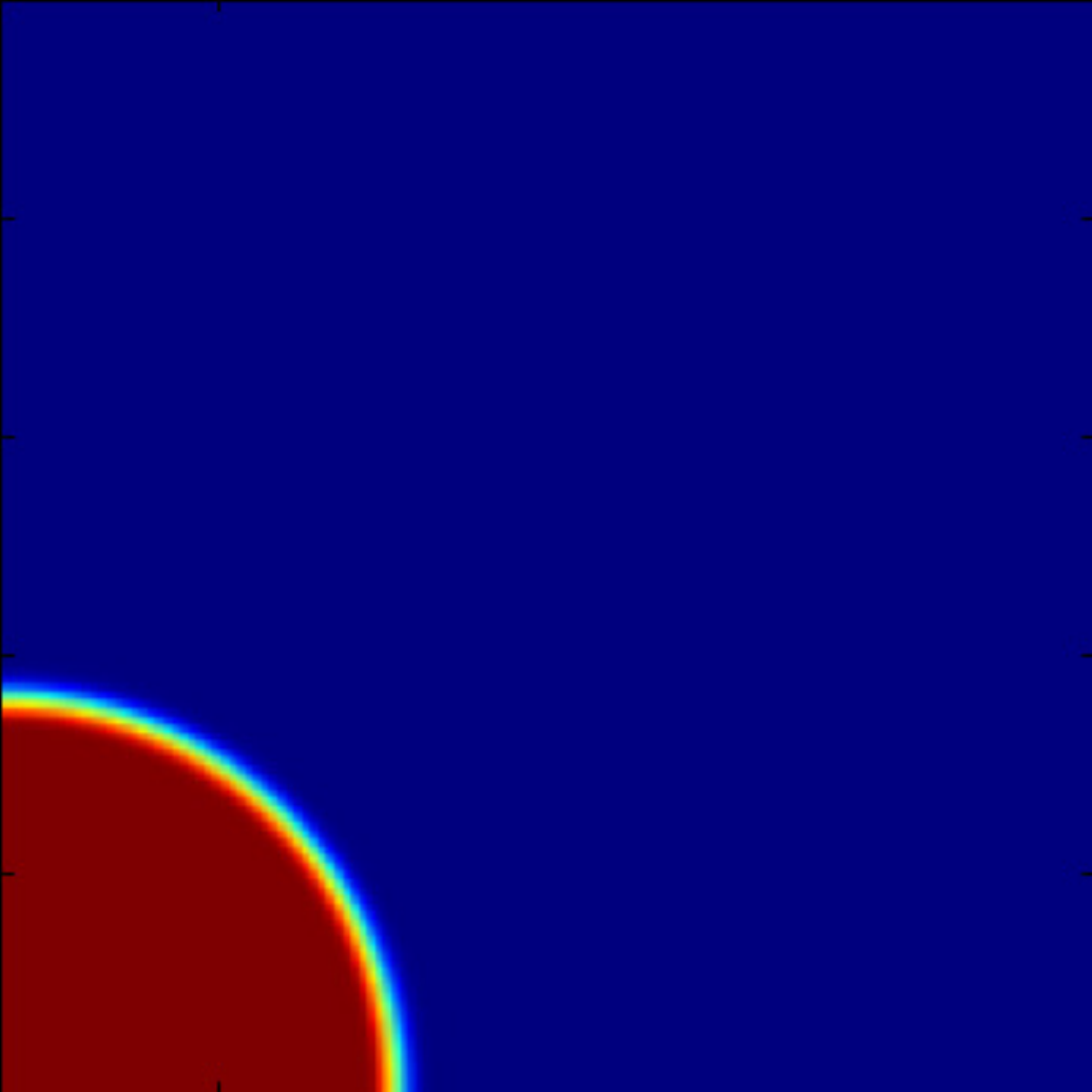}
  \hspace{0.1cm}
  \includegraphics[scale=0.3]{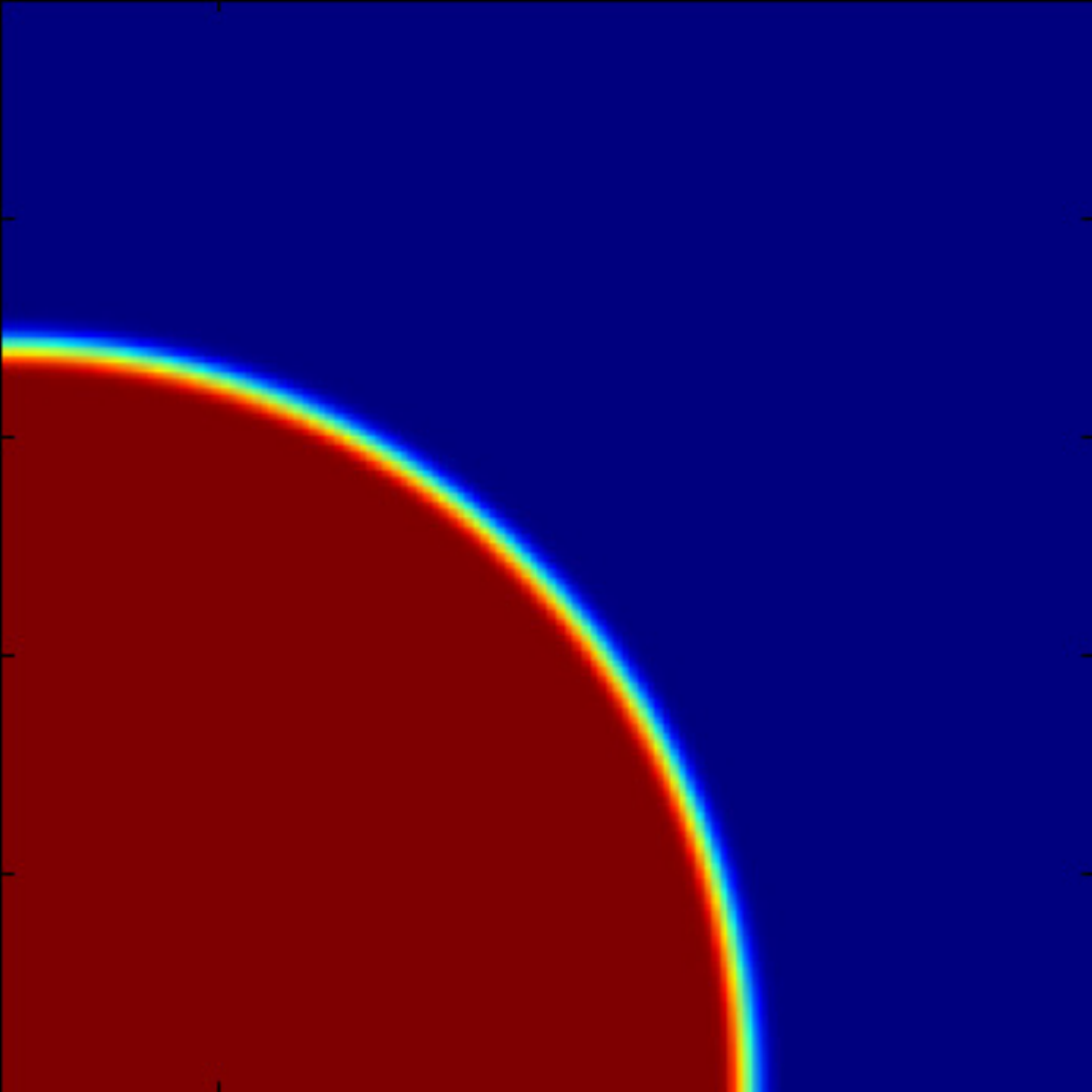}
  \hspace{0.1cm}
  \includegraphics[scale=0.3]{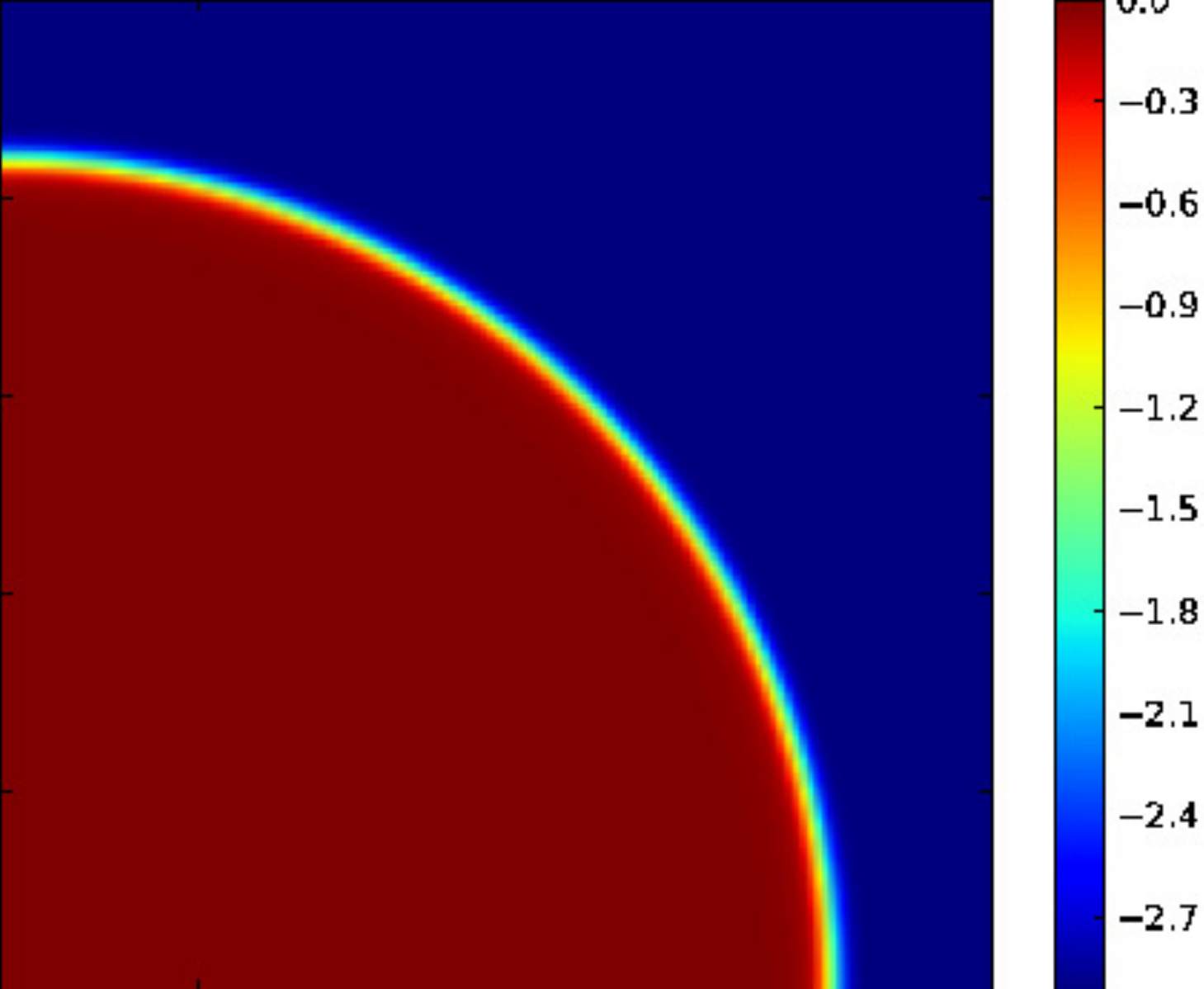}
  \hfill}
  \caption{HII slices perpendicular to the $z$ axis (log$_{10}$
    scale).  We plot the evolution of the ionized region at times of
    10, 100 and 500 Myr (columns), and using spatial meshes of $16^3$,
    $32^3$, $64^3$ and $128^3$ (rows), to demonstrate the convergence
    to a spherical bubble.}
  \label{fig:i1_contours}
\end{figure}

Finally, in Figure \ref{fig:i1_contours} we plot a slice of the
computed HII region through the center of the domain, perpendicular to
the $z$-direction (other directions are equivalent), to show the
convergence of the ionized region to a sphere at varying spatial
resolutions.

\subsubsection{Cosmological radiative ionization}
\label{subsec:test2}

This verification test is a slight variation of the previous problem,
adding only the additional complication of a cosmologically expanding
universe.  Due to the cosmological expansion, the Str{\"o}mgren radius
itself increases due to the expansion of space, that reduces the
hydrogen number density $n_H$ as time proceeds,
\begin{equation}
  \label{eq:rs_cosmo}
  r_s(t) = \left(\frac{3\,\dot{N}_{\gamma}}{4\pi\,\alpha_B\,n_H(t)^2}\right)^{1/3}.
\end{equation}
This causes the I front to initially approach $r_s$, but eventually
fall behind as the expansion drives $r_s$ outward.  The analytical
solution to this problem is given by \cite{ShapiroGiroux1987},
\begin{align}
  \label{eq:radius_cosmo}
  r(t) &= r_{s,0} \left(\lambda e^{-\tau(a(t))} \int_{1}^{a(t)}
    e^{\tau(\tilde{a})}\left(1-2q_0+\frac{2q_0}{\tilde{a}}(1-z_0)\right)^{-1/2}\,\mathrm
    d\tilde{a} \right)^{1/3}, \\
  \notag \mbox{where} \qquad & \\
  \label{eq:tau}
  \tau(a) &= \lambda\left(F(a) - F(1)\right) \left(6q_0^2(1+z_0)^2\right)^{-1}, \\
  \label{eq:Fa}
  F(a) &= \left(2 - 4q_0 - \frac{2q_0}{a}(1+z_0)\right) 
     \left(1-2q_0+\frac{2q_0}{a}(1+z_0)\right)^{1/2}.
\end{align}
Here, the parameter $\lambda = \alpha_B n_{H,0} / H_0 / (1+z_0)$, with
$r_{s,0}$, $z_0$ and $n_{H,0}$ as the Str{\"o}mgren radius, redshift
and hydrogen number density at the beginning of the simulation.
Additionally, $q_0$ is the cosmological deceleration parameter, $H_0$
is the Hubble constant, and $a(t) = (1+z(t))^{-1}$ is the cosmological
expansion parameter. 

We run this problem using the parameters $q_0 = 0.5$, 
domain $[0,80\,\mbox{kpc}]^3$ (comoving), time/redshift domain $z =
[4,0]$, $H_0=0.5$, energy density contributions $\Omega_m = 1$,
$\Omega_A = 0$ and $\Omega_b = 0.2$.  Our initial conditions are $\rho
= 1.175\times 10^{-28}$ g cm$^{-3}$, $T = 10^4$ K and $E = 10^{-45}$
erg cm$^{-3}$.

\begin{figure}[t]
\centerline{\hfill
  \includegraphics[scale=0.3, trim=1.0cm 1.0cm 1.0cm 0.5cm]{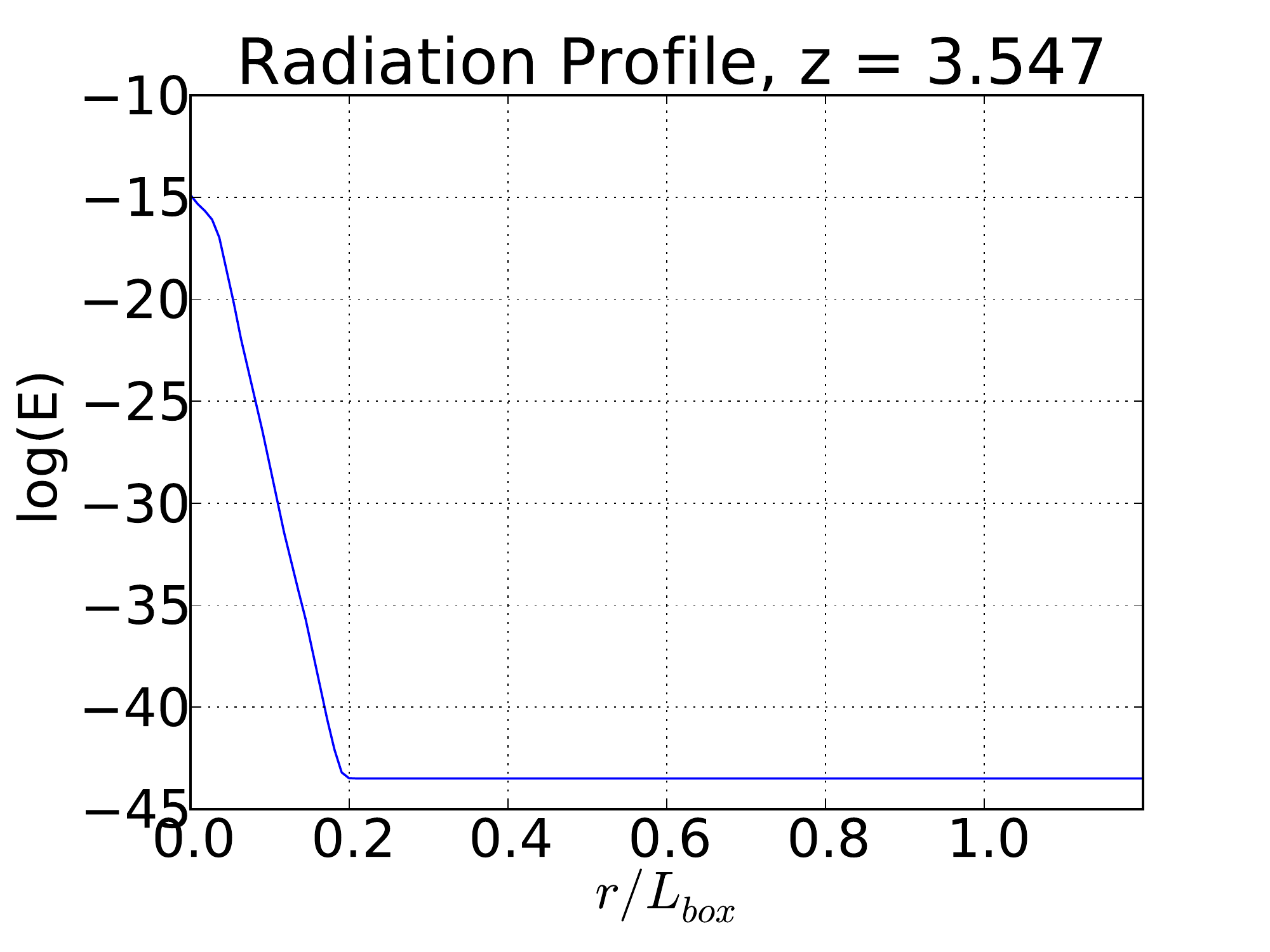}
  \includegraphics[scale=0.3, trim=1.0cm 1.0cm 1.0cm 0.5cm]{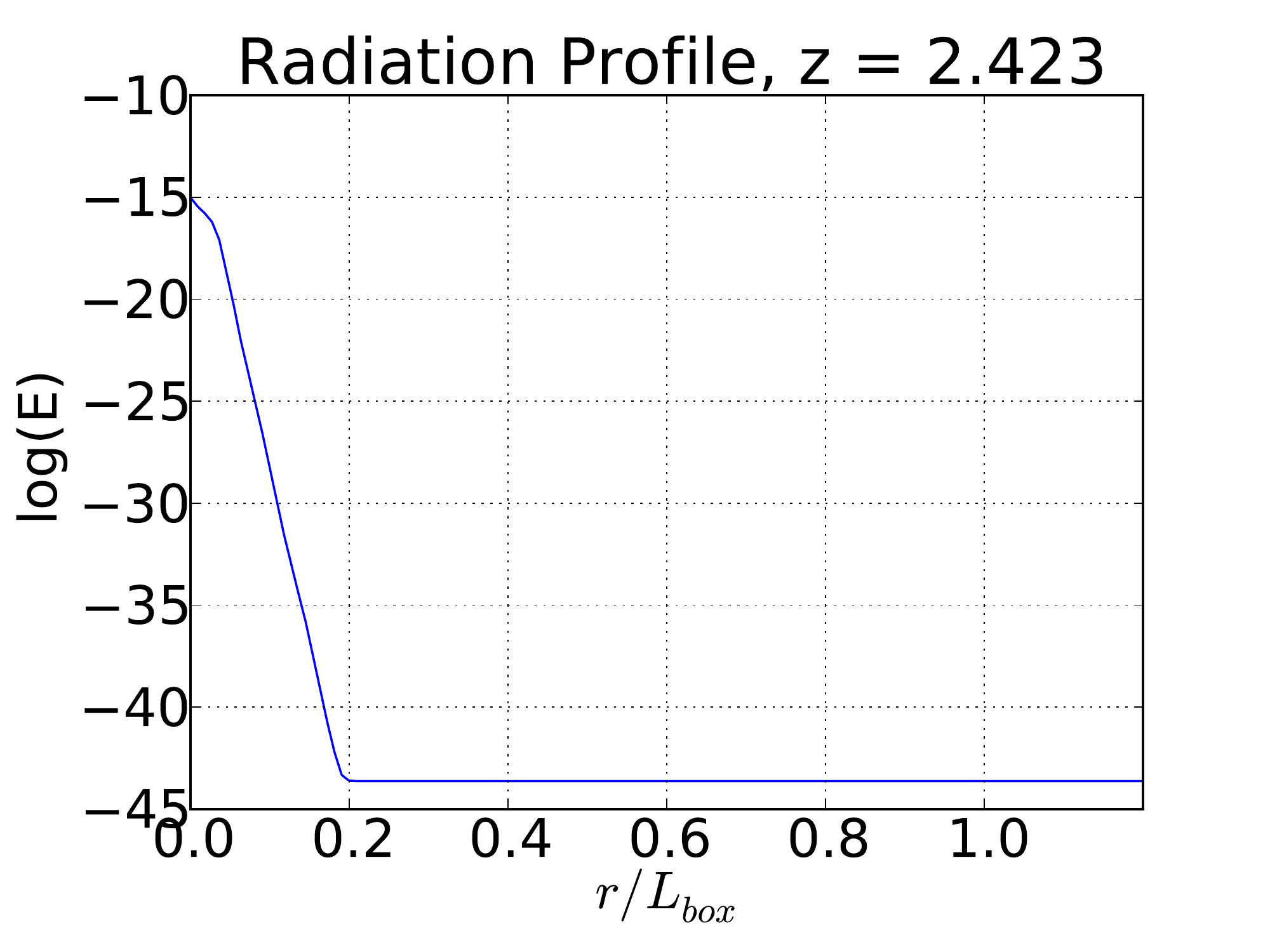}
  \includegraphics[scale=0.3, trim=1.0cm 1.0cm 1.0cm 0.5cm]{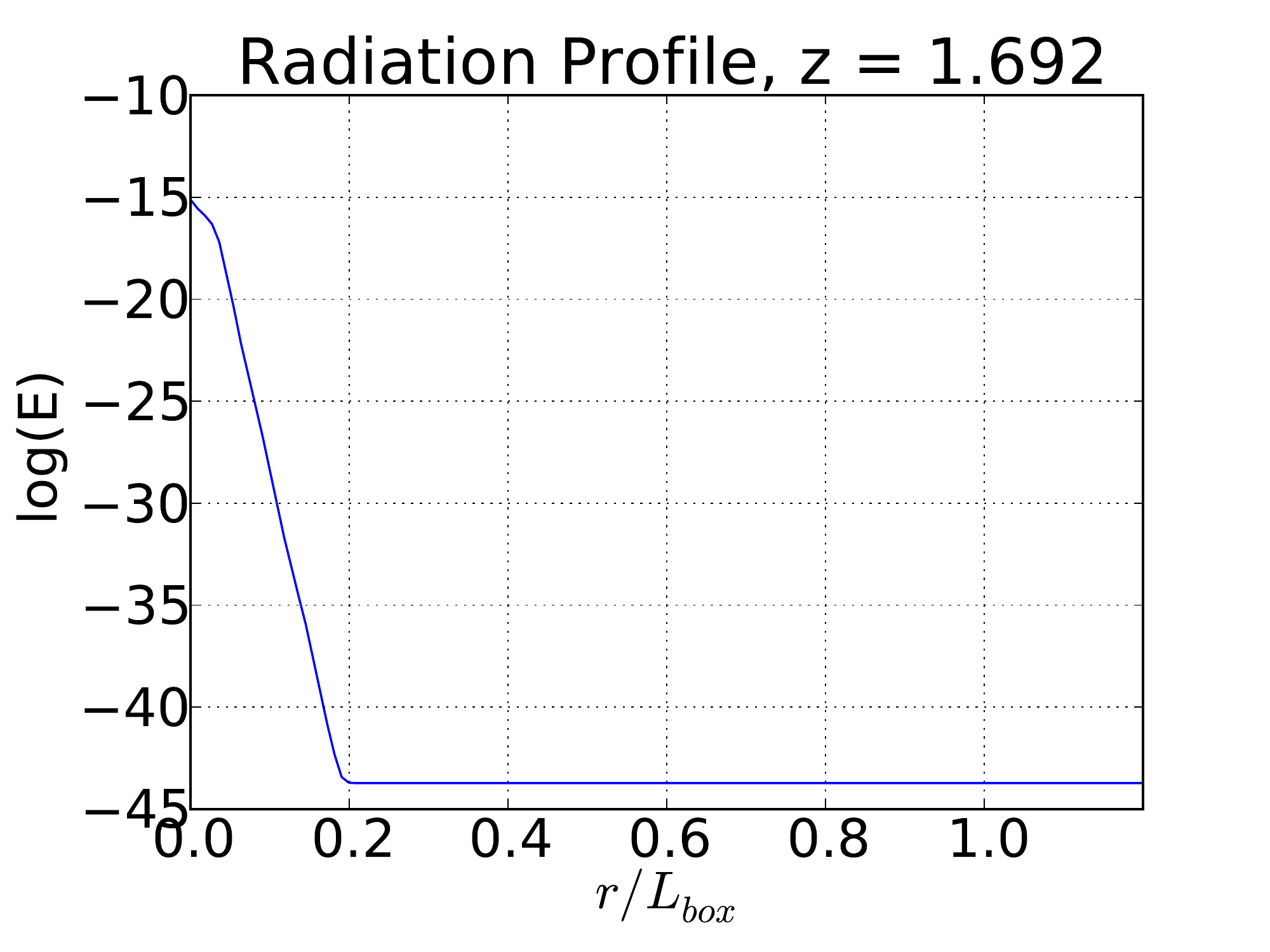}
  \hfill}
\centerline{\hfill
  \includegraphics[scale=0.3, trim=1.0cm 0.5cm 1.0cm 0.5cm]{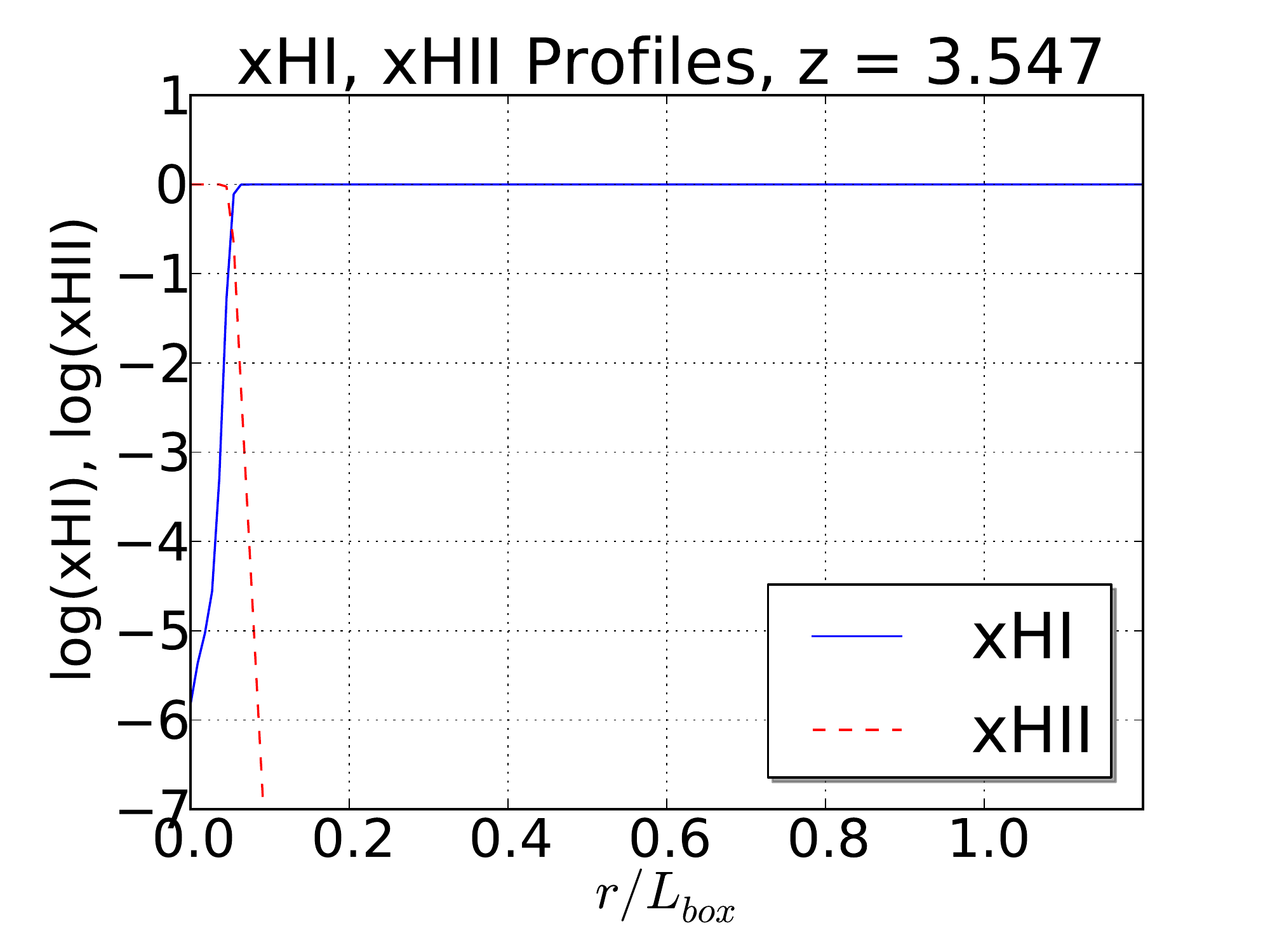}
  \includegraphics[scale=0.3, trim=1.0cm 0.5cm 1.0cm 0.5cm]{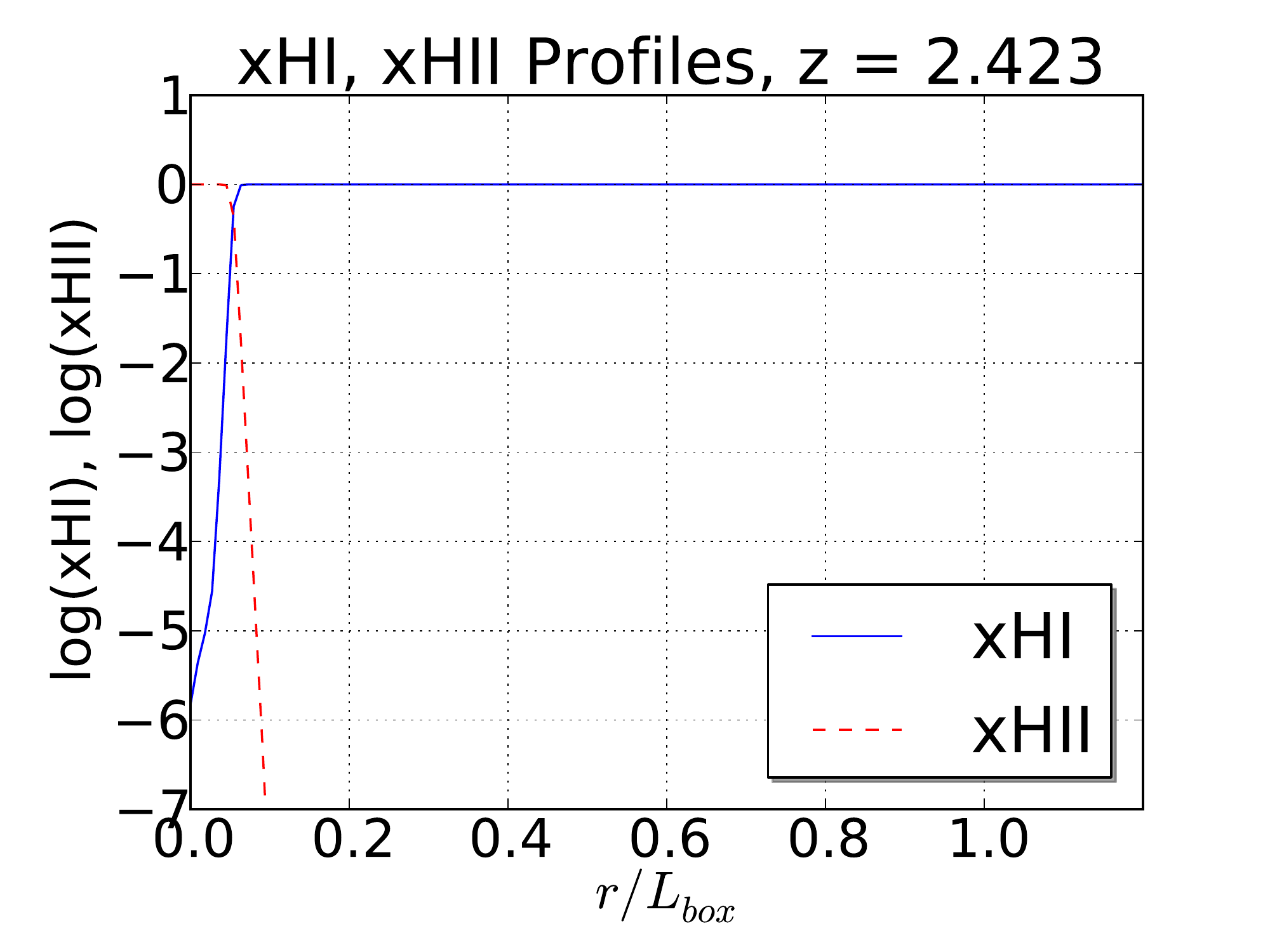}
  \includegraphics[scale=0.3, trim=1.0cm 0.5cm 1.0cm 0.5cm]{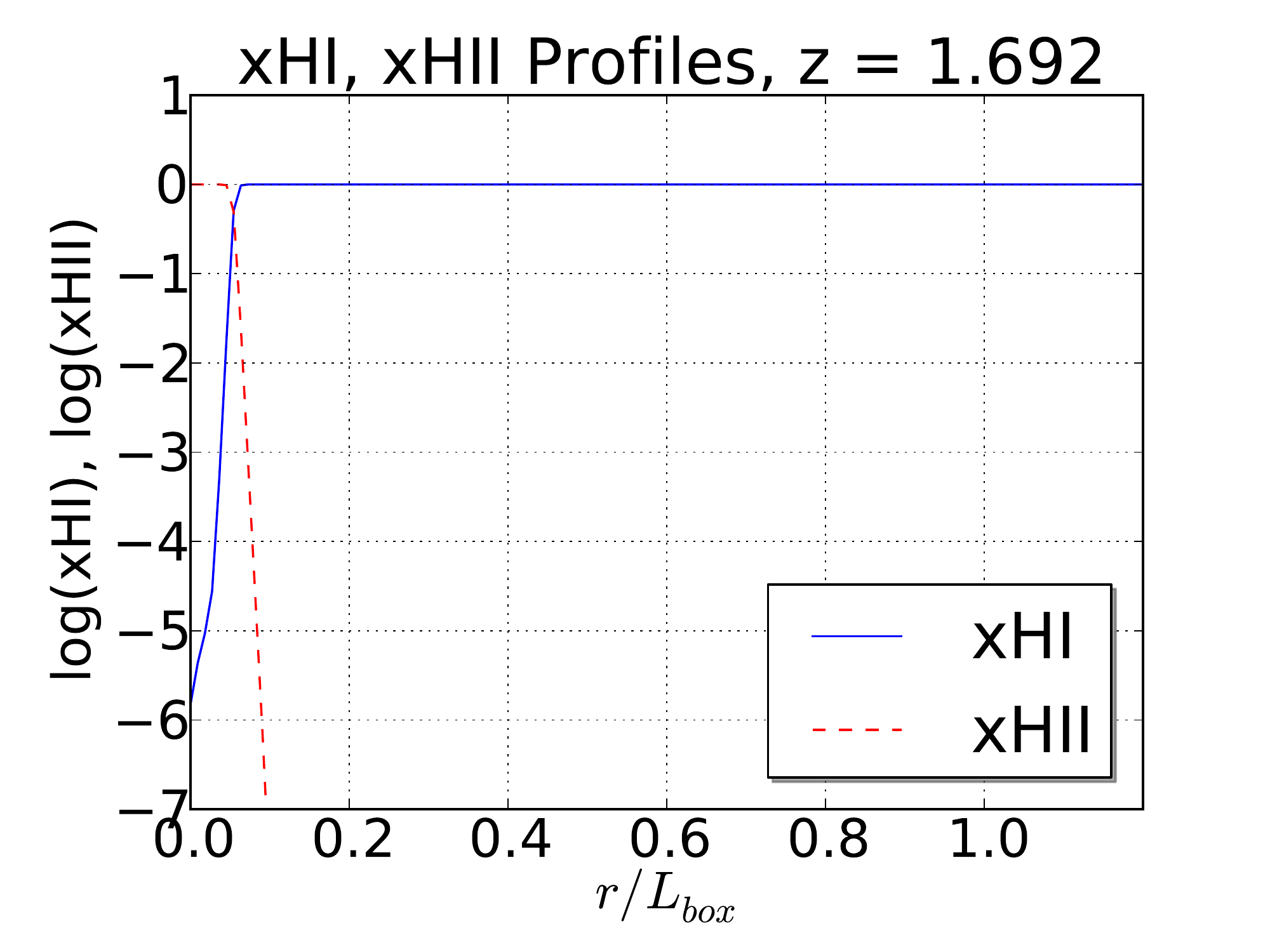}
  \hfill}
  \caption{Spherically-averaged radial profiles of radiation energy
    density and ionization fractions for the cosmological ionization
    test in section \ref{subsec:test2} using a $128^3$ mesh and
    time step tolerance $\tau_{tol} =10^{-4}$.  Plots are shown at z=3.547, 2.423 and 1.692
    (left to right), with the radiation energy density on
    the top row and ionization fractions on the bottom row.}
  \label{fig:sg_results}
\end{figure}

\begin{figure}[t]
\centerline{\hfill
  \includegraphics[scale=0.45, trim=1.0cm 0.0cm 1.0cm 0.5cm]{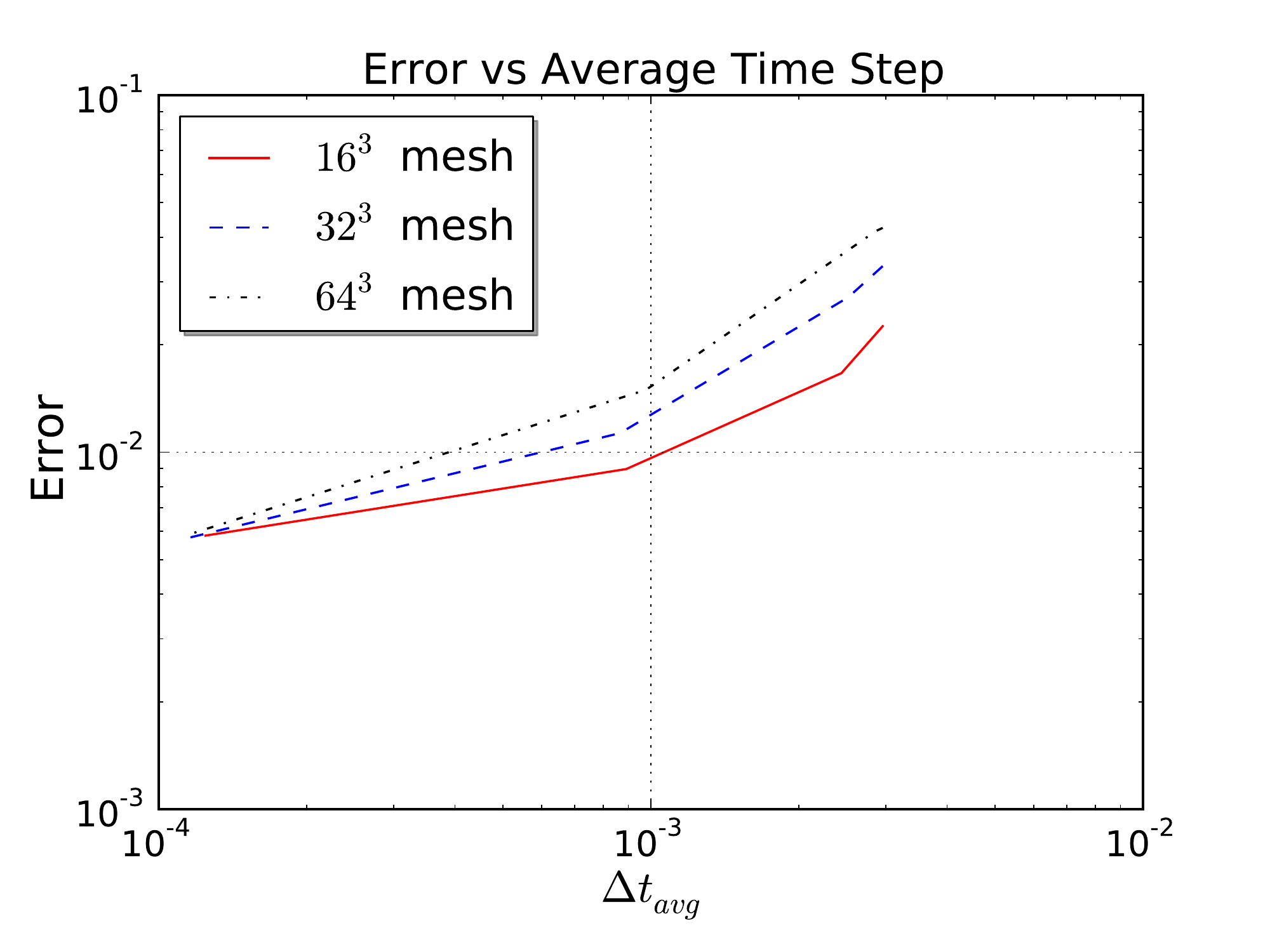}
  \includegraphics[scale=0.45, trim=1.0cm 0.0cm 1.0cm 0.5cm]{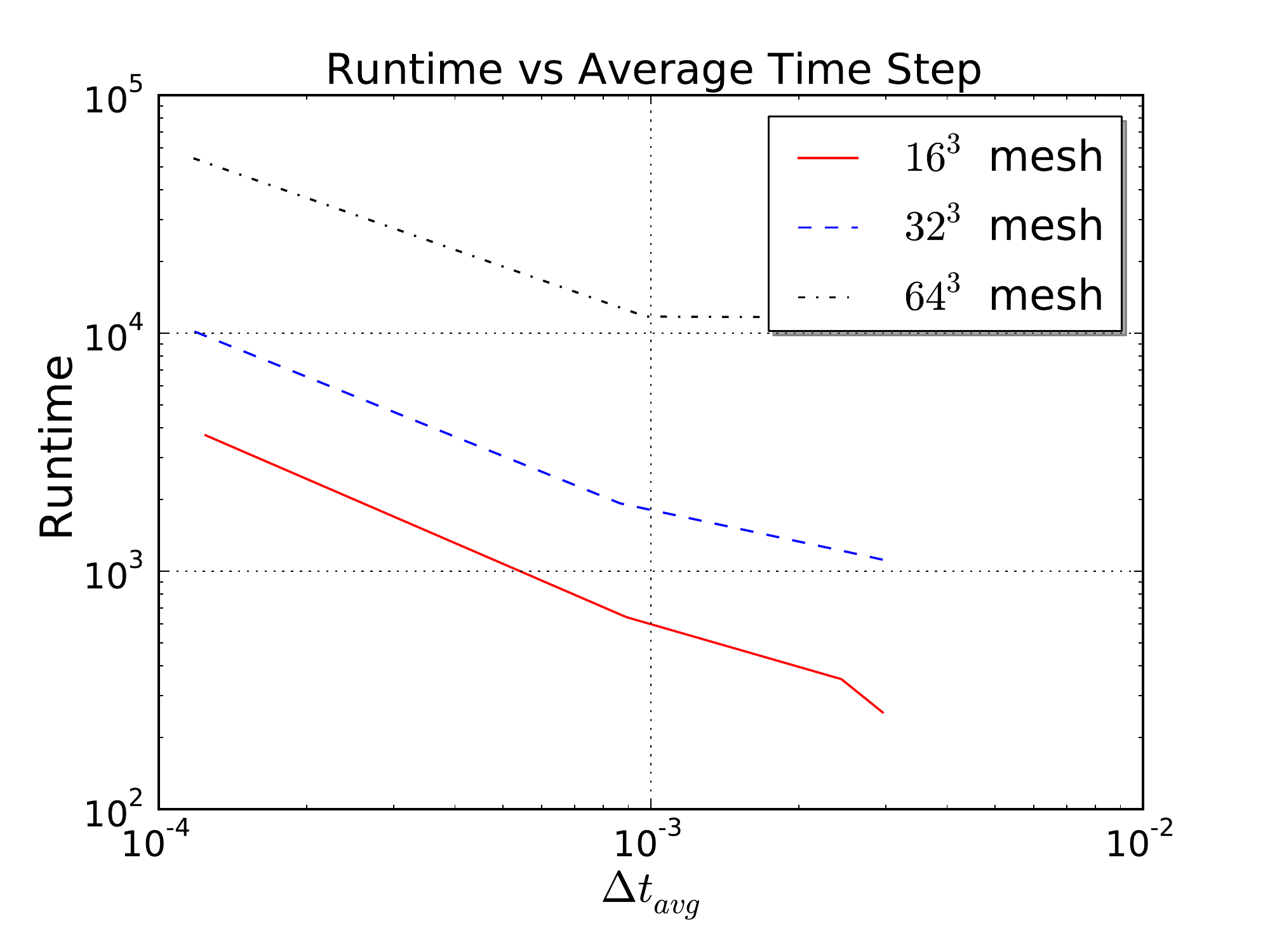}
  \hfill}
  \caption{We ran tests using mesh sizes of
    $16^3$, $32^3$ and $64^4$, and time step tolerances of
    $10^{-2}$, $10^{-3}$, $10^{-4}$ and $10^{-5}$, and plot the I
    front position error as a function of the average time step size.
    As with Figure \ref{fig:i1_stats}, the runtime scales linearly
    with the inverse $\Delta t_{avg}$, and the error scales linearly
    with $\Delta t_{avg}$, at least until other sources of error
    dominate the calculation.} 
  \label{fig:sg_stats}
\end{figure}

We again plot spherically-averaged radial profiles of the radiation energy
density and the ionization fractions at redshifts 3.547, 2.423 and 1.692 from a
simulation using a 128$^3$ spatial grid and time step tolerance
$\tau_{tol} = 10^{-4}$ in Figure \ref{fig:sg_results}, showing the
expected propagation and eventual stalling of the radiation front and
resulting I-front in time.
As with the previous test, we investigated the accuracy of our new
splitting approach between the radiation and chemistry solvers using
the same set of mesh sizes and time step tolerances as the test in
section \ref{subsec:test1}.  Figure \ref{fig:sg_stats} contains the
corresponding plots of the solution error and total runtime as a
function of the average time step size.  
Our results are similar to those from the previous test, indicating
that the modified time evolution approach employed in this work
successfully achieves accurate solutions of our coupled radiation and
ionization system.

\subsection{Validation Tests}
Validation tests are tests without analytic solutions that nonetheless serve as a
useful point of comparison between codes implementing different physical 
models and numerical methods \citep{IlievEtAl2006,IlievEtAl2009}. Our purpose is not to run all possible tests, but rather to validate the application of FLD to large scale reionization simulations, and in particular to investigate the well-known inability of FLD to cast a shadow on the general progress of reionization.
In this section we test our algorithm against 
four validation tests that are most relevant to the problem of cosmological 
reionization. The first two are radiation hydrodynamic tests studied by \cite{IlievEtAl2009} (hereafter RT09). They are the propagation of an I-front in a $r^{-2}$ density gradient, and the photoevaportion of a dense cloud irradiated from one side. The third validation test is the consolidated \hii region produced by two sources of equal luminosity introduced by \cite{Petkova09}. These three tests were chosen from a larger number of tests in the literature because they form a natural sequence with regard to the expansion and merging of isolated \hii regions in a clumpy IGM.  Finally, in a fourth test of our own design, we perform a direct comparison between FLD and ray tracing on a fully coupled reionization simulation in a small box. These tests demonstrate that although FLD ionizes dense clouds somewhat faster than methods that cast shadows, this affects primarily the earliest phases of cosmic reionzation when  a piece of neutral IGM is irradiated by the brightest nearby source. Later on, when multiple sources ionize the gas from multiple directions, FLD and ray tracing produce very similar evolutions.

\subsubsection{Test 6 -- I-front expansion in a $r^{-2}$ density profile}
\label{subsec:test6}

As our first validation test, we investigate Test 6 in
\cite{IlievEtAl2009} (hereafter RT09), that focuses on a full
radiation-hydrodynamics simulation of an ionized hydrogen (HII)
region in a spherically-symmetric density field.  Here, the center of
the region has constant number density, but at a specified core radius
$r_0$ the density rapidly decreases with radius.  Denoting this
functional relationship as $n_H(r)$, 
\[
   n_H(r) = \begin{cases}
     n_0,\quad&\text{for}\; r\le r_0\\
     n_0\left(\frac{r_0}{r}\right)^{2},\quad&\text{for}\; r> r_0.
   \end{cases}
\]
We follow the parameters choices from \cite{IlievEtAl2009}:  cubic
simulation domain of $[0,L]^3$ with $L=0.8$ kpc, core number density
$n_0 = 3.2$ cm$^{-3}$, core radius $r_0 = 91.5$ pc, zero initial
ionization fraction, ionization source at the origin with strength
$\dot{N}_{\gamma}=10^{50}$ photons s$^{-1}$ and a $T=10^5$ K blackbody
SED, initial temperature $T = 100$ K, reflective boundaries that touch
the origin and transmissive boundaries elsewhere, and a total
simulation time of $0\le t\le 25$ Myr.  Under these choices the
the I-front transitions from R-type to D-type within the core.  Once
the I-front reaches the beginning of the density gradient it begins to
accelerate, subsequently transitioning back to R-type.  Unfortunately,
for these simulation parameters, the problem exhibits no analytical
solution, so we refer to RT09 and WA11 for reference solutions to
compare against our own.

\begin{figure}[t]
\centerline{\hfill
  \includegraphics[scale=0.42, trim=0.0cm 0.0cm 0.0cm 0.0cm]{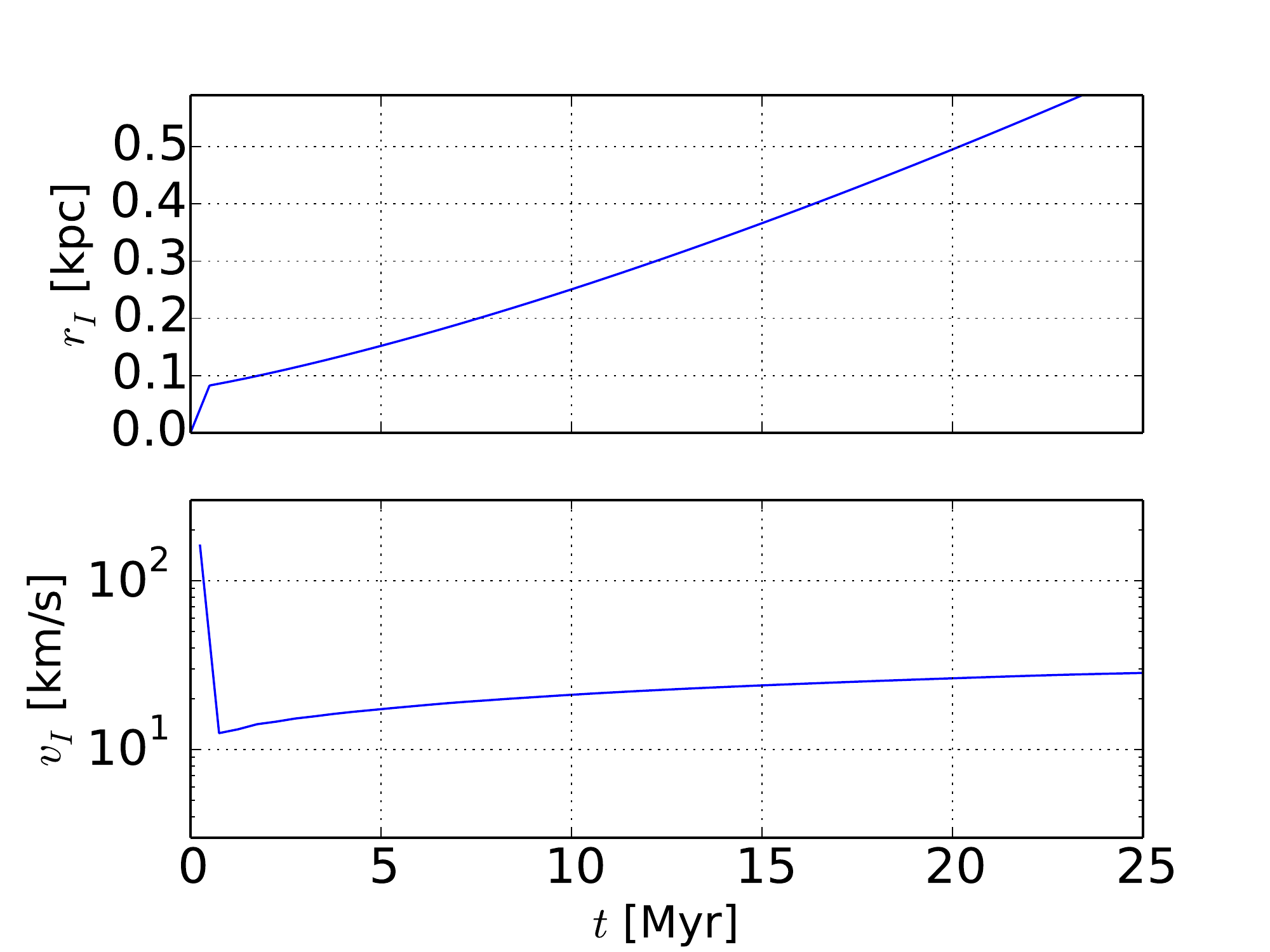}
  \includegraphics[scale=0.42, trim=0.0cm 0.0cm 0.0cm 0.0cm]{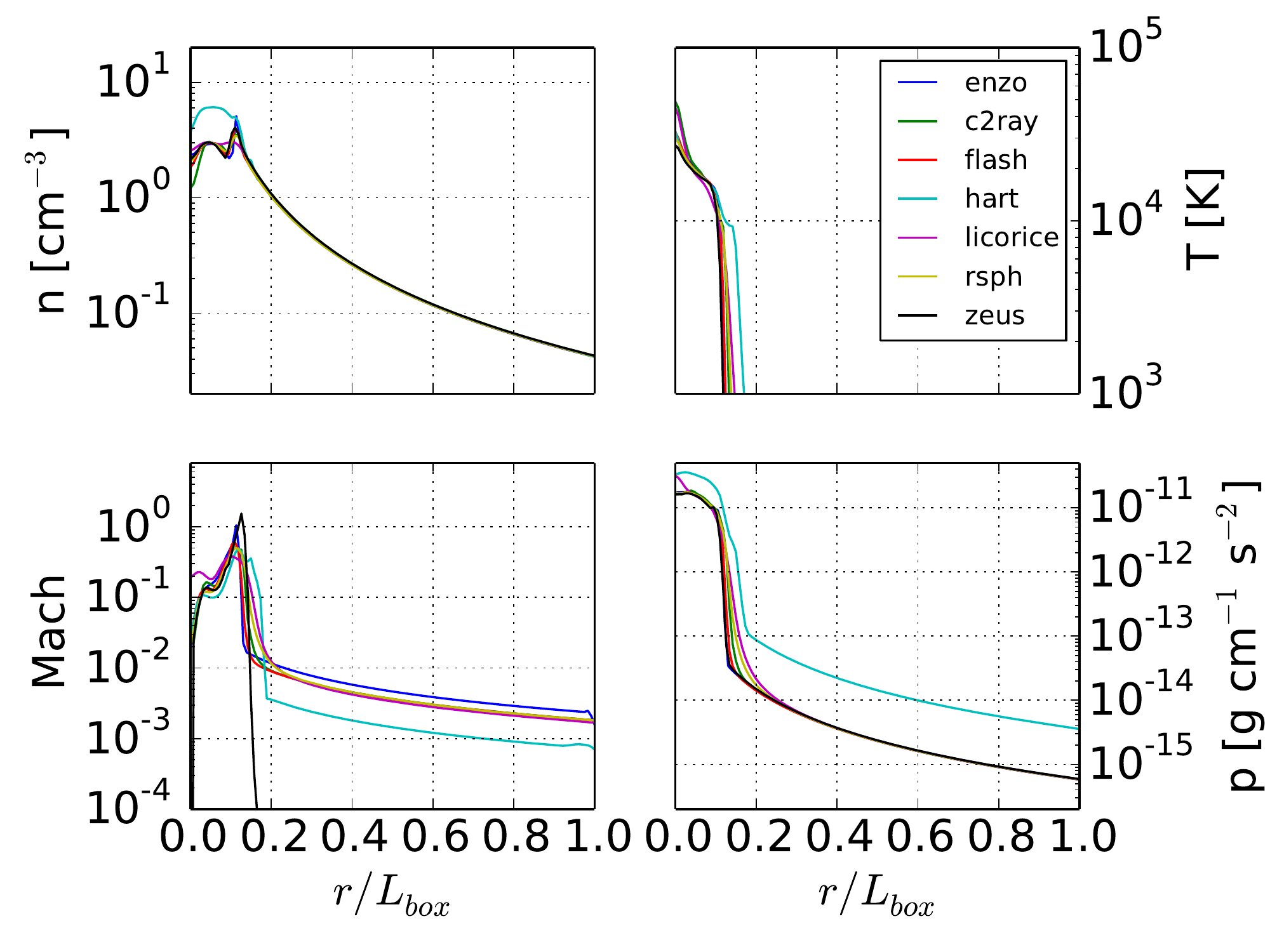}
  \hfill}
\centerline{\hfill 
  \includegraphics[scale=0.42, trim=0.0cm 0.0cm 0.0cm 0.0cm]{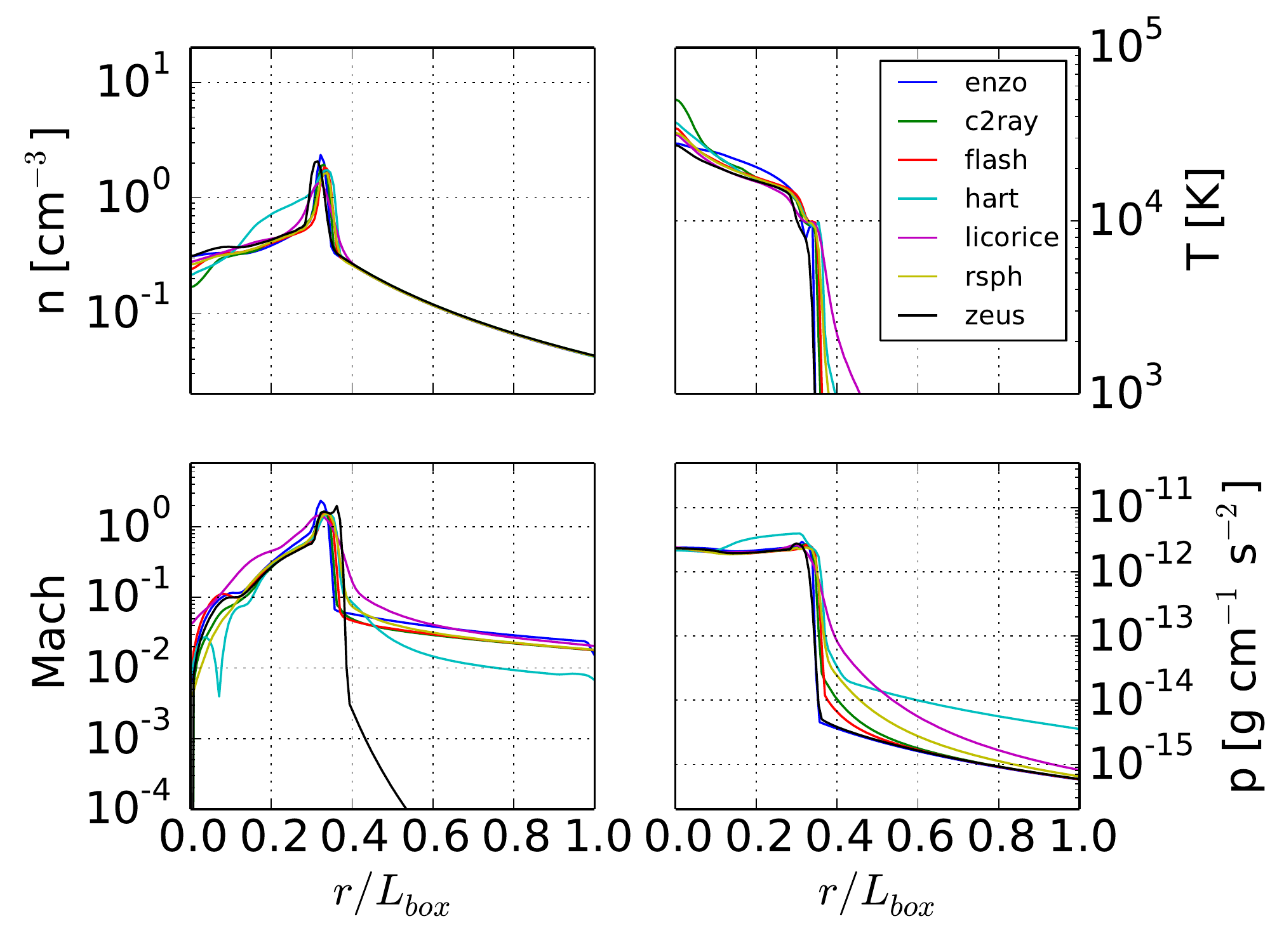}
  \includegraphics[scale=0.42, trim=0.0cm 0.0cm 0.0cm 0.0cm]{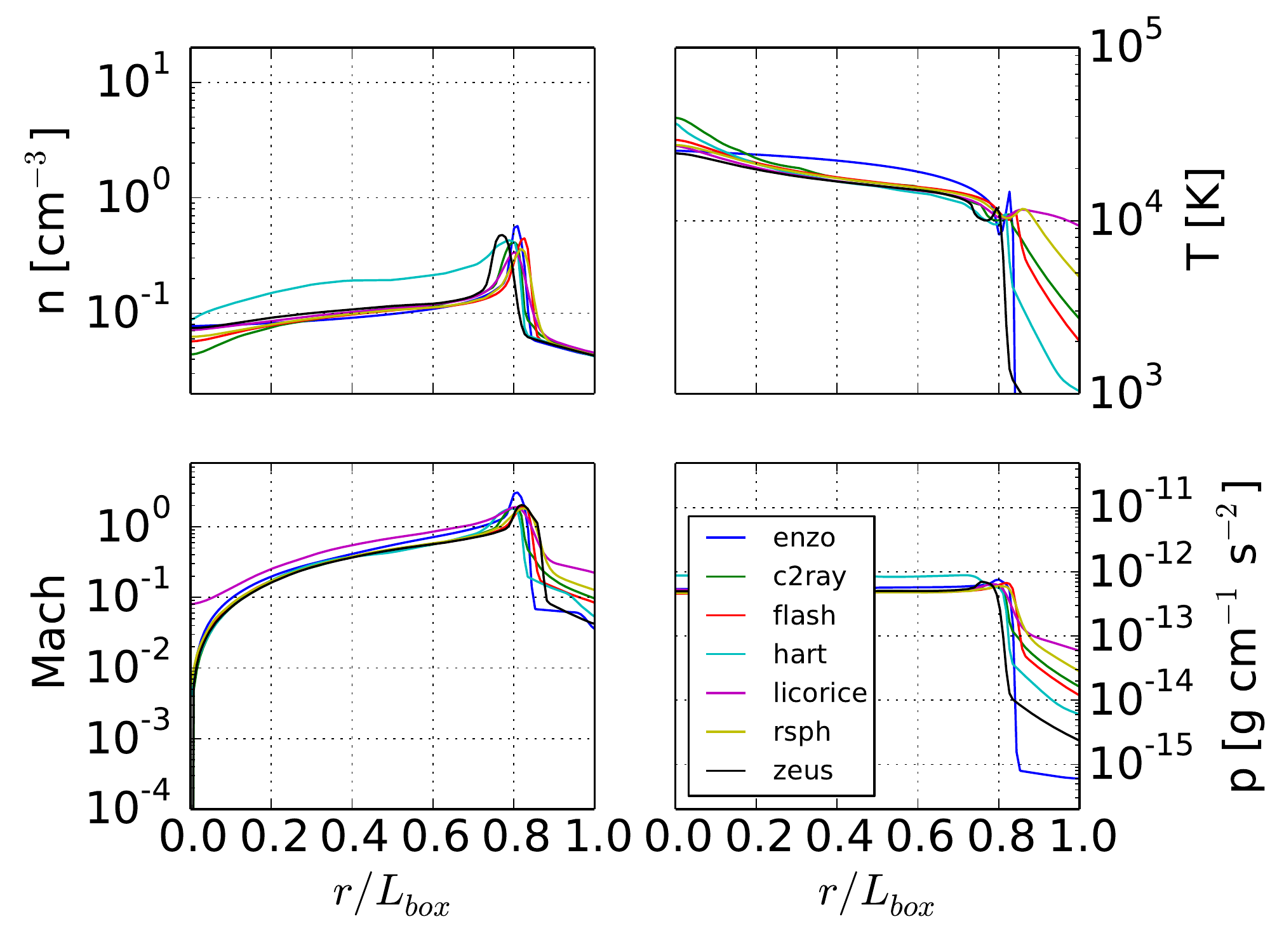}
  \hfill}
  \caption{Test 6 (HII region in a $r^{-2}$ density profile).
    (a) Top: growth of the computed I-front radius, computed as the
    radius with 50\% ionized fraction. Bottom: velocity of
    I-front radius, as computed from the upper plot at 0.5 Myr
    intervals.  (b-d) radial profiles at 3, 10 and 25 Myr, clockwise
    from top left as number density, temperature, pressure and Mach
    number. The curve labeled ``enzo" is computed with the FLD method
    presented in this paper. The other curves are taken from \cite{IlievEtAl2009}
    and labeled as in that reference. The lack of a multifrequency treatment in our FLD algorithm
    accounts for the lack of preheating and pressurization
    ahead of the I-front. Otherwise our results in are in good agreement with the majority
    of the results compared here.
    } 
  \label{fig:test6_plots}
\end{figure}

In Fig.~\ref{fig:test6_plots}a we show time histories
of the I-front radius and velocity, that exhibit strong agreement with
the results from both RT09 and WA11.  In 
Figs.~\ref{fig:test6_plots}b-d we plot radial profiles of the number
density, temperature, ionized fraction and pressure in the simulation
at 3, 10 and 25 Myr. We overplot the curves from Figs. 25-28 in 
\cite{IlievEtAl2009}, and label the curves as they do. 
Our results agree with reference
results from RT09 and WA11.  Perhaps the most notable difference may
be seen in the temperature and pressure profiles in comparison with
RT09, where our use of a grey approximation does not capture gas
preheating and pressurization ahead of the I-front.

\begin{figure}[t]
\centerline{\hfill
  \includegraphics[scale=0.42, trim=0.0cm 0.0cm 0.0cm 0.0cm]{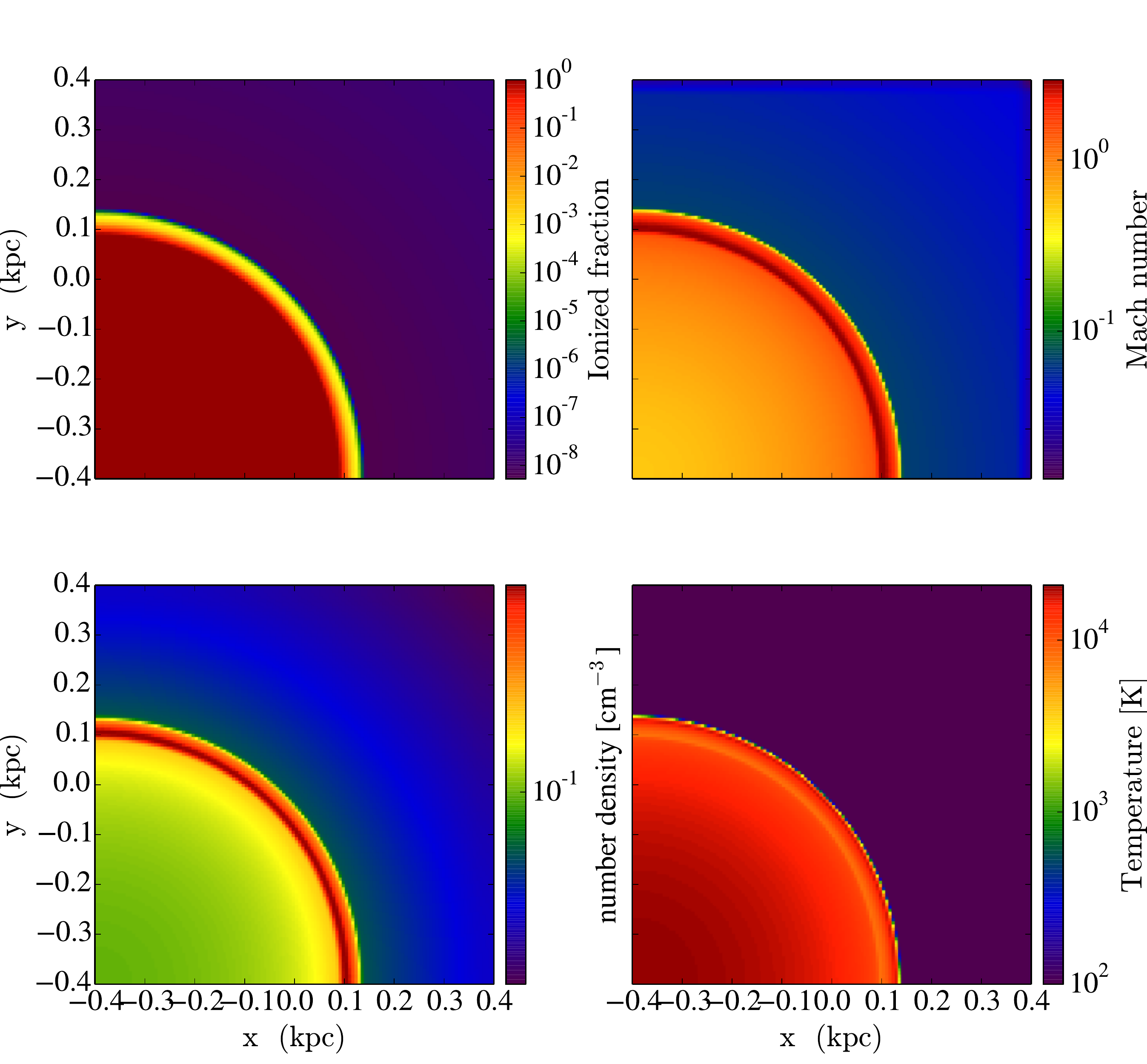}
  \hfill}
  \caption{Test 6 (HII region in a $r^{-2}$ density profile). Slices
    through the origin at 25 Myr.  Clockwise from top left: ionized
    fraction, Mach number, temperature and number density.}
  \label{fig:test6_slices}
\end{figure}

In Fig.~\ref{fig:test6_slices} we plot slices through the origin
of the ionized fraction, neutral fraction, temperature and number
density in the simulation at 25 Myr.  As it evident in these plots,
the FLD radiation approximation maintains a nearly spherical solution
profile throughout the simulation, with a slight anisotropy in the
non-coordinate aligned directions.  However, even with this minor
deviation of spherical symmetry, the results compare well against the
those in RT09 and WA11, many of which exhibit much more significant
anisotropy than that shown here.

\subsubsection{Test 7 -- Photoevaporation of a dense clump}
\label{subsubsec:test7}
As our second validation test we run Test 7 of RT09, also studied by WA11 using {\em Enzo+Moray}. 
This is a radiation hydrodynamic test involving ionizing radiation impinging on a dense, opaque spherical cloud
which is subsequently photoevaporated. RT09 set this up as a plane wave of ionizing radiation sweeping over the 
cloud, whereas WA11 illuminated the cloud with a single point source whose luminosity was adjusted to produce the same ionizing flux at the cloud. To facilitate comparison with the RT09 results, we implement the plane-wave illumination scheme, as described below. If run without hydrodynamics, the I-front is trapped in the dense cloud,
and the cloud casts a sharp shadow \citep{IlievEtAl2006}.  In reality, recombination radiation which would partially fill in the shadow zone \citep{AubertTeyssier2008}, but this effect has not been included in validation tests to date. With hydrodynamics engaged, the side of the cloud facing the source photoheats and expands, permitting a deeper
penetration of radiation into the cloud. Eventually, the entire cloud is photoevaporated. It is important to check
what FLD will do in this circumstance, in particular how the lack of a shadow affects the
photoevaporation time for the cloud. As we will now show, the effect is weak, validating our use of FLD for
large scale reionization problems.

The setup is as follows. A cubic domain 6.6 kpc on a side is employed, filled with an ambient medium of 
$n_H = 2 \times 10^{-4}$ cm$^{-3}$ and T=8000 K. The cloud is in pressure equilibrium with the intercloud
medium with density $n_H = 0.04$ cm$^{-3}$ and T=40 K. The cloud is a top-hat sphere with radius
$r_c$ = 0.8 kpc, and is centered at (x,y,z) = (5, 3.3, 3.3) kpc. The ionized fraction is initially zero everywhere. 
We implement the plane-wave illumination setup of RT09 by initializing an array of point sources on the left domain boundary, each emitting $10^6 A_{cell}$ ionizing photons/sec, where $A_{cell}$ is the area of the cell face. We have verified that this results in the correct flux of ionizing photons inside the domain. 

\begin{figure}[t]
\centerline{\hfill
  \includegraphics[width=0.75\textwidth]{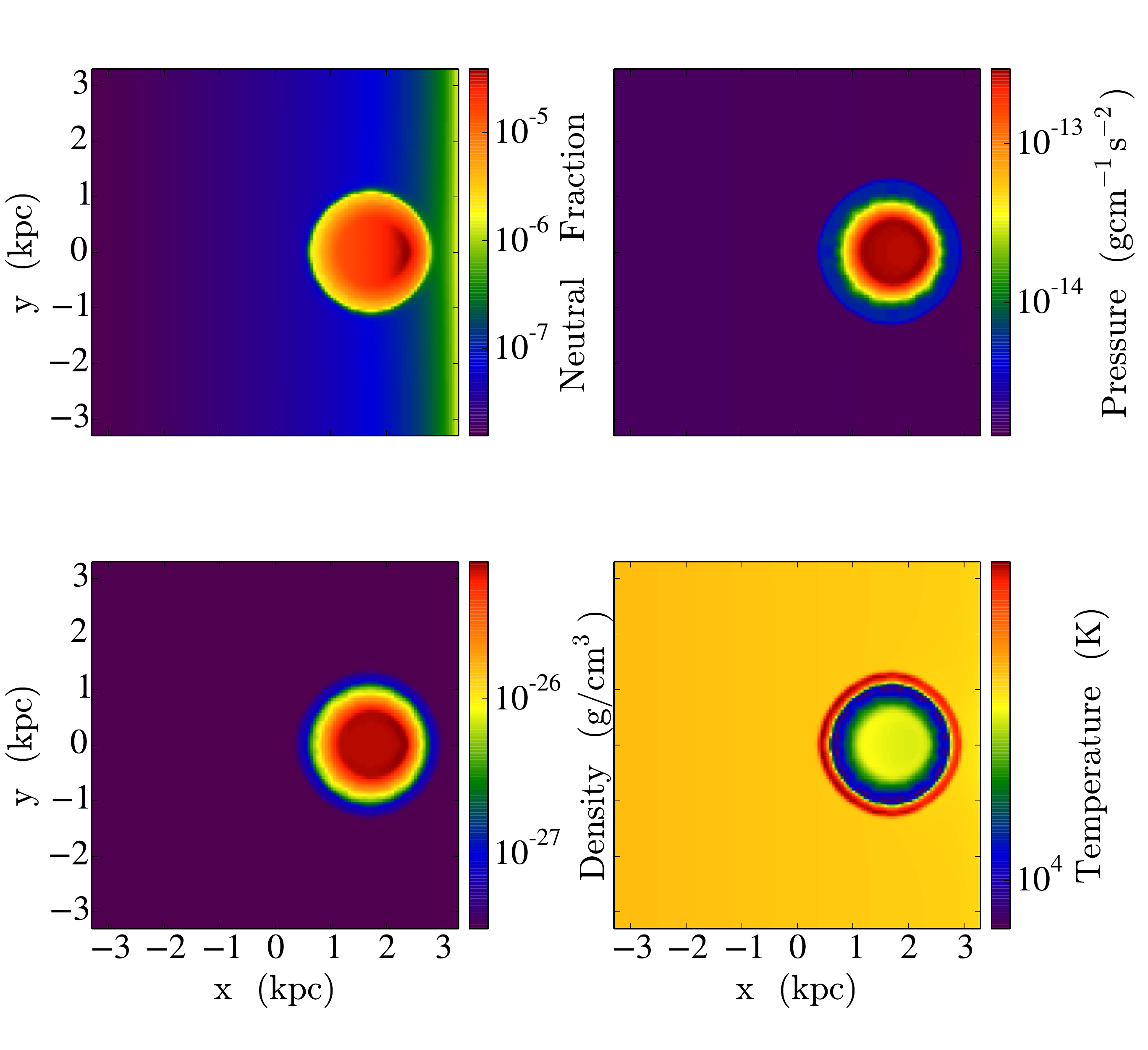}
  \hfill}
  \caption{Test 7. Photo-evaporation of a dense clump. Clockwise from upper left: Slices through the clump midplane of neutral fraction, pressure, temperature, and density at time $t=10$ Myr.}
  \label{fig:test7_slices_10}
\end{figure}

\begin{figure}[t]
\centerline{\hfill
  \includegraphics[width=0.75\textwidth]{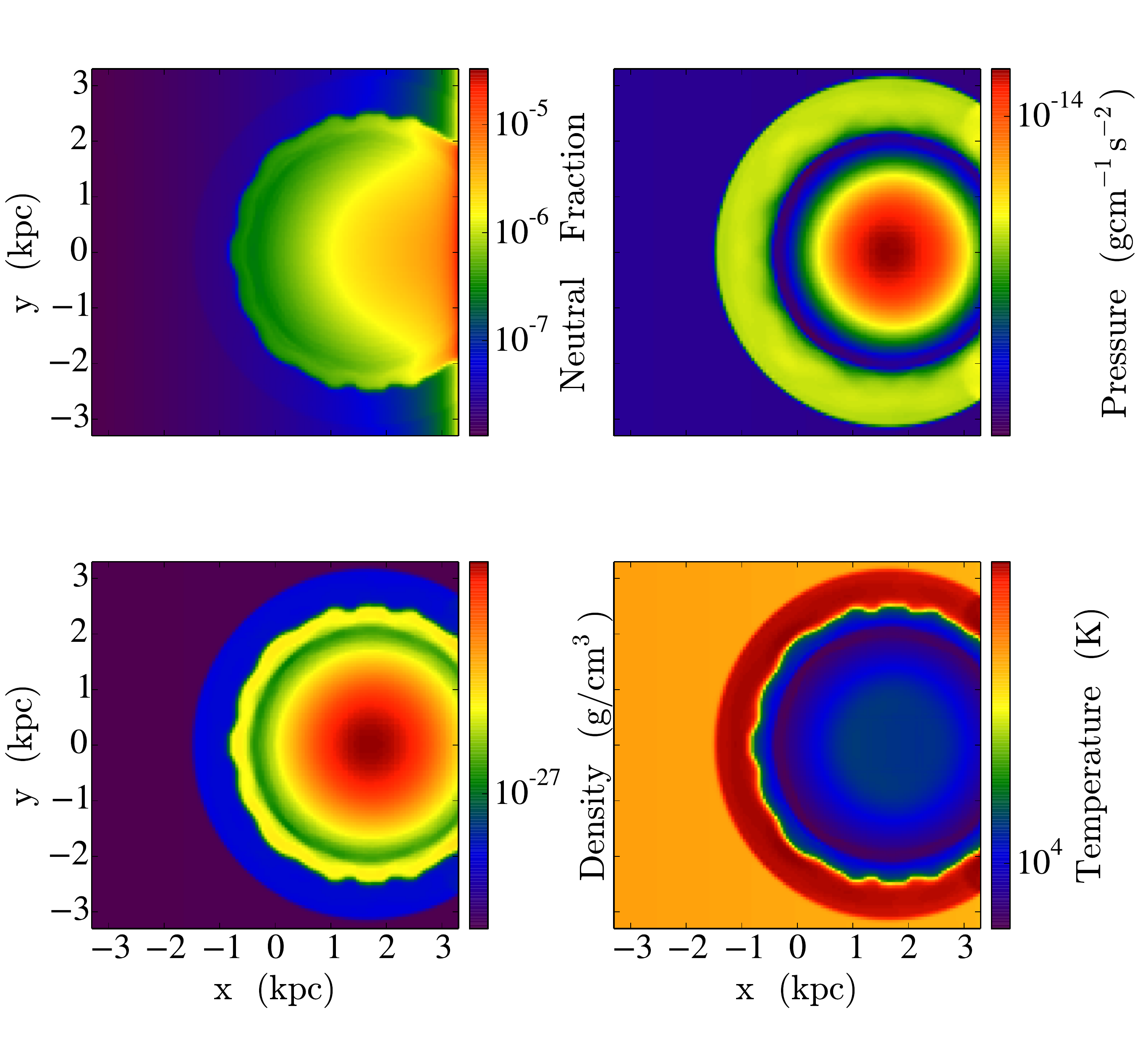}
  \hfill}
  \caption{Test 7. Photo-evaporation of a dense clump. Same as Fig. \ref{fig:test7_slices_10} at time $t=50$ Myr.}
  \label{fig:test7_slices_50}
\end{figure}

Figs. \ref{fig:test7_slices_10} and \ref{fig:test7_slices_50}  shows slices through the cloud midplane of neutral fraction, pressure, temperature, and 
density at time $t=$10 and 50 Myr, respectively. 
By 10 Myr the cloud is fully ionized, whereas the results in RT09 show for the 6 codes capable of casting shadows that the cloud is only partially ionized at this time. Inspection of Figs. 32 and 42 in RT09 suggest that the I front has only reached the center of the cloud after 10 Myr. The explanation for this large discrepancy is that within the FLD approximation, the radiation quickly flows around the cloud, filling in the shadow region within 1 Myr. The cloud is thus irradiated from all sides, not just one side. Therefore the time to ionize the cloud should be roughly the time it takes for the I-front to traverse a distance equal to the radius of the cloud. Ignoring attenuation, recombinations, and hydrodynamic effects the speed of the I-front in the dense gas is $v_{IF}=2.5 \times 10^7$ cm/s.  At this speed, it should take the I-front roughly 3.2 Myr to  reach the center of the cloud.

\begin{figure}[t]
\centerline{\hfill
  \includegraphics[width=0.7\textwidth]{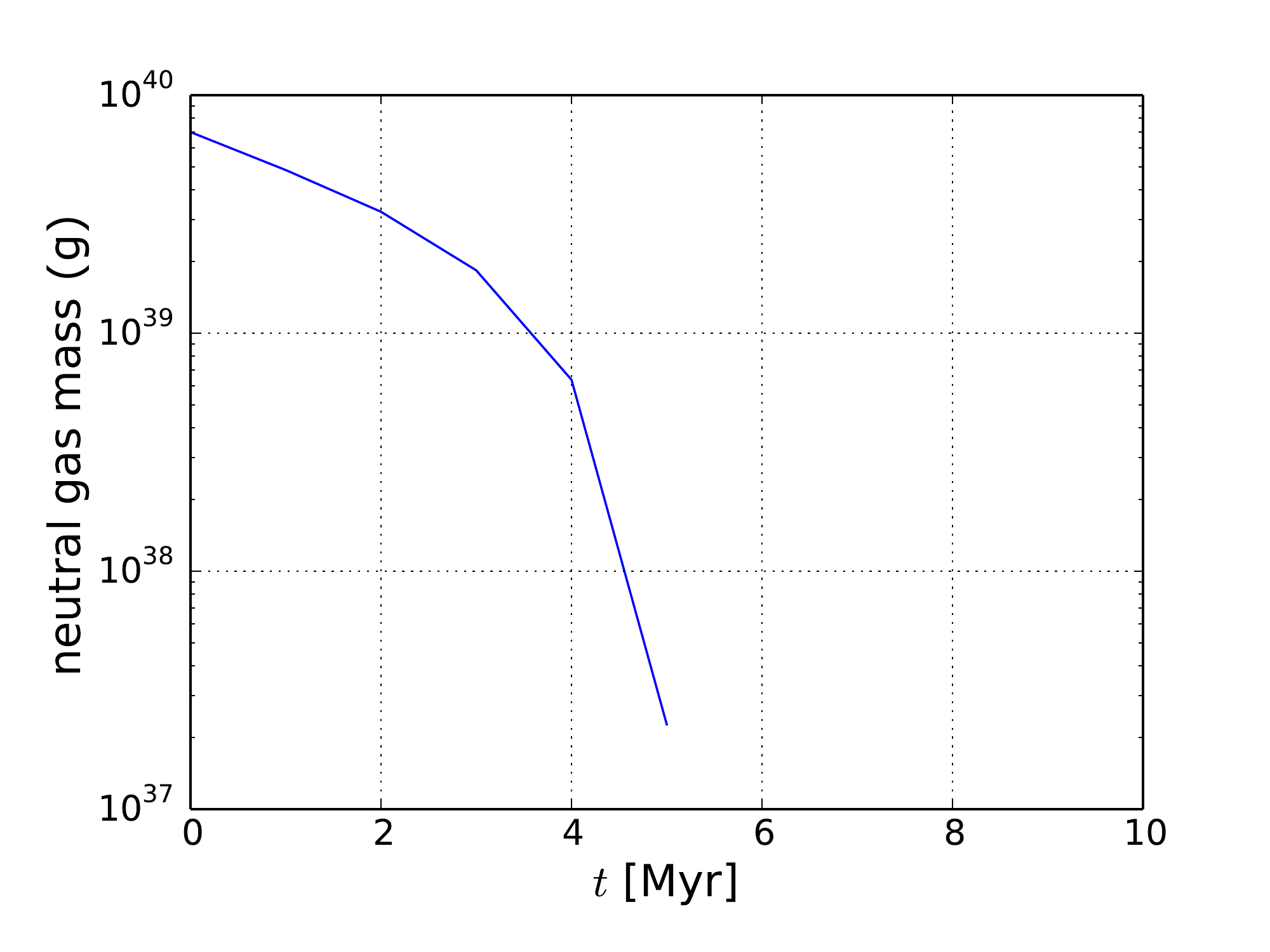}
  \hfill}
  \caption{Test 7. Time evolution of the mass of neutral hydrogen in the cloud, where neutral is defined as gas with an ionization fraction of less than 10\%.}
  \label{fig:time-evol}
\end{figure}

In Fig. \ref{fig:time-evol} we plot the time evolution of the mass of neutral hydrogen in the cloud, where neutral is defined as gas with an ionization fraction of less than 10\%. We see that the cloud becomes ionized on a timescale of about 5 Myr, somewhat longer than the estimate above due to attenuation, recombinations, and hydrodynamic effects. Naively one would expect a cloud irradiated from only one side to take twice as long 
($\sim 10$ Myr) to ionize since the I-front must propagate across the diameter. However this ignores attenuation of the ionizing flux between the edge and center of the cloud, which is
significant. The fact that the I-front becomes trapped in the static ionization test reinforces this point.

By 50 Myr the cloud has
expanded considerably due to photoevaporation, exhibiting a roughly spherical shape. 
The results of RT09 and WA11 are similar, except for a small wedge-shaped neutral patch on the back of the cloud, which casts 
a small shadow into the diffuse intercloud medium. 
In reality, ionizing recombination radiation from the denser cloud gas would partially fill in this shadow and ionize the 
diffuse gas there \citep{AubertTeyssier2008}, making it more like the FLD solution. 

\begin{figure}[t]
\centerline{\hfill
  \includegraphics[scale=0.42, trim=0.0cm 0.0cm 0.0cm 0.0cm]{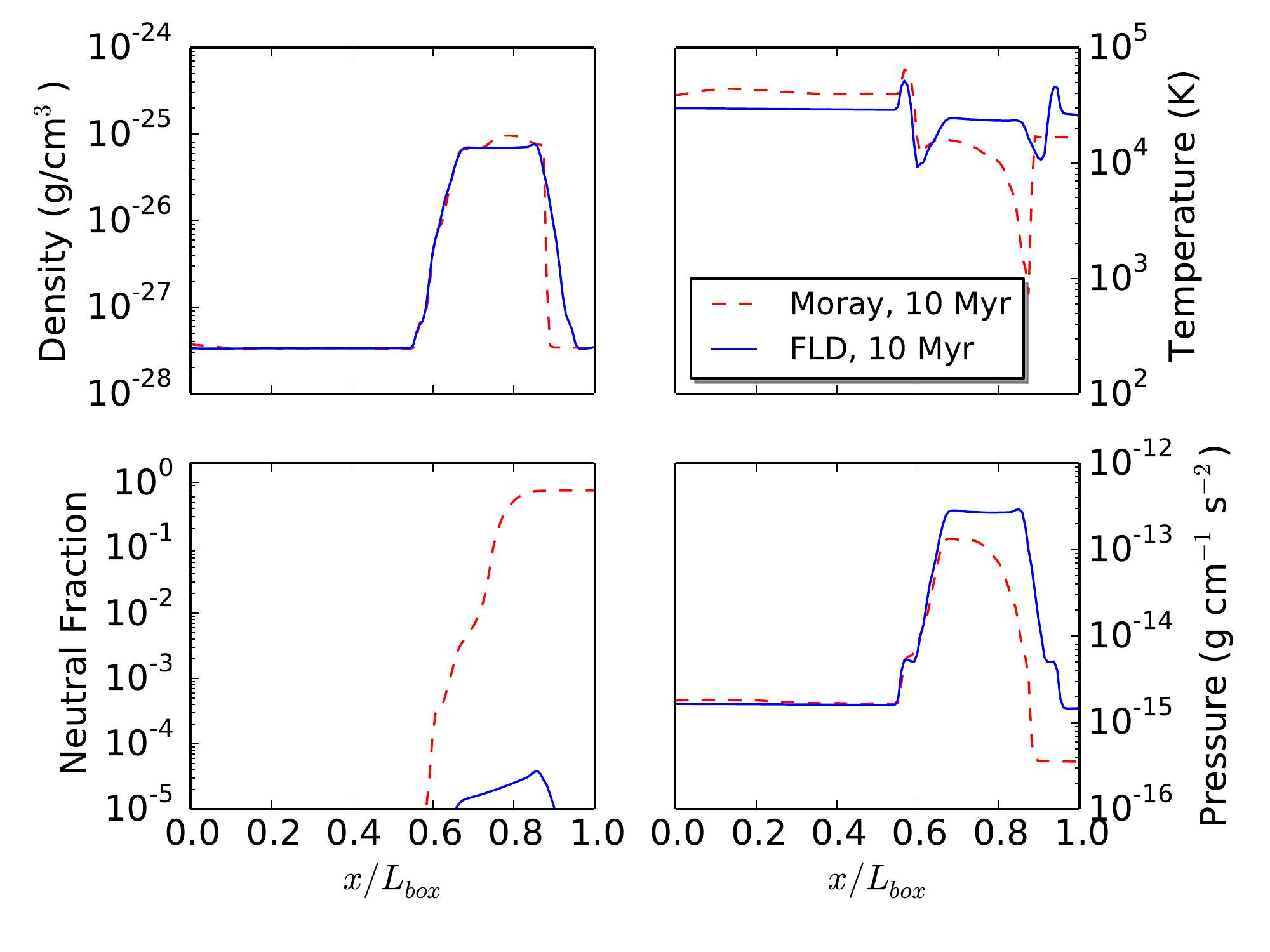}
  \includegraphics[scale=0.42, trim=0.0cm 0.0cm 0.0cm 0.0cm]{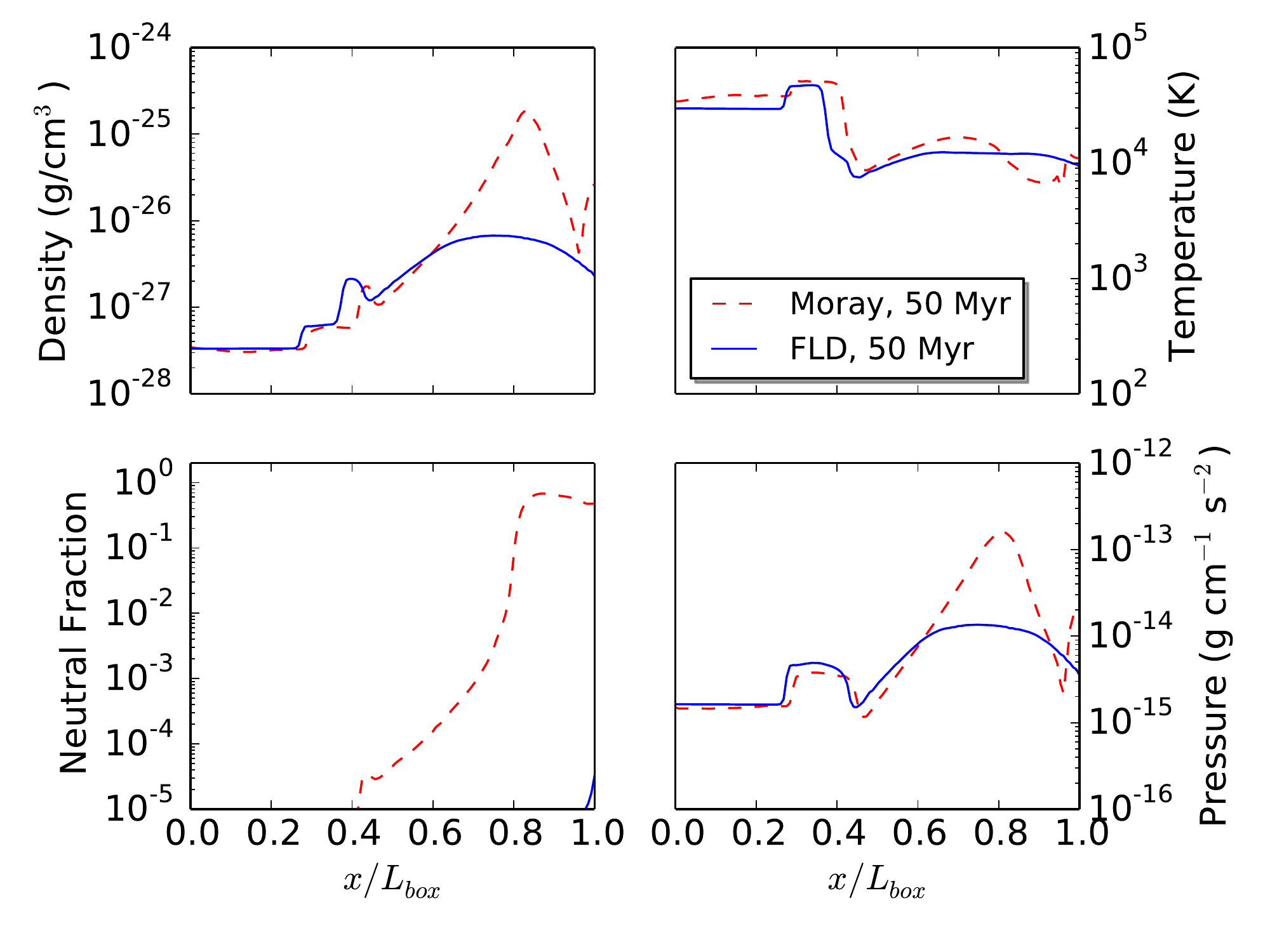}
  \hfill}
  \caption{Test 7. Photo-evaporation of a dense clump. Line cuts through the center of the cloud at 10 Myr (left figure) and 50 Myr (right figure). Lines labeled FLD and Moray are, respectively, results from our method and the {\em Enzo+Moray} code from \cite{WiseAbel11}. Each 4-panel figure plots (clockwise from upper left) density, temperature, pressure, and neutral fraction. }
  \label{fig:test7_profiles}
\end{figure}

To enable a more quantitative comparison with RT09 and WA11,
we plot in Figs. \ref{fig:test7_profiles}a-c line cuts from the point source through the center of the cloud at
$t=$ 10 and 50 Myr. For ease of illustration we only overplot the solution from WA11 which is in good agreement with the results presented in RT09.
Because the FLD method ionizes the cloud from all sides with only a small delay between
dayside and nightside irradiation, the FLD line cuts are basically symmetric about the center of the cloud, whereas the Moray line cuts show a strong asymmetry. This is particulary apparent in the temperature and neutral fraction line cuts at 10 Myr. At 50 Myr the Moray results show an appreciable amount of dense neutral gas remains on the backside of the cloud due to shadowing, whereas this is entirely absent in the FLD results. Both methods show good agreement on the position and structure
of the dense shell swept up by the expanding cloud at 50 Myr at $x/L_{box} \sim 0.4$.

\begin{figure}[t]
\centerline{\hfill
  \includegraphics[scale=0.42, trim=1.0cm 0.5cm 1.0cm 0.5cm]{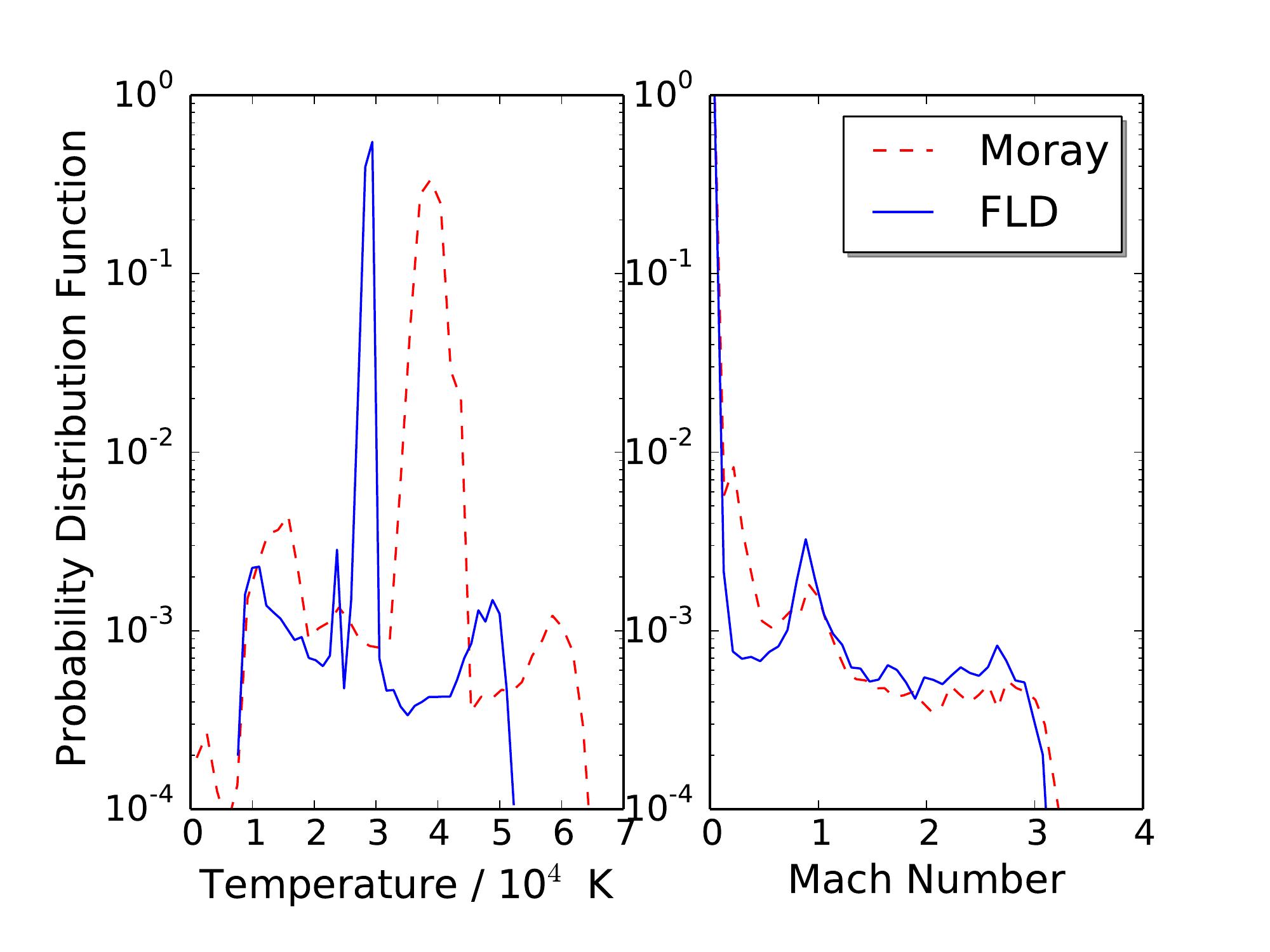}
  \includegraphics[scale=0.42, trim=1.0cm 0.5cm 1.0cm 0.5cm]{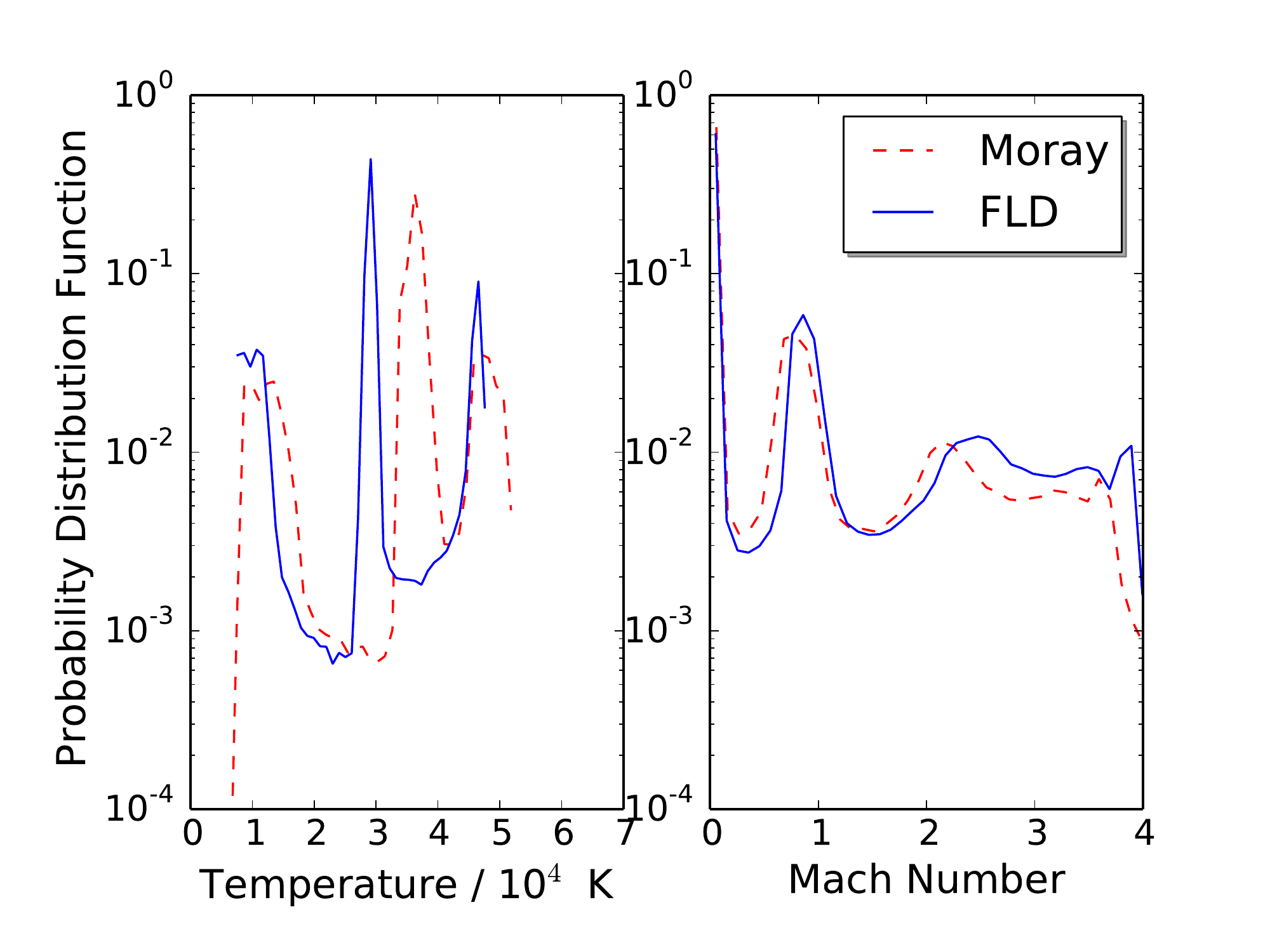}
  \hfill}
  \caption{Test 7. Photo-evaporation of a dense clump. Probability distribution functions for temperature and Mach number at 10 Myr (left figure) and 50 Myr (right figure). Lines labeled as described in Fig. \ref{fig:test7_profiles}. }
  \label{fig:test7_PDF}
\end{figure}

In Figs. \ref{fig:test7_PDF}a-b we show probability distribution functions for gas temperature and flow Mach number at 10 and 50 Myr for Test 7. As above, we overplot the Moray results to enable a direct comparison. The temperature plot shows a slightly narrower range and a lower value for the most common temperature ($3 \times 10^4$ K vs. $4 \times 10^4$ K) compared to the Moray results due to the different spectral treatments.
Otherwise the features are quite similar. At a given time, the flow Mach number distribution shows a somewhat higher maximum Mach number for the FLD results compared to the Moray results, presumably because the cloud ionizes sooner. Otherwise the distributions show similar features, especially at $t=50$ Myr. 

\begin{figure}[t]
\centerline{\hfill
  \includegraphics[width=0.7\textwidth]{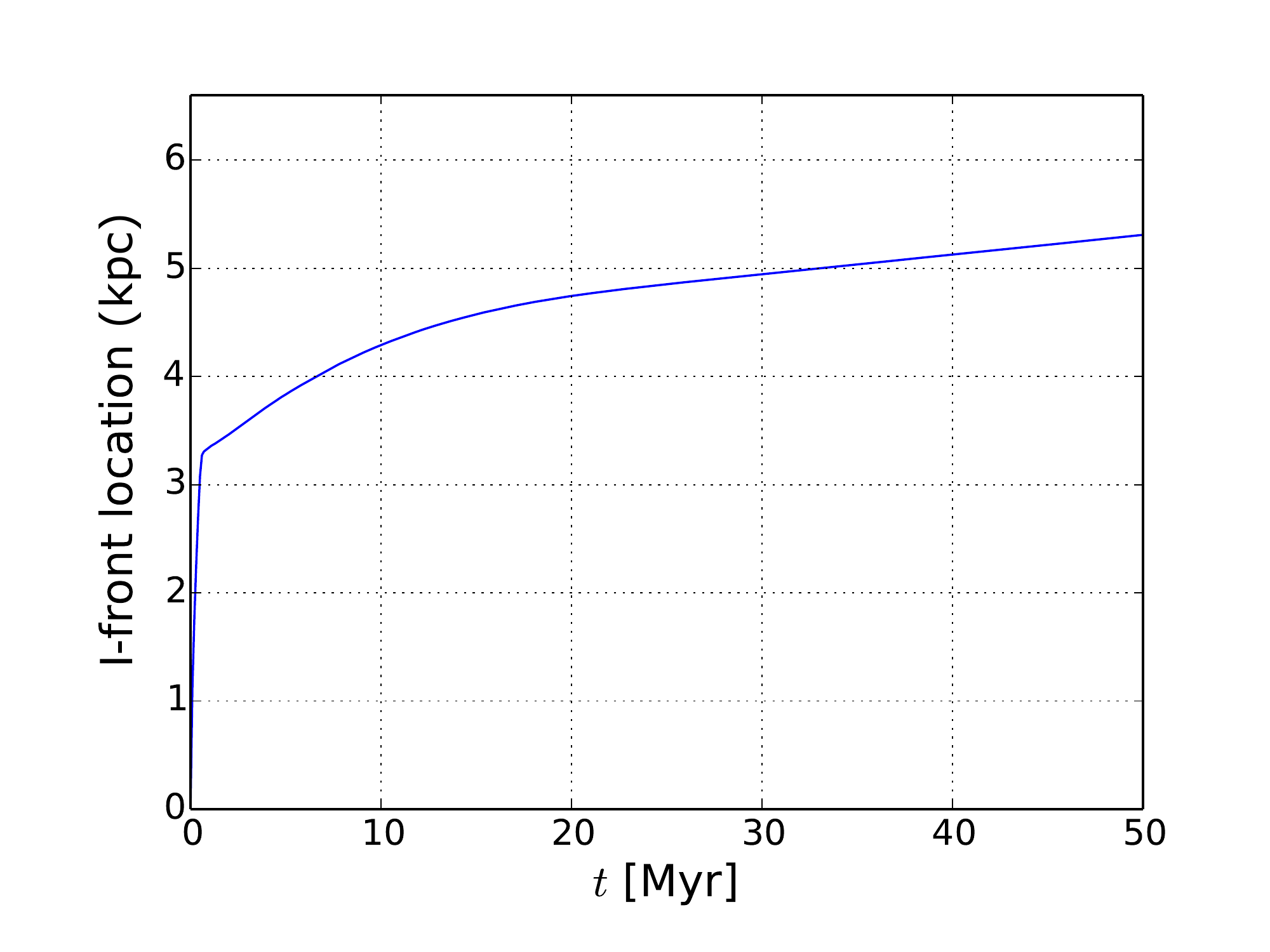}
  \hfill}
  \caption{I-front position vs. time in the 1D cloud photoevaporation problem described in the    text. The I-front takes negligible time to reach the front edge of the cloud at $x \approx  3.2$ kpc. It decelerates instantaneously as it enters the dense gas, becoming a D-type I-front. It  continuously decelerates thereafter as the ionizing flux is attenuated within the cloud. This trajectory is in good agreement with Fig. 30 in \cite{IlievEtAl2009} for the 3D  cloud photoevaporation problem (Test 7), indicating that our method is behaving properly in optically thick media, as a diffusion approximation must. }
  \label{fig:box_wall_hydro-Ifront_history}
\end{figure}

To  confirm that the rapid ionization of the cloud we see in the FLD calculation is not due to an incorrect propagation speed for the I-front in the cloud, but rather due to omnidirectional irradiation, we ran a 1D version of the opaque cloud test using our code. We placed a step function jump in density from $2 \times 10^{-4}$ cm$^{-3}$ to a value 200 times that at the center of the box; $x=3.3$ kpc. The position of the I-front versus time is shown in Fig. \ref{fig:box_wall_hydro-Ifront_history}. The I-front takes negligible time to reach the front edge of the cloud. It decelerates instantaneously as it enters the dense gas, becoming a D-type I-front. The I-front decelerates continuously as it traverses the front half of the cloud, and propagates very slowly after $\sim 15$ Myr. At 50 Myr, the I-front position is about 5.3 kpc, in excellent agreement with results presented in \cite{IlievEtAl2009}, Fig. 30.

Overall, the FLD calculation ionizes the cloud faster than predicted by methods that cast shadows. However by 50 Myr all methods
produce a cloud which is either fully ionized or nearly so, and there is good agreement on the size of the expanding cloud.  
The most significant difference is that the ray-tracing calculation predicts a small, neutral wedge-shaped patch on the nightside
of the cloud which is absent in the FLD calculation. It is unlikely that this difference will be important in large scale reionzation
simulations since clouds will be irradiated from multiple directions during the overlap phase. However, in the early stages of reionization (isolated \hii region expansion phase) we might expect FLD to ionize the IGM somewhat faster than methods that cast shadows as opaque clouds would be irradiated primarily by a single dominant source. We verify this conjecture in Sec. \ref{subsec:FLDvsRT}.

\subsubsection{Consolidated HII region with two sources}
As our third validation test we demonstrate the performance of our FLD radiative transfer method on a consolidated \hii region with two nearby point sources of equal luminosity. This problem was introduced by Petkova \& Springel (2009) (hereafter PS09) and included in the extensive suite of tests carried out by WA11. This is a validation problem because it has no analytic solution (that we know of.) PS09 studied it because their method uses the variable tensor Eddington factor moment method \citep{StoneMihalasNorman92,NormanAbelPaschos98,HayesNorman2003}  where the Eddington tensor is computed assuming the medium is optically thin everywhere \citep{GnedinAbel2001}. As discussed in \cite{GnedinAbel2001} it is known that the shapes of consolidated \hii regions are slightly inaccurate due to the optically thin assumption, in the sense that the \hii region is  more elongated in the axial direction, and less expanded in the transverse direction than in reality. 
The solution presented by PS09 shows this elongation. The solution presented
in WA11, which shows rounder but still slightly elongated I-fronts, should be a closer approximation to truth since it is calculated using adaptive
ray tracing which in principle gets the geometric effects correct. However the omission of 
diffuse ionizing recombination radiation which becomes dominant near a stalled I-front
means that even the WA11 solution is an approximation to the true shape. It is thus interesting
to see what FLD produces for this problem. 

The setup is as follows. Two sources with luminosities of $5 \times 10^{48}$ photons/s
are separated by 8 kpc. The ambient medium is static with uniform density $10^{-3}$
cm$^{-3}$ and T = $10^4$ K. The computational domain is 20 kpc in width and 10 kpc
in height and depth, and resolved with mesh of $128 \times 64 \times 64$ cells. The 
problem is evolved for 500 Myr, which is long enough for the consolidated HII region to
evolve to a steady state. 

\begin{figure}[t]
\centerline{\hfill
  \includegraphics[width=0.75\textwidth]{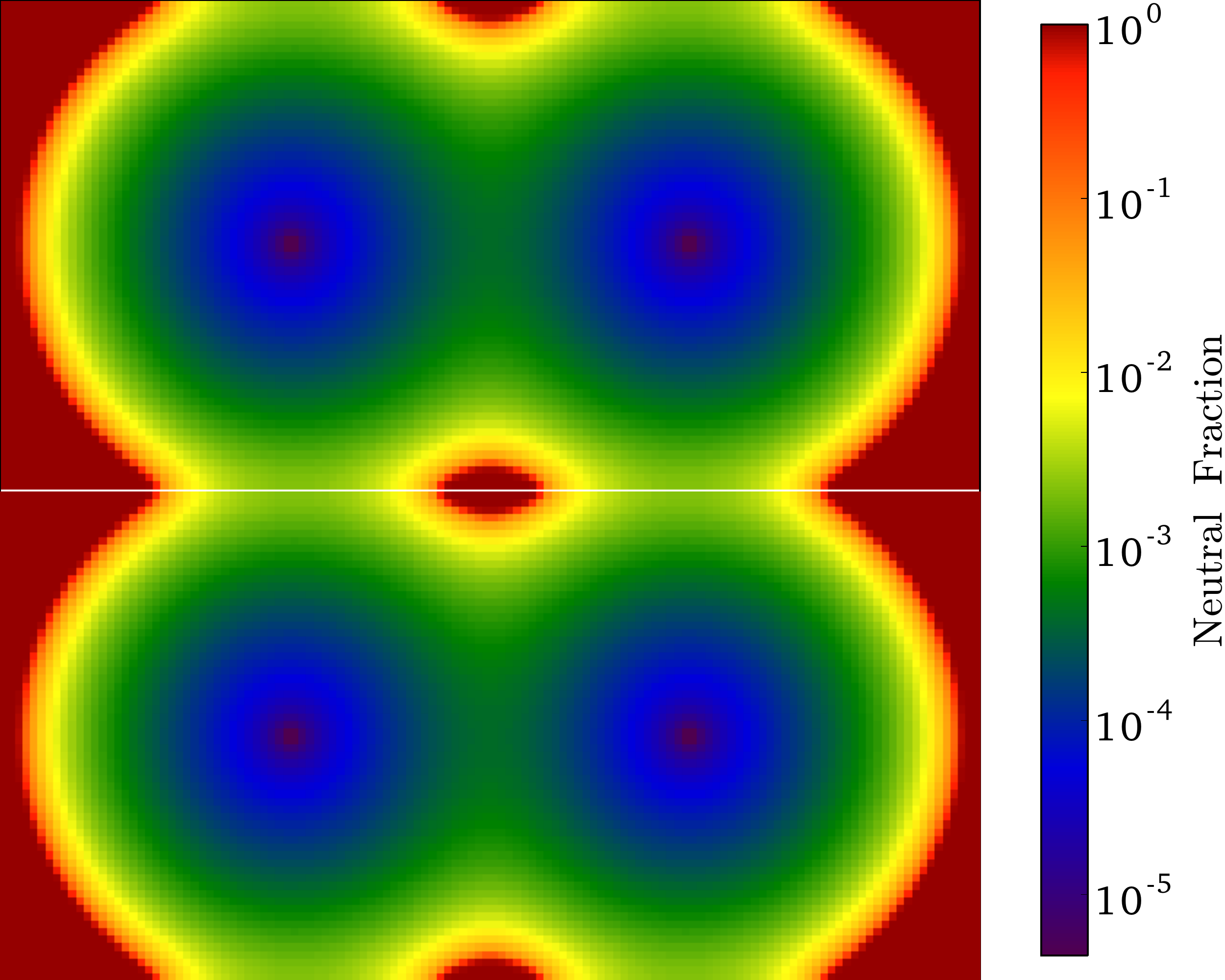}
  \hfill}
  \caption{Consolidated HII region test - slices of the neutral fraction through the $x-y$ and $x-z$ planes at $t=500$ Myr.}
  \label{fig:consolidated}
\end{figure}

Fig. \ref{fig:consolidated} shows slices of neutral fraction on $x-y$ and $x-z$ planes 
through the axis connecting the sources. The consolidated
\hii region is similar in size to the solution presented by WA11, but noticeably rounder
near its extremities. We do not include diffuse ionizing recombination radiation in our
formalism, and thus this must be a consequence of FLD. Since we are solving the same
problem, we expect the WA11 solution is closer to the truth, but note that the FLD solution is an acceptable approximation to truth given our intended application to large scale reionization.

\subsubsection{Comparing FLD and Ray Tracing on a Cosmological Reionization Test Problem}
\label{subsec:FLDvsRT}
In  \cite{So2014} we introduced a new test problem to directly compare the results of Enzo's adaptive ray tracing scheme {\em Moray} with FLD in order to assess the importance of shadowing. We examined the evolution of the ionized hydrogen volume and mass fractions in a fully-coupled cosmological reionization simulation in a 6.4 comoving Mpc box. We also examined PDFs of temperature and ionization fraction as a function of baryon overdensity. We showed that while FLD ionizes somewhat faster than ray tracing at early times ($Q_{\hii}<<1$), the late time evolutions are very similar. We hypothesized that the reason for this is that early times, shadowing is more important since opaque clouds are primarily irradiated by a single dominant source driving the expansion of isolated \hii regions. At late times, as the ionized volume fraction approachs unity, a given opaque neutral cloud will be irradiated from multiple directions--a situation that FLD approximates well. Here we present additional analysis. 

The test problem is a scaled down version of the 20 Mpc simulation described in Paper II. It is identical in physics model, mass resolution, and spatial resolution, but is in a volume $(20/6.4)^3 \approx 30\times$ smaller. The simulation is carried out on a uniform mesh of $256^3$ grid cells and an equal number of dark matter particles. The reader is referred to Paper II for additional details.

\begin{figure}[t]
\centerline{\hfill
  \includegraphics[width=0.5\textwidth]{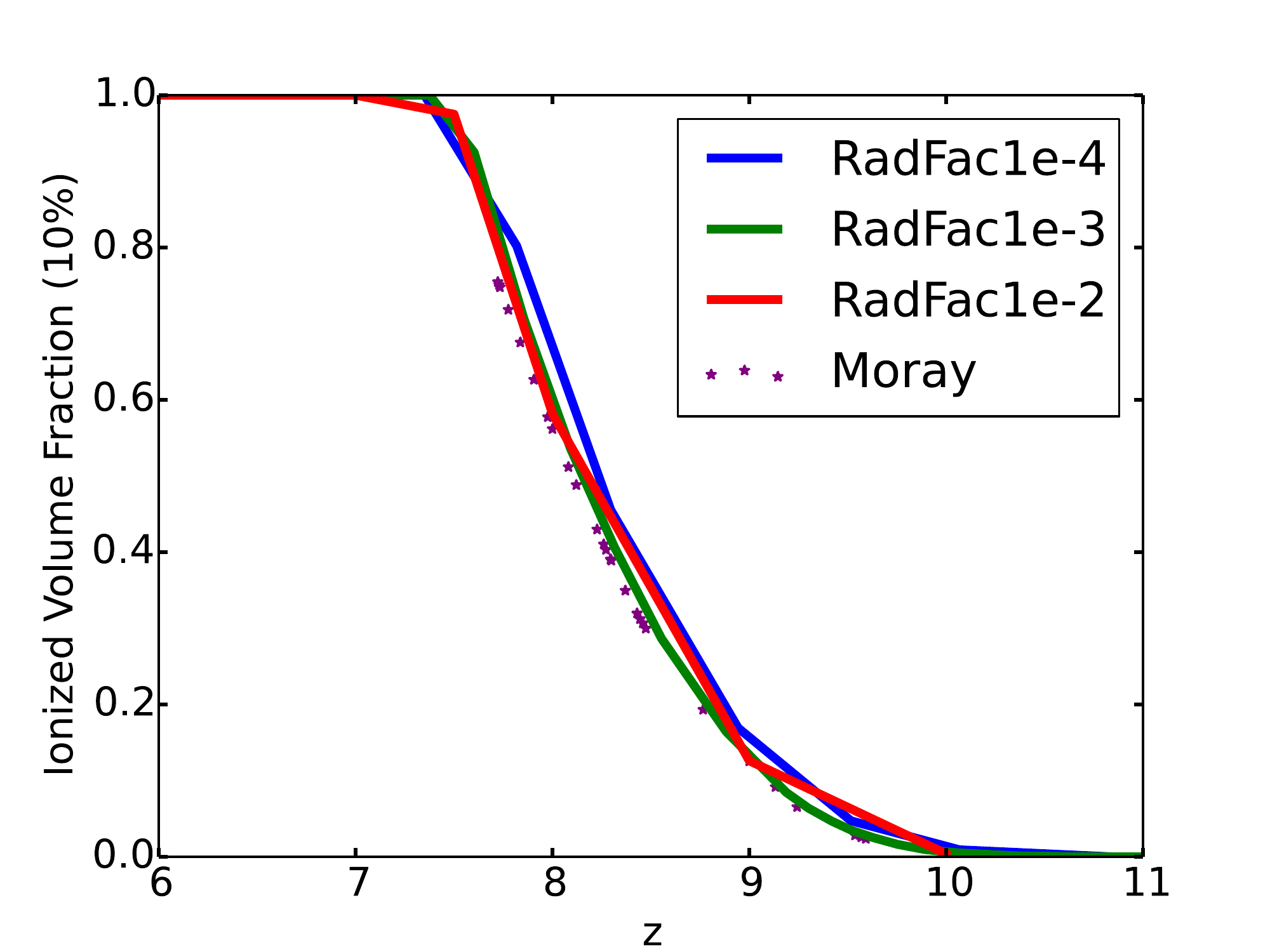}
  \includegraphics[width=0.5\textwidth]{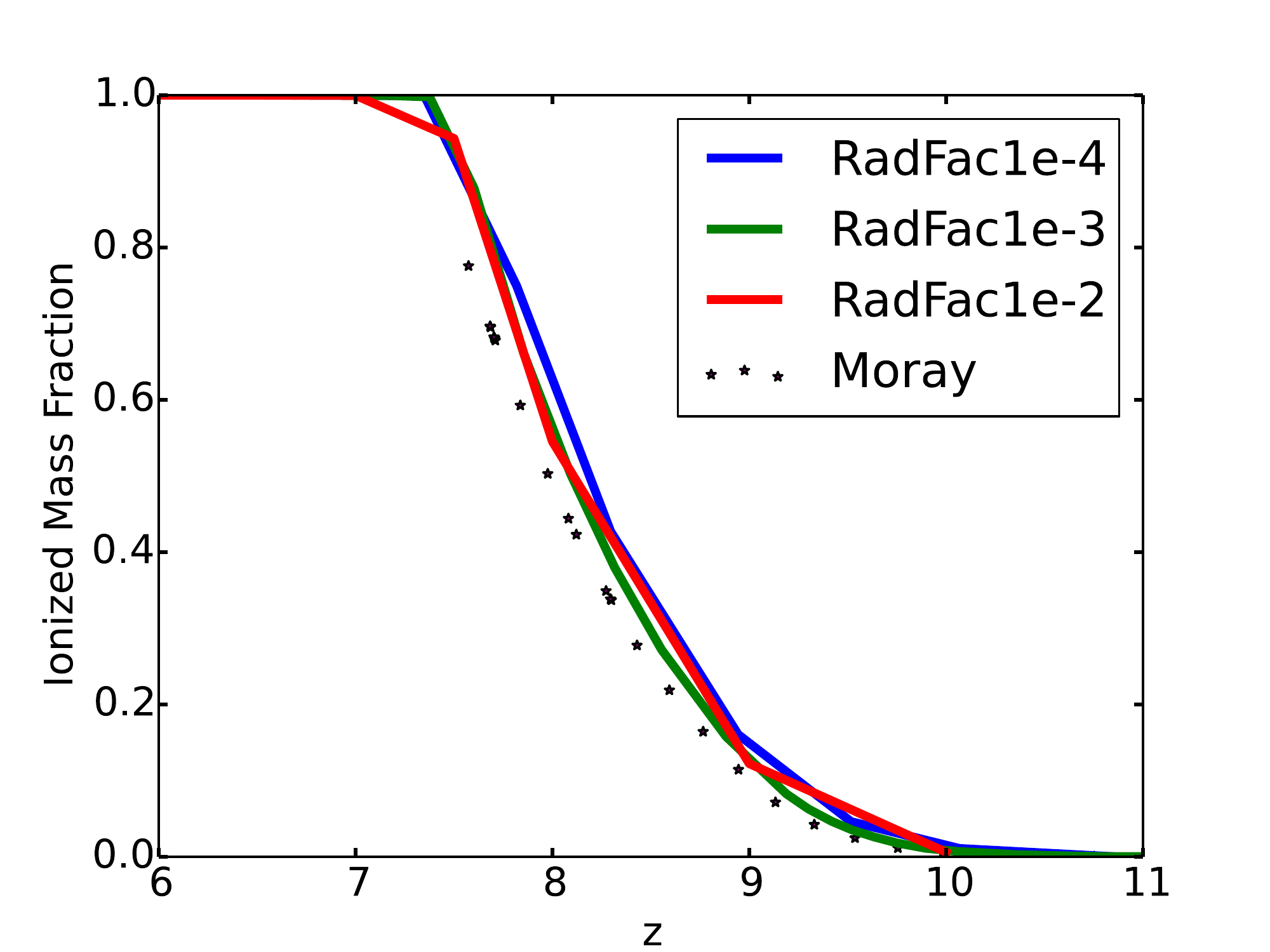}
  \hfill}
  \caption{Comparing FLD and ray tracing solutions on a cosmological reionization test problem. (a) evolution of ionized hydrogen volume fraction, and (b) evolution of ionized hydrogen mass fraction for gas that is at least 10\% ionized. In each figure we plot the {\em Moray} curve and three curves from FLD simulations with three different radiation timestep tolerance parameters $\tau_{tol}$. Note that the sampling intervals are different for each simulation, which accounts for the jagged FLD curves. For a smoother representation, see Figure 32 in Paper II.}
  \label{fig:ionization_evolution}
\end{figure}

In Fig. \ref{fig:ionization_evolution} we plot the evolution of the ionized hydrogen volume and mass fractions for the {\em Moray} simulation, and three FLD simulations differing in the choice for the radiation transport timestep control factor $\tau_{tol} = 10^{-2}, 10^{-3}, 10^{-4}.$ Note that the sampling intervals are different for each simulation, which accounts for the jagged FLD curves. For a smoother representation for the $\tau_{tol} = 10^{-3}$ case see Figure 32 in Paper II. Generally we see that the FLD simulations ionize slightly faster than the ray tracing simulation, with this being more evident in the mass-weighted curves as compared with the volume-weighted curves. This is qualitatively what we would expect given the results of the Test 7 described above, wherein we showed that FLD completely ionizes an opaque cloud irradiated from one side whereas ray tracing leaves a small neutral patch on the ``nightside" of the cloud. 

\begin{figure}[t]
\centerline{\hfill 
  \includegraphics[width=0.5\textwidth]{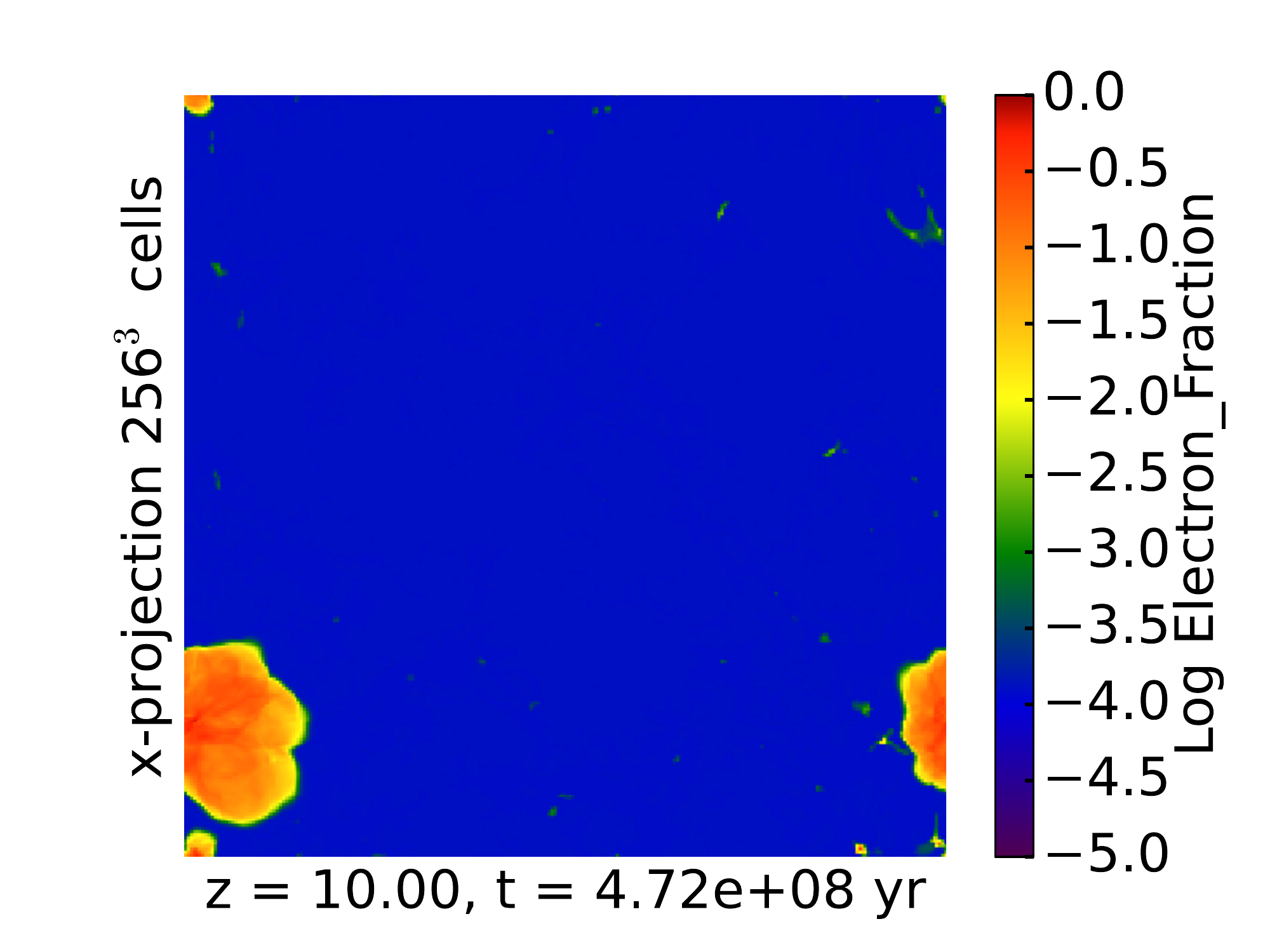}
  \includegraphics[width=0.5\textwidth]{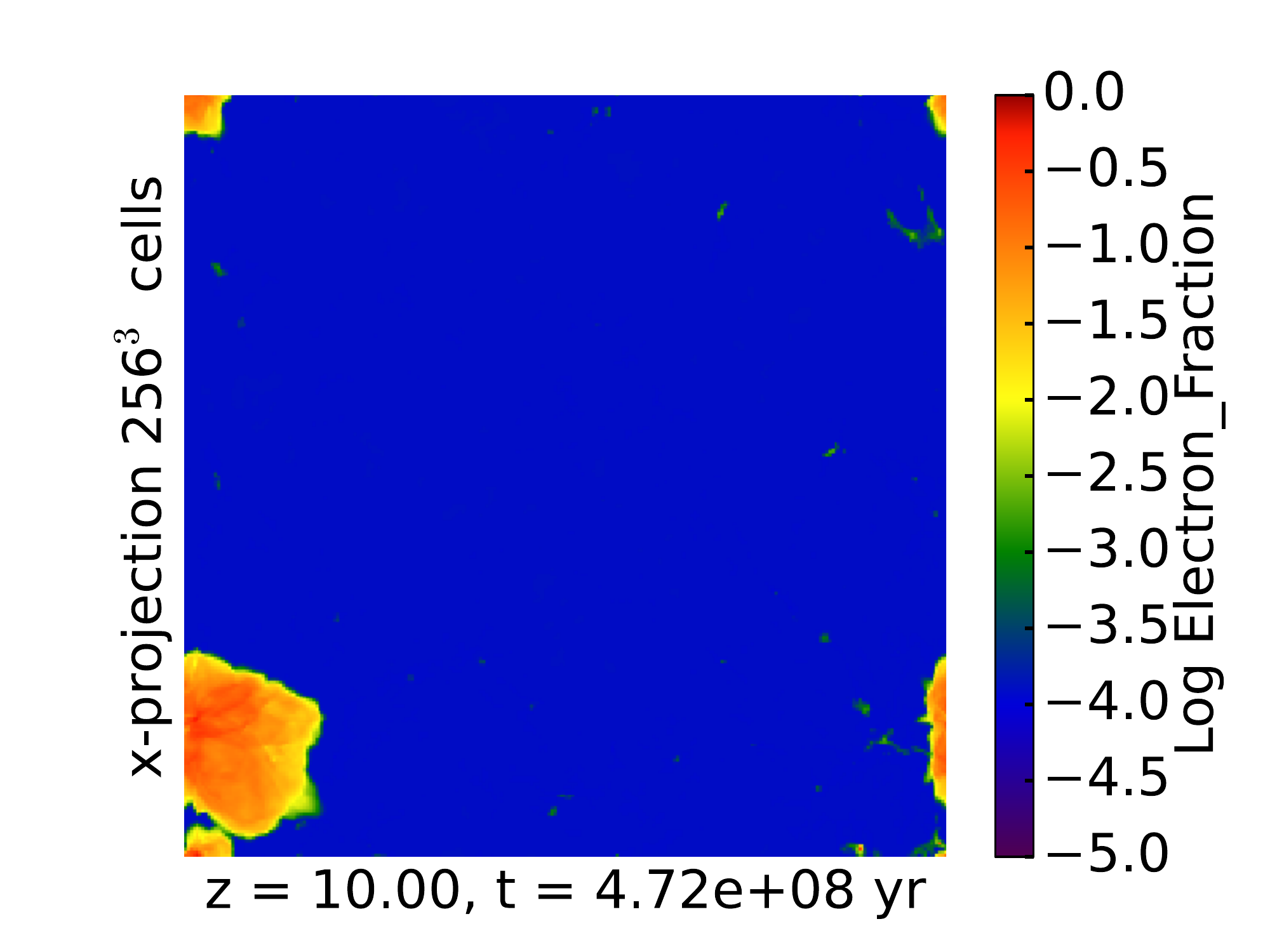}
  \hfill}
  \centerline{\hfill
  \includegraphics[width=0.5\textwidth]{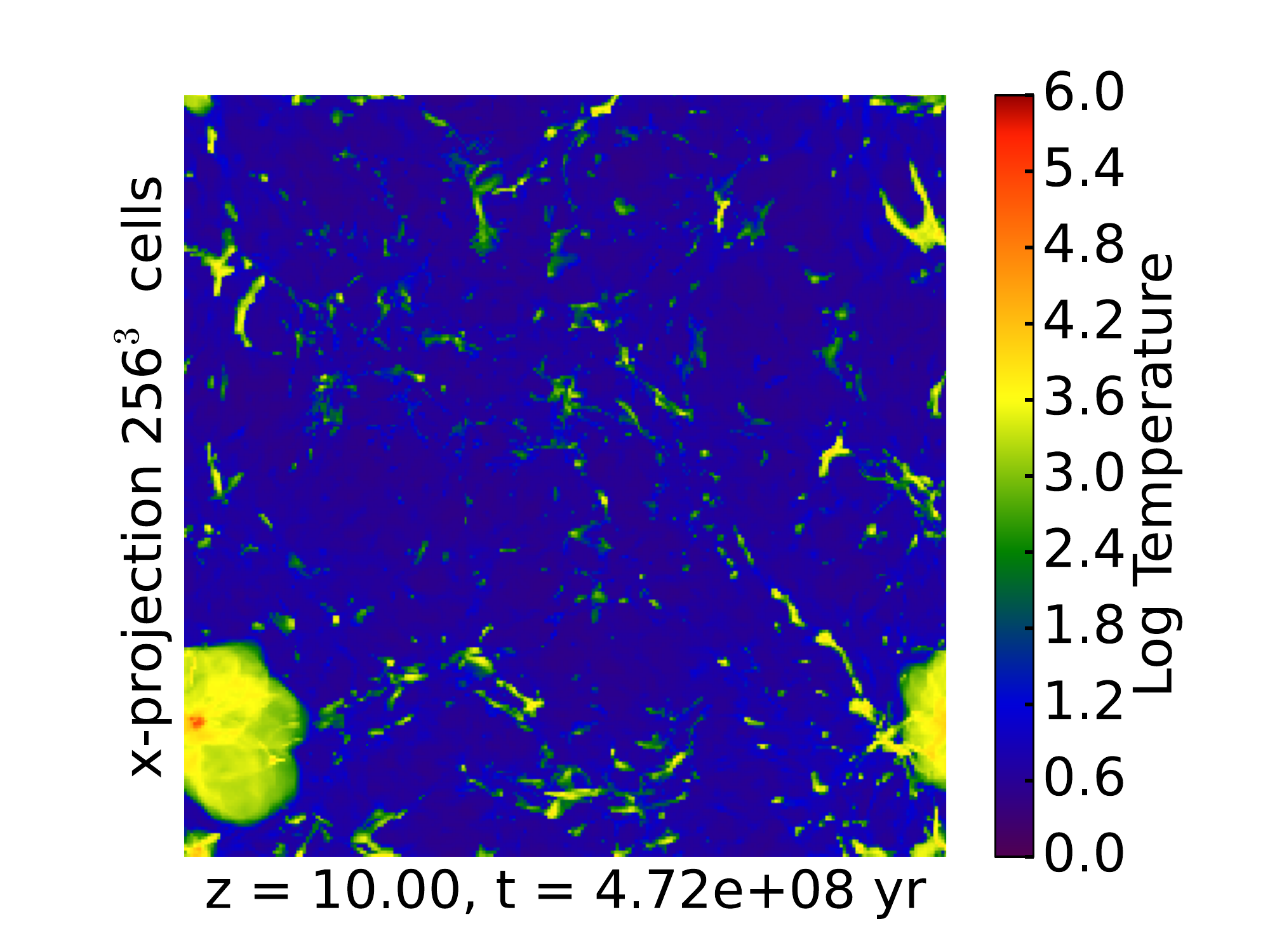}
  \includegraphics[width=0.5\textwidth]{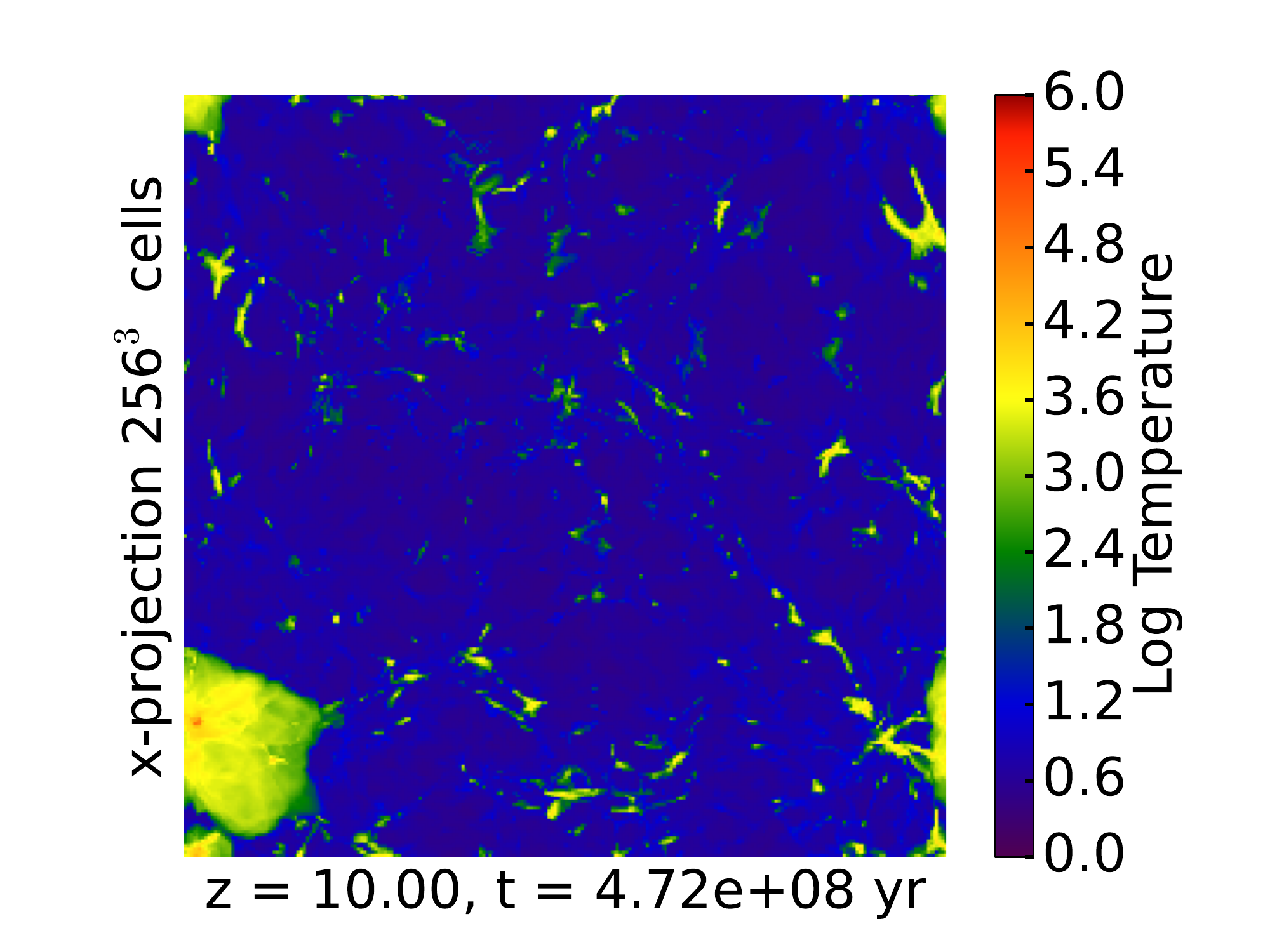}
  \hfill}
  \caption{Comparing FLD and ray tracing solutions on a cosmological reionization test problem. Shown are projections of density-weighted electron fraction (top row) and temperature (bottom row) for FLD (left column) and {\em Moray} adaptive ray tracing solutions (right column). Results are shown for $z=10$.}
  \label{fig:FLD_Moray_z10}
\end{figure}

\begin{figure}[t]
\centerline{\hfill
  \includegraphics[width=0.5\textwidth]{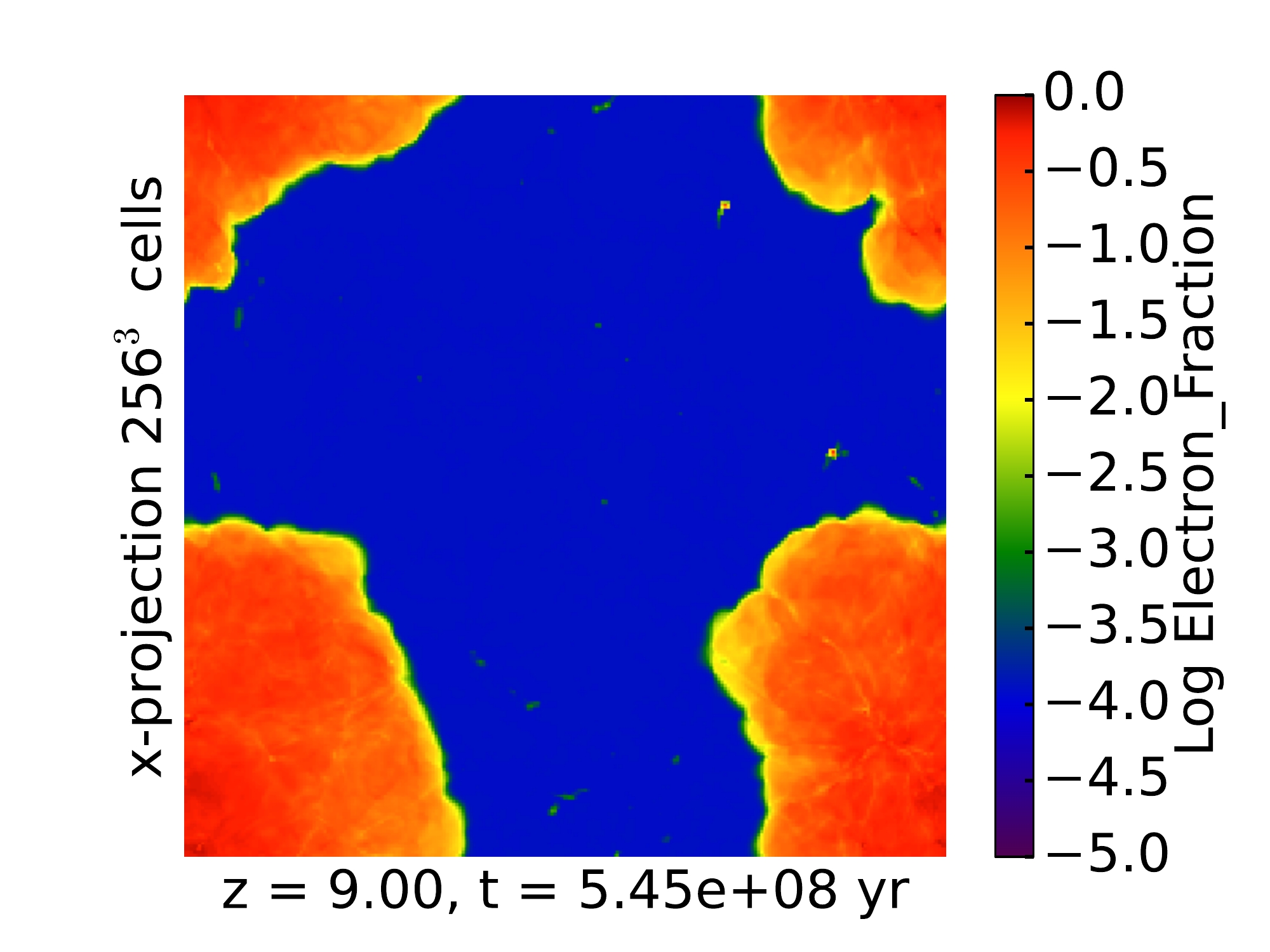}
  \includegraphics[width=0.5\textwidth]{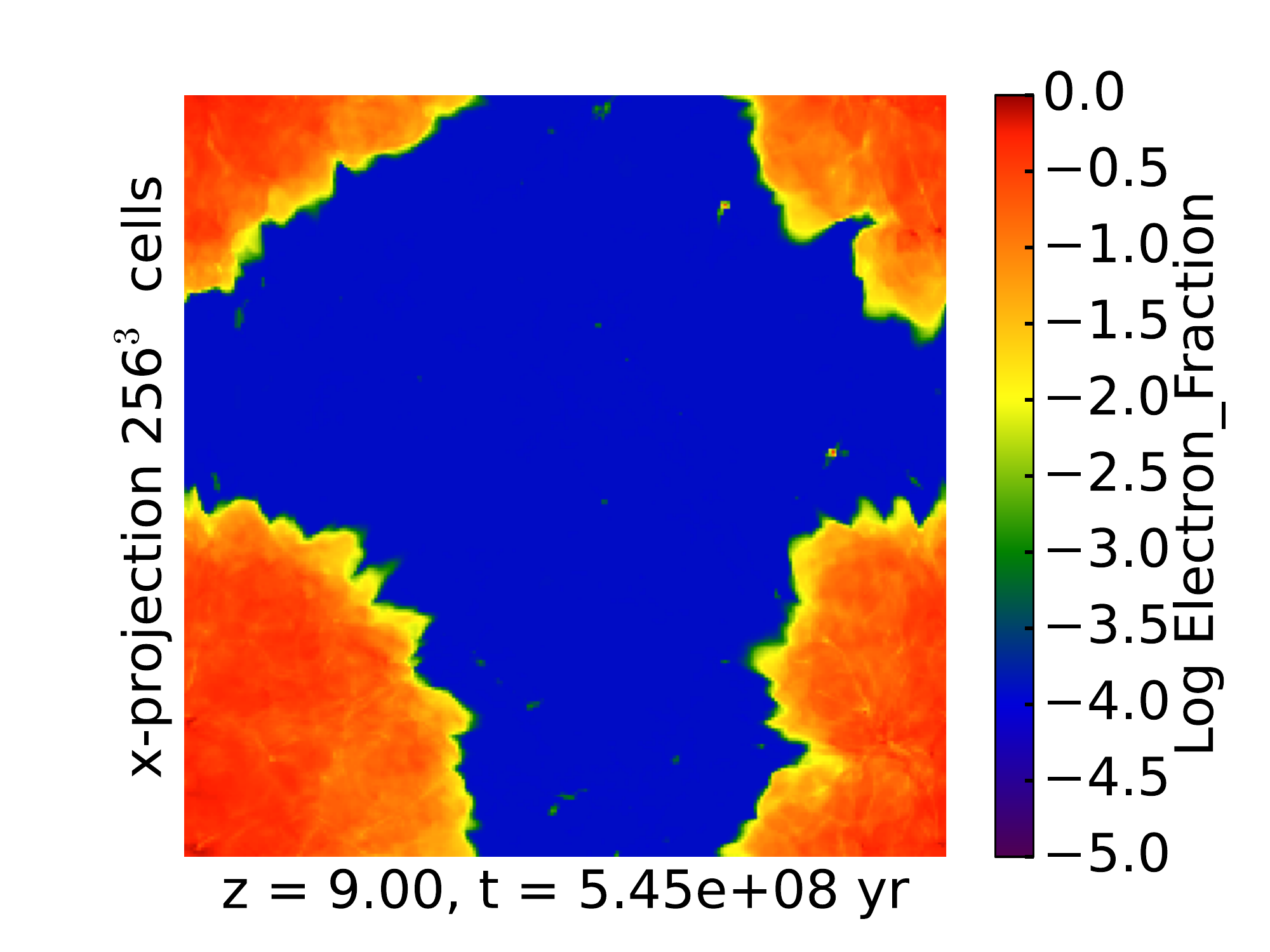}
  \hfill}
\centerline{\hfill
  \includegraphics[width=0.5\textwidth]{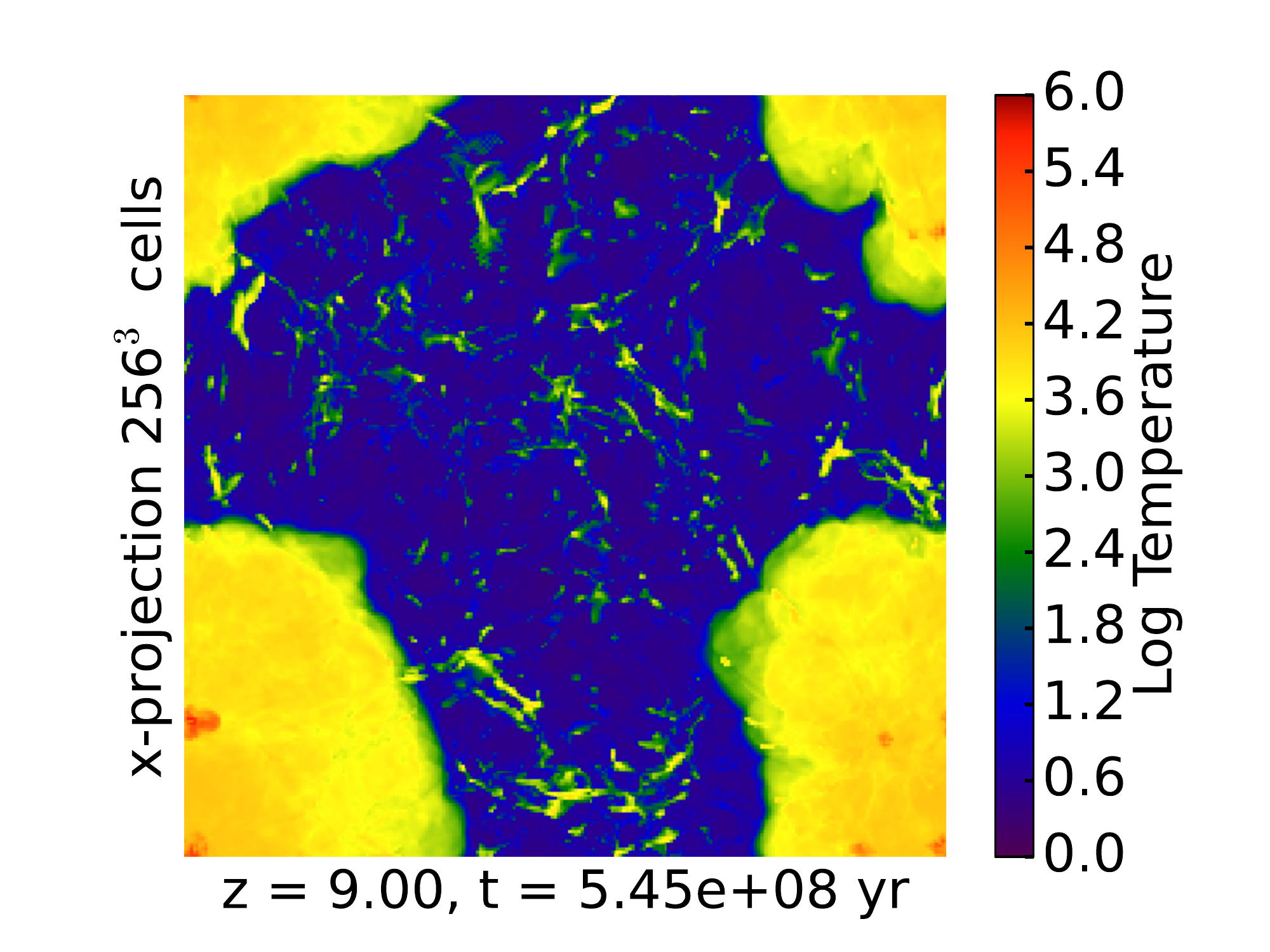}
  \includegraphics[width=0.5\textwidth]{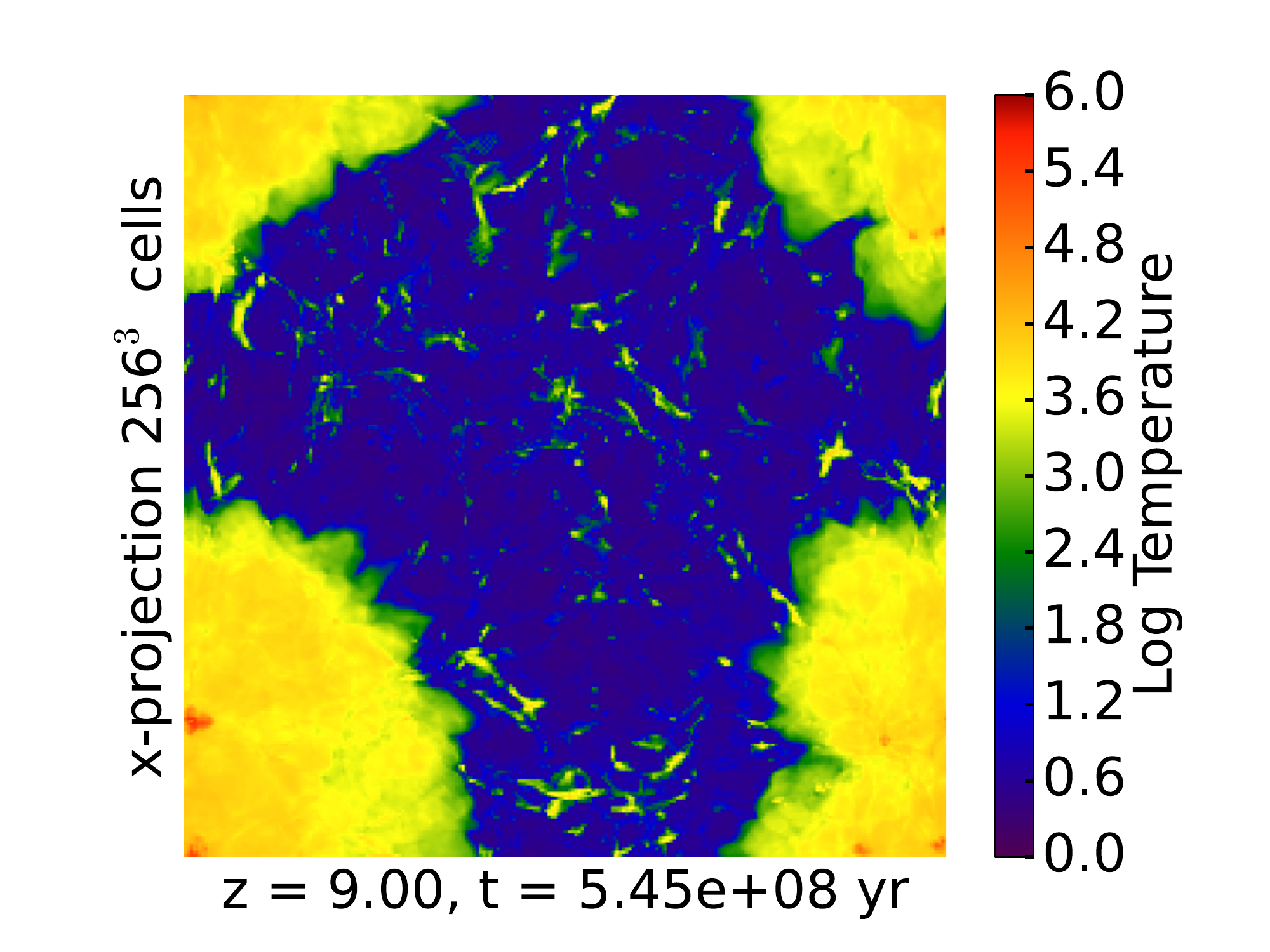}
  \hfill}
  \caption{ Same as Fig. \ref{fig:FLD_Moray_z10} except for $z=9$.}
  \label{fig:FLD_Moray_z9}
\end{figure}

\begin{figure}[t]
\centerline{\hfill
  \includegraphics[width=0.5\textwidth]{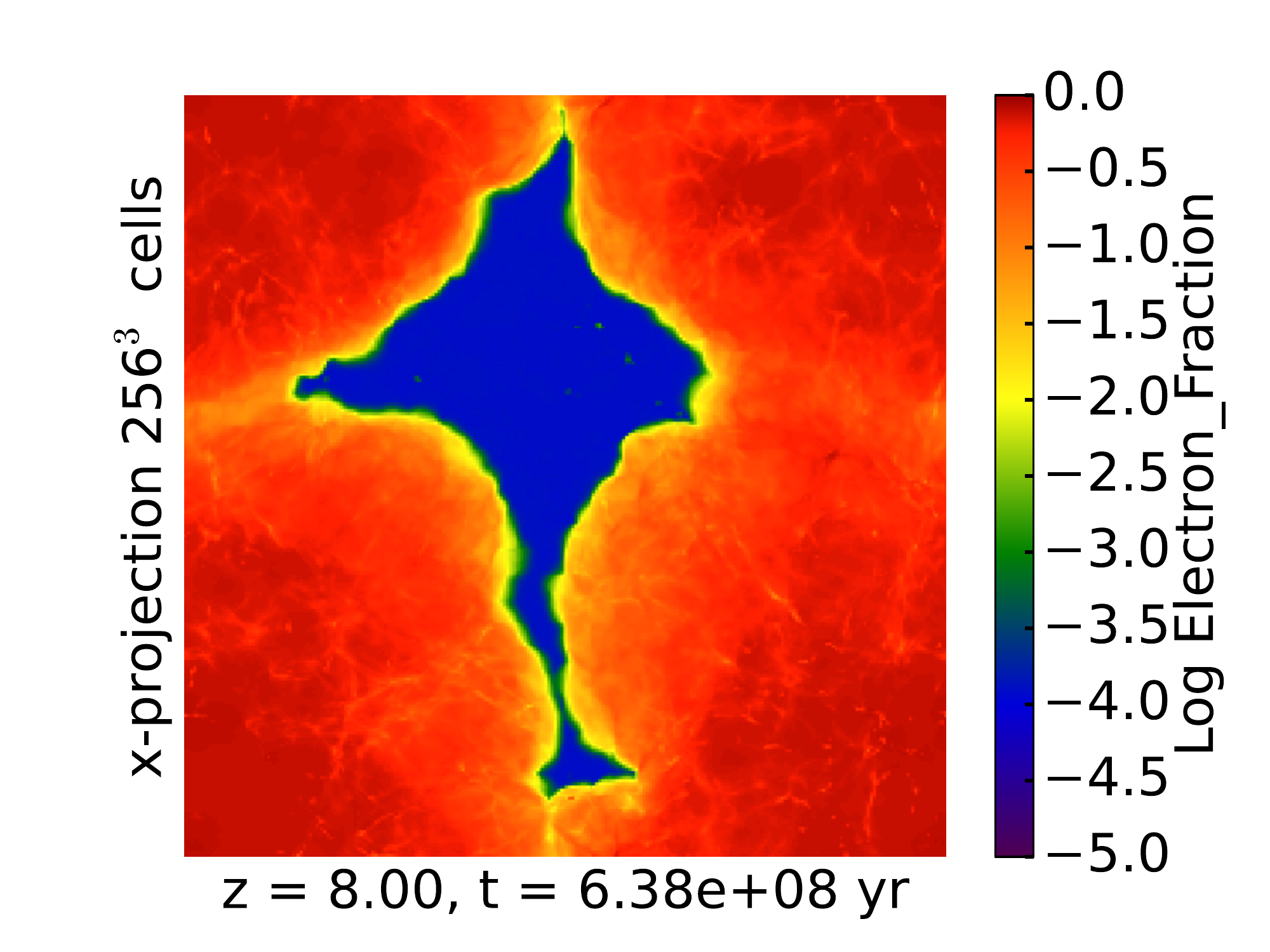}
  \includegraphics[width=0.5\textwidth]{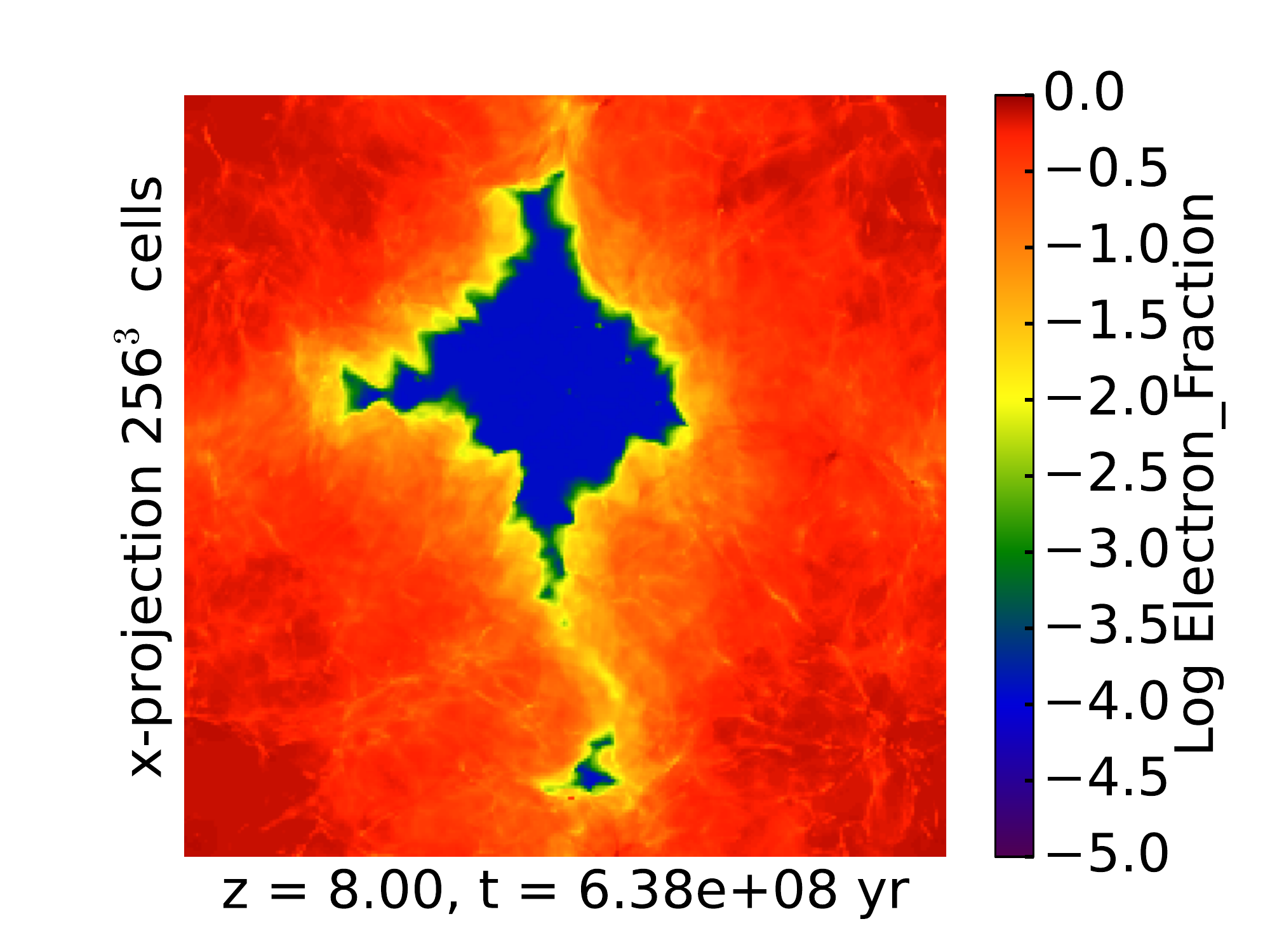}
  \hfill}
\centerline{\hfill
  \includegraphics[width=0.5\textwidth]{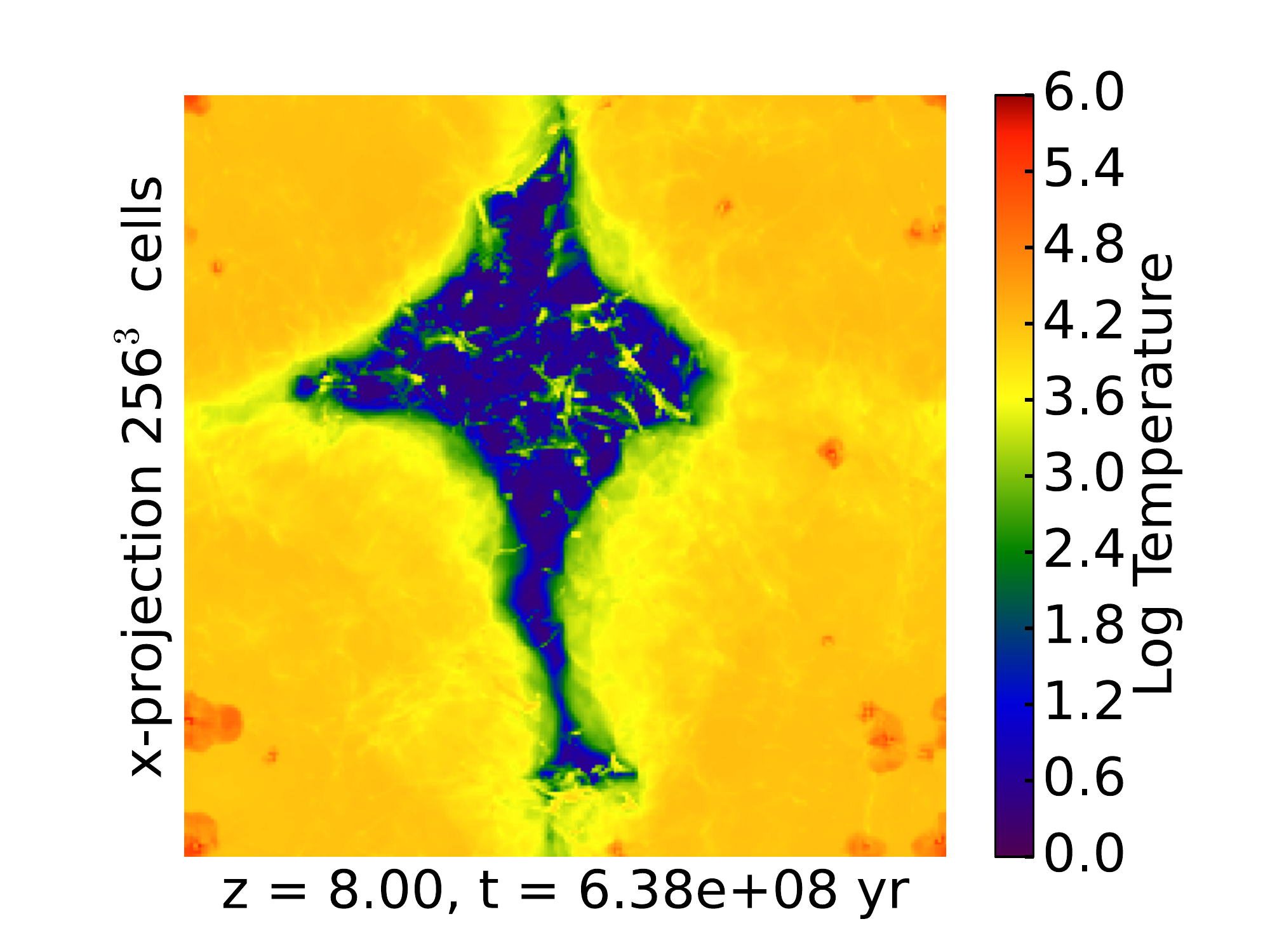}
  \includegraphics[width=0.5\textwidth]{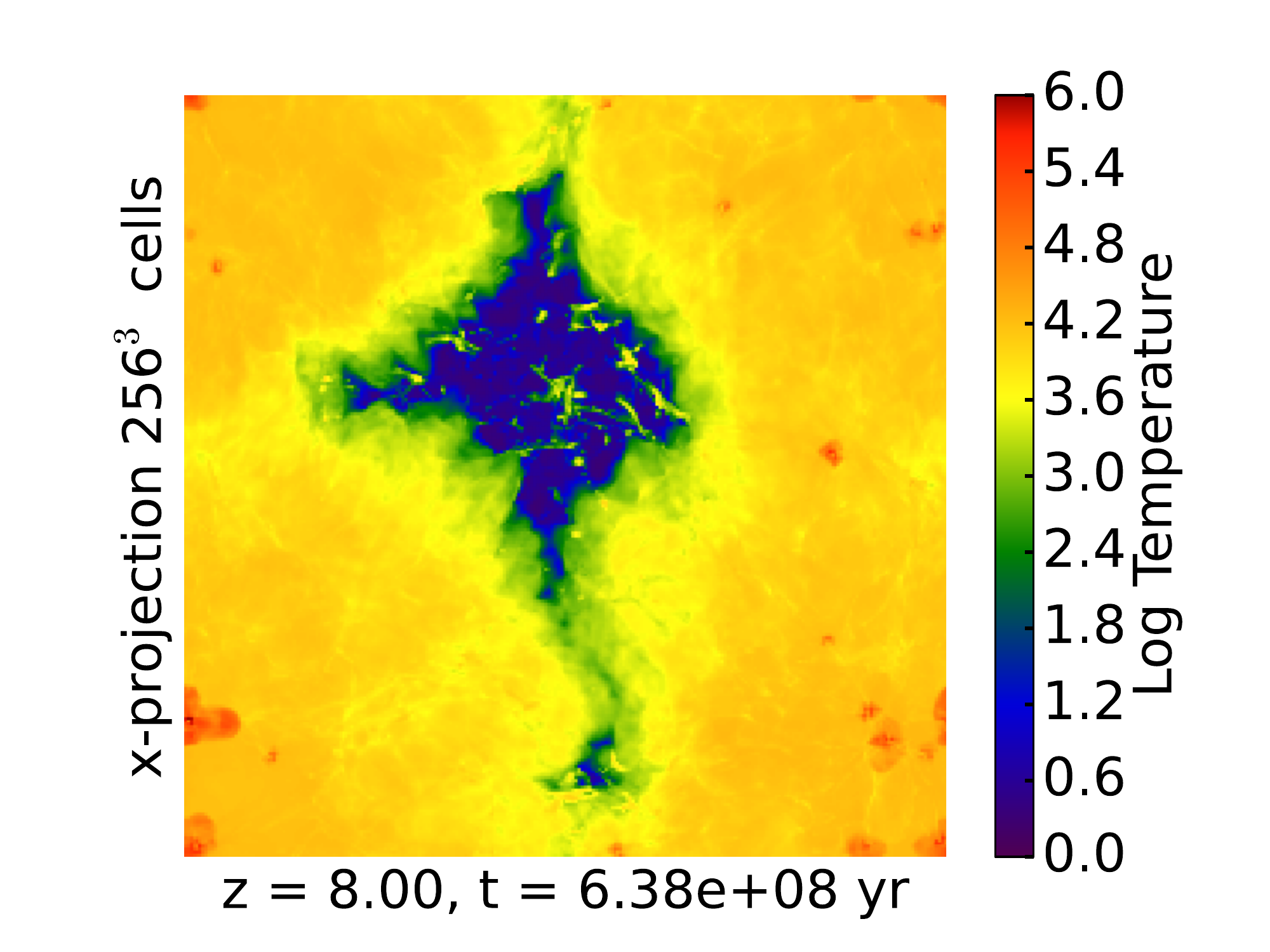}
  \hfill}
  \caption{Same as Fig. \ref{fig:FLD_Moray_z10} except for $z=8$.}
  \label{fig:FLD_Moray_z8}
\end{figure}

Figures \ref{fig:FLD_Moray_z10}, \ref{fig:FLD_Moray_z9}, and \ref{fig:FLD_Moray_z8} show projections through the box of the density-weighted electron fraction and gas temperature for the FLD and ray tracing simulations at three redshifts: $z=10, 9, 8$. Overall, we see very good correspondance between ionized regions at all redshifts. The one noticeable difference is the smoothness of the FLD I fronts versus the jaggedness of the ray tracing I fronts. Without further detailed study we cannot say whether this difference is the result of shadowing, ray discreteness effects, or a manifiestation of I front instabilities \citep{WhalenNorman2008b,WhalenNorman2008a,WhalenNorman2011}. 

\begin{figure}[t]
\centerline{\hfill
  \includegraphics[width=0.5\textwidth]{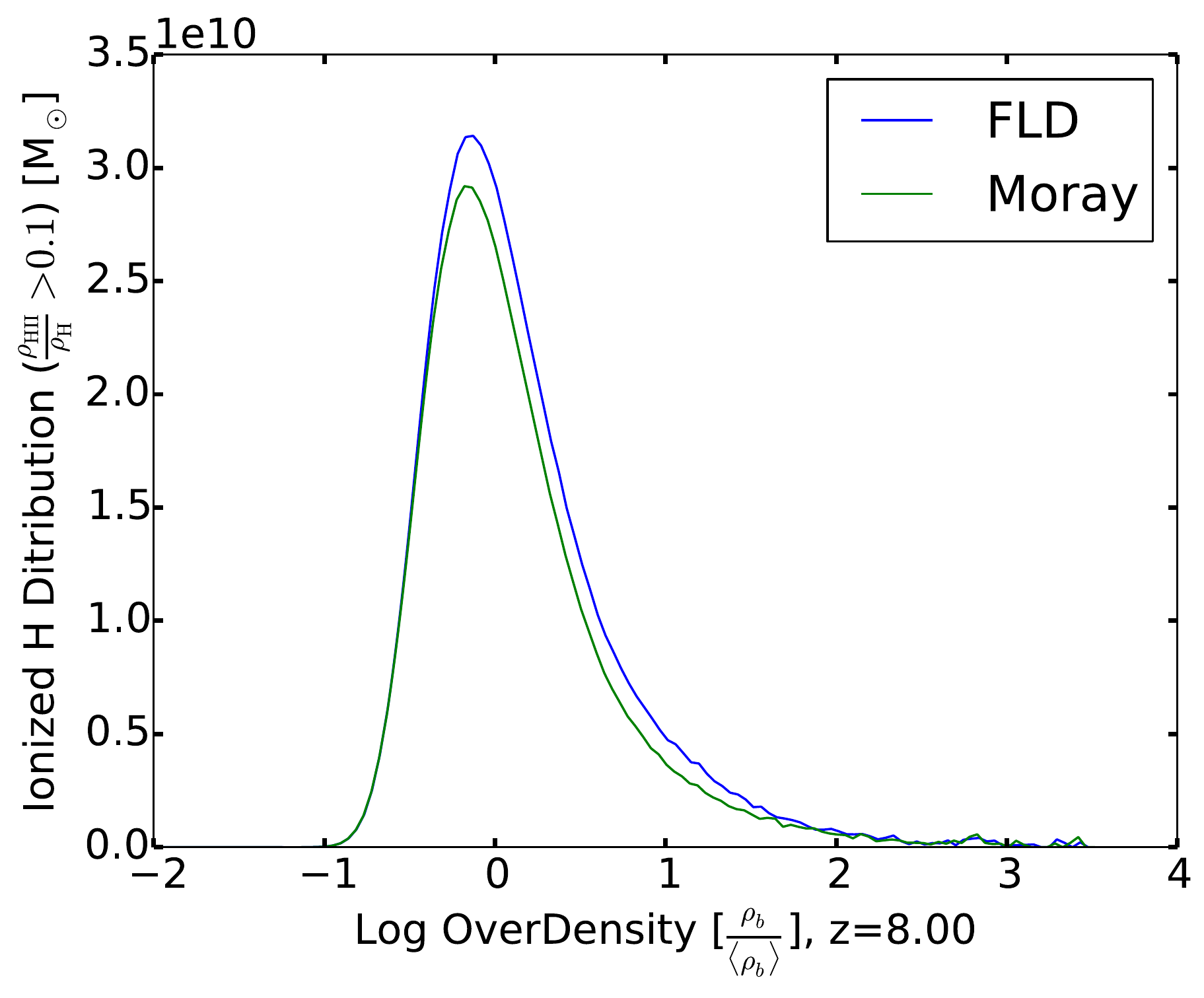}
  \includegraphics[width=0.5\textwidth]{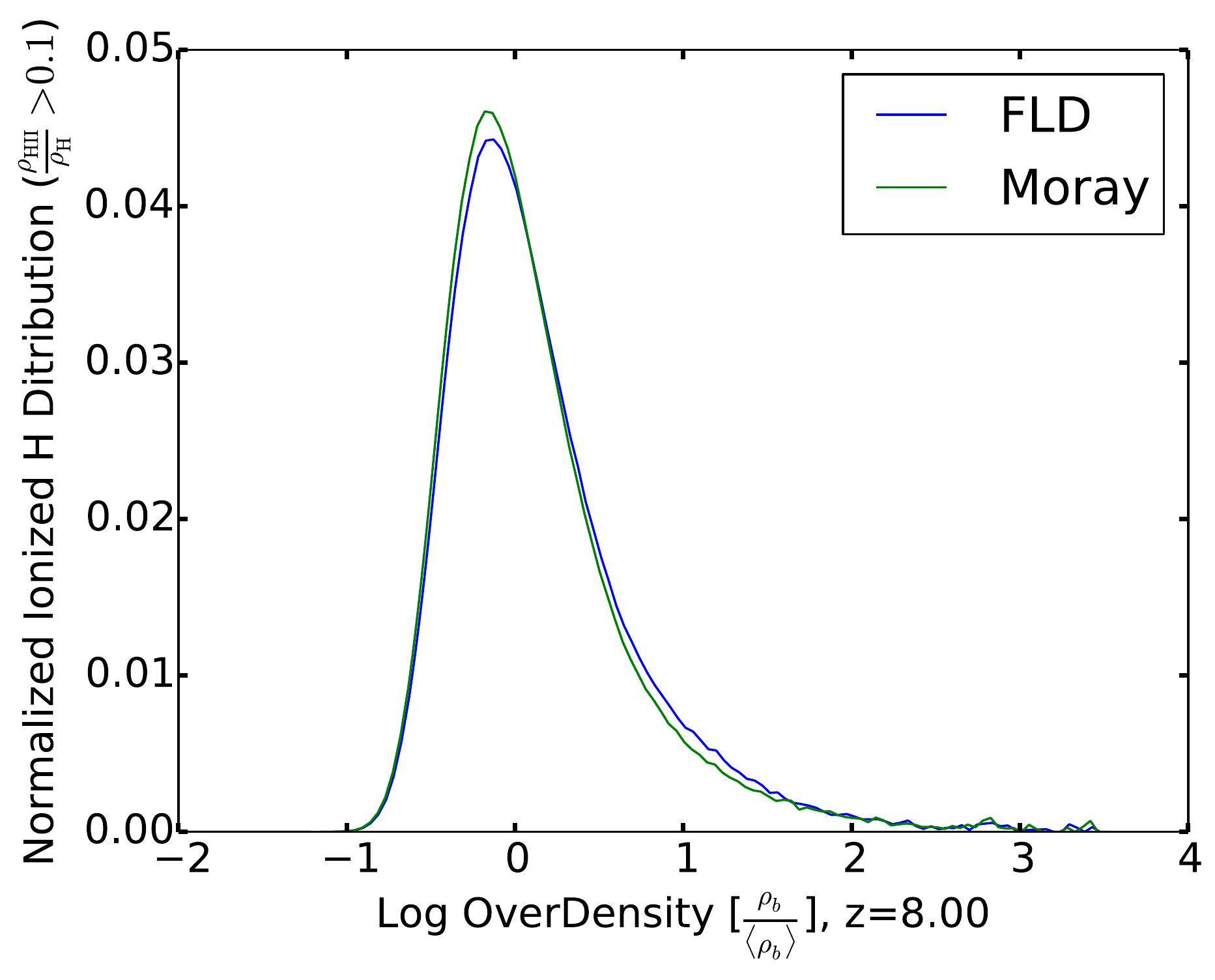}
  \hfill}
  \caption{Distribution functions of ionized hydrogen versus baryon overdensity for the FLD and ray-tracing simulations at $z=8$. (a) total mass of \hii; (b) normalized mass of \hii .}
  \label{fig:ionized_DFs}
\end{figure}

Finally we show in Figure \ref{fig:ionized_DFs} distribution functions of ionized hydrogen versus baryon overdensity for the FLD ($\tau_{tol}=10^{-3}$) and ray-tracing simulations at $z=8$. Figure \ref{fig:ionized_DFs}a plots the total mass of \hii versus overdensity. We see than FLD ionizes about 10\% more gas at mean density and above compared to ray tracing. The integral of this difference over all densities is somewhat larger, as can be seen from the evolution of the ionized mass fraction in Fig. \ref{fig:ionization_evolution}. At z=8, the difference is closer to 20\%.  Figure \ref{fig:ionized_DFs}b plots the normalized mass of \hii versus overdensity. This scales out the total mass and allows us to see how the ionized gas is distributed at various overdensities. The two distribution functions agree at $\log (\Delta_b) < -0.5$ and $\log (\Delta_b) > 2$. However in the intermediate regime $-0.5 \leq \log (\Delta_b) \leq 2$ we see that FLD has somewhat less gas at near mean density relative t ray tracing, consistent the greater ionized volume fraction, and has somewhat more gas with overdensities of $\sim 10$ in the (partially) ionized state.  This is consistent with the interpretation stated above and discussed in more detail in Paper II that gas of moderate overdensities is over-ionized relative to ray tracing at the level of about 20\% in mass, and a smaller amount in volume.
We expect this number to be resolution-dependent, and are in the process of doing a resolution study on this problem to see if convergence can be achieved at higher resolution.

\subsection{Parallel Scalability}
\label{subsec:scalability}


Our numerical method is highly scalable. The scalability of the radiation diffusion solver has already been presented in \cite{ReynoldsHayesPaschosNorman2009}. There we show that our geometric multigrid--based solver exhibits the expected $\log p$ scaling on an idealized weak scaling test involving an array of ionizing point sources. 

More interesting and relevant is the weak scaling of the entire reionization computation combining dynamics, gravity, and radiation transport. This is the regime relevant to cosmic reionization. The weak scaling of our combined calculation can be illustrated very straight-forwardly. We have performed two fully-coupled reionization simulations differing only in the box size. One is performed in a volume 20 comoving Mpc on a side resolved by $800^3$ cells and dark matter particles \citep{So2014}. The other, described below, is performed in a volume 80 comoving Mpc on a side, resolved by $3200^3$ cells and dark matter particles. The larger simulation is 64 times the volume of the smaller simulation, but is computed using 64 times as many cells and particles. Thus, the mass and spatial resolutions are identical. Both simulations were computed on Cray XT5 systems at ORNL with identical node designs and interconnects. The compute nodes had dual hex-core AMD Operton processors with 16 GB RAM and a clock speed of 2.6 GHz. The smaller simulation was computed on 512 cores, and consumed 255,000 core-hrs. The larger simulation was computed on 31,250 cores and consumed 38,000,000 core-hrs. Since the execution time is dominated by the radiation diffusion solver, and this uses a multigrid solver with demonstrated $\log p$ scaling, the ratio of costs would be predicted to be 255,000 core $\times 64 \log (31,250) / \log(512) = 27.2$M core-hrs. Our large simulation is thus running at 72\% parallel efficiency relative to the small simulation. The origin of this inefficiency is the lack of particle load-balancing among the processors, as well as the global gravity solve which is performed using less than optimal FFTs at this scale.

\subsection{Execution Speed Tests: Hydro vs. Rad-Hydro}
\label{subsec:speed}

Here we examine the relative execution speed between a pair of {\em Enzo} cosmological simulations with and without
FLD radiative transfer engaged, henceforth referred to as RHD and HD models, respectively. The RHD model is a simulation of
inhomogeneous cosmic reionization in a 80 Mpc comoving volume resolved by a uniform mesh of $3200^3$ cells and the same
number of dark matter particles. The problem is partitioned into $25^3 = 15,625$ MPI tasks, each of which evolves
a $128^3$ tile of the global mesh and is assigned to a different processor core. The physics model is as described in Sec. 2. 
The RHD model includes star formation and feedback (radiative, thermal, and chemical) as described in Sec. \ref{subsec:starform},
FLD radiative transfer, and 6-species primordial gas chemistry and ionization. 
The simulation was
carried out on the Cray XT5 supercomputer architecture {\em ORNL Jaguar}. The HD model is identical in all 
respects to the RHD model except that the FLD solver is not called each timestep. The HD simulation corresponds
to a primordial  6-species hydro-cosmological simulation with star formation and supernova feedback which is similar in
all respects to a standard Lyman alpha forest simulation in which the IGM is ionized by a homogeneous UV 
background, treated in the optically thin limit (e.g., \citet{Jena05}). 

\begin{figure}[t]
\centerline{\hfill
  \includegraphics[width=0.6\textwidth, angle=-90]{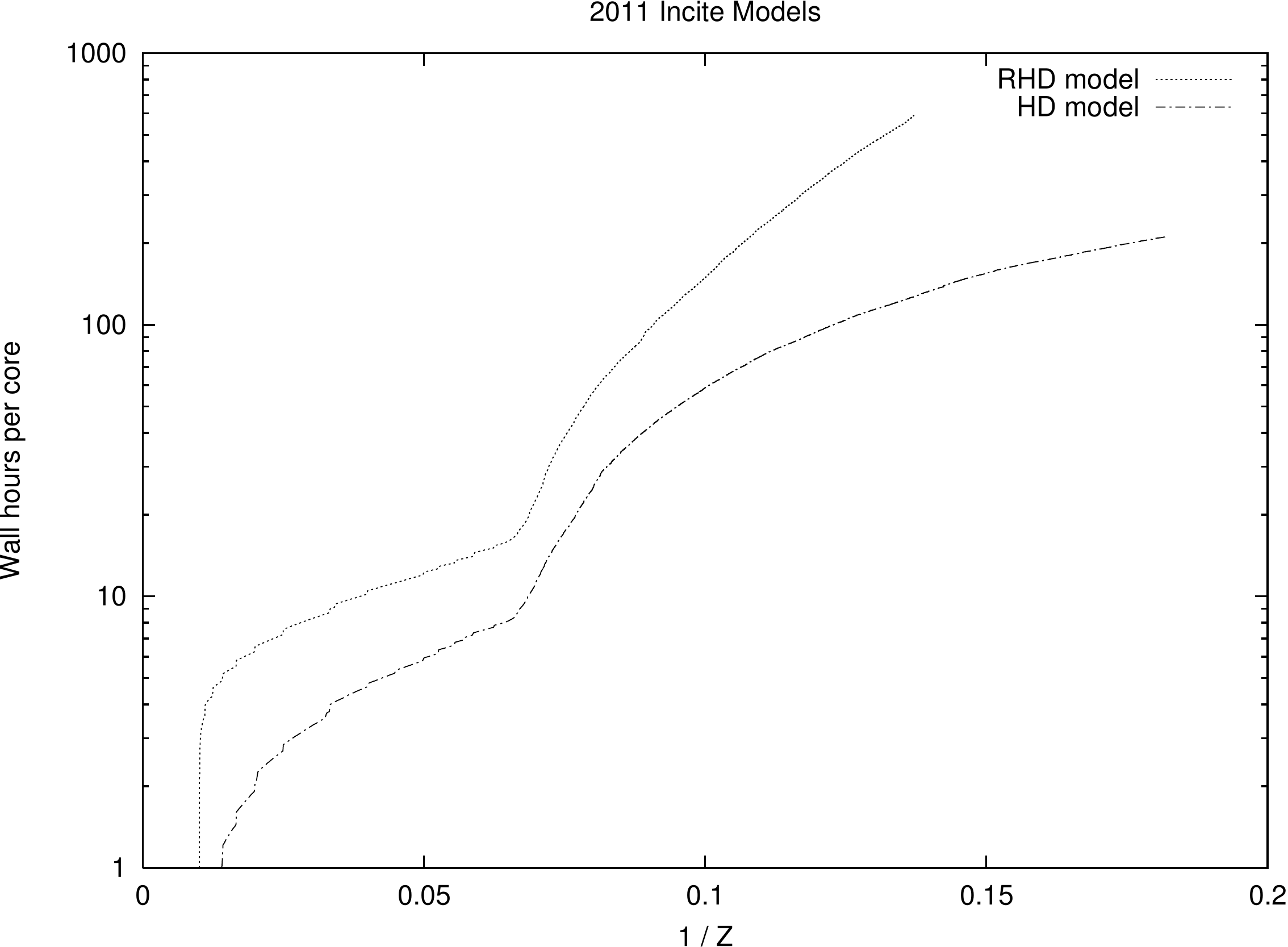}
  \hfill}
  \caption{Cumulative wall time per processor core as a function of 1/z for the HD and RHD models, which differ only in whether the FLD 
radiative transfer solver is called (RHD) or not (HD). The inflection in the curves at 1/z $\approx 0.07$ corresponds to the onset of Pop II
star formation in dwarf galaxies. }
  \label{fig:logcost}
\end{figure}

Fig. \ref{fig:logcost} shows the cumulative wall time per core for the HD and RHD models plotted as a function of 1/z. 
The inflection in the curves at 1/z $\approx 0.07$ corresponds with the onset of star formation at z=14. Subsequently 
hot $10^6-10^7$ K gas is produced by supernova feedback in growing amounts which Courant limits the timestep (see Fig. \ref{fig:logcost})
and increases the cost of the HD simulation per unit time. The RHD model is more costly
than the HD model by a factor which growns from $\sim 2x$ at early times to $\sim 8x$ at late times. The reason for this
is discussed next. 

\begin{figure}[t]
\centerline{\hfill
  \includegraphics[width=0.6\textwidth, angle=-90]{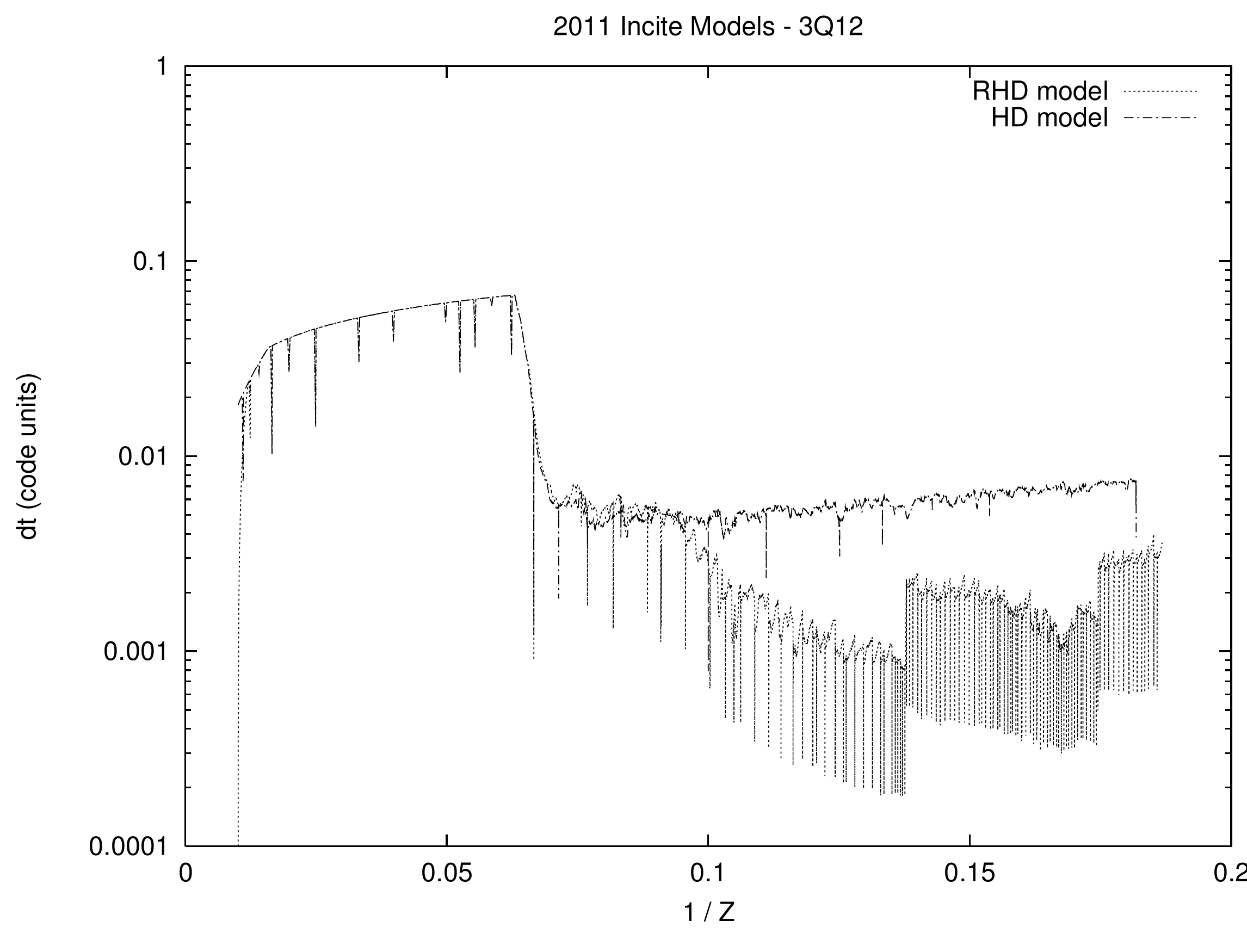}
  \hfill}
  \caption{Timestep history, measured in code units, for two large cosmological simulations with (RHD) and without (HD) radiative transfer. The simulations use identical cosmological initial conditions within a 80 Mpc periodic box resolved with $3200^3$ cells and dark matter particles. The decrease in the RHD timestep at 1/z $\approx$ 0.1 corresponds to the expansion of the first isolated HII regions. Downward spikes in the curves are transient artifacts resulting from restarting the calculation.}
  \label{fig:logdt}
\end{figure}

In Fig. \ref{fig:logdt} we plot the timestep size versus 1/z for the two models. Focusing on the curve labeled HD, we see that the timestep drops suddenly by roughly an order of magnitude at $z \approx 0.07$, which marks the onset of star formation. This is due to a more stringent
Courant limit on the timestep arising from shock-heated gas surrounding star forming halos. The sharp downward spikes in the timestep curve 
are short duration transients associating with restarting the calculation. Upon restart, the timestep is set to a low value, and then allowed 
to float upward at a certain geometric rate per timestep until it again becomes globally Courant-limited. Focussing on the curve labeled RHD, we see that it tracks the HD timestep curve until 1/z $\approx 0.1$, and thereafter slowly decreases until 1/z $\approx 0.13$ where it is about
1/8 the size of the HD timestep. This means that at this time, the RHD simulation is taking $8\times$ as the HD simulation to evolve forward in time. The smaller timestep is a consequence of the radiation subcycling algorithm described in Sec. 3.4, which takes as input the relative change tolerance parameter $\tau_{tol}$, taken to be 0.01. The steady decrease in the timestep is understood to be the consequence of the growth in the number of grid points in the I-front transition region, which is proportional to the total area of the I-fronts times some skin depth of the transition region. 

In order to speed up the simulation, we increased the accuracy parameter $\tau_{tol}$ to 0.02 at 1/z  $\approx 0.1$. This resulted in an approximately $3\times$ increase in the timestep, as can be seen in Fig. \ref{fig:logdt}. Through separate tests on a smaller (1/64) volume test at the same resolution (i.e., an $800^3$ simulation), we determined that this change had a $< 5\%$ change on the redshift of overlap, which in the full simulation is $z_{reion} \approx 5.8$. After overlap, we increased the accuracy parameter to $\tau_{tol}$ to 0.03, resulting in a RHD timestep which is smaller than the HD timestep by a factor of about 2.5. Using the small box tests, we have determined that if we raise the radiation solve accuracy parameter to $\tau_{tol}$ to 0.05, the timestep becomes equal to the Courant-limited HD timestep. 

\section{Example Simulation: Cosmic Reionization by Stellar Sources}
\label{sec:example_simulation}

To illustrate the application of our radiation hydrodynamic cosmological code we
simulate hydrogen reionization due to stellar sources in a comoving volume of (80 Mpc)$^3$ with a grid
resolution of $3200^3$ and the same number of dark matter particles (this is the RHD simulation referred to above.) This yields a comoving spatial resolution of 25 kpc and
dark matter particle mass of $4.8 \times 10^5 M_{\odot}$. This resolution yields a dark matter halo mass function that
is complete down to $M_h = 10^8 M_{\odot}$. This is by design, since this is the halo mass scale below which gas cooling becomes inefficient and therefore we expect a negligible fraction of the ionizing flux to be emitted by such halos (but see, however, \cite{Wise2014}.) The box is large enough to contain the rare, luminous galaxies at the bright end of the high-z galaxy luminosity function (Norman et al., in prep.)

We simulate a $\Lambda$CDM cosmological model with the following parameters: 
$\Omega_{\Lambda} = 0.73$, $\Omega_m = 0.27$, $\Omega_b = 0.047$, 
$h = 0.7$, $\sigma_8 = 0.82$, $n_s = 0.95$, where these are, respectively, the fraction of the 
closure density at the present epoch in vacuum energy, matter, baryons;
the Hubble contant in units of 100 km/s/Mpc; the power spectrum normalization; 
and the slope of the scalar fluctuations of the primordial power spectrum. These are
consistent with the 7-year WMAP measurements \cite{WMAP7}. A Gaussian random
field is initialized at z=99 using the {\em Enzo} initial conditions generator {\em init} 
using the Eisenstein \& Hu (1999) fits to the transfer functions. The star formation efficiency
parameter $f_*$ is adjusted to match the observed star formation rate density in the interval $6 \leq z \leq 10$ from
Bouwens et al. (2011).  Further details of the simulation input parameters and assumptions are described in \cite{So2014} as they are identical to the smaller 20 Mpc box simulation analyzed there.
The simulation was run to a stopping redshift of z = 5.5 and consumed 38 Million core-hrs running on 31,264 cores of the Cray XT5 system {\em Jaguar} operated by the 
National Center for Computational Science at ORNL. 

\begin{figure}[t]
\centerline{\hfill
  \includegraphics[scale=0.75, trim=0.5cm 0.5cm 0.5cm 0.5cm]{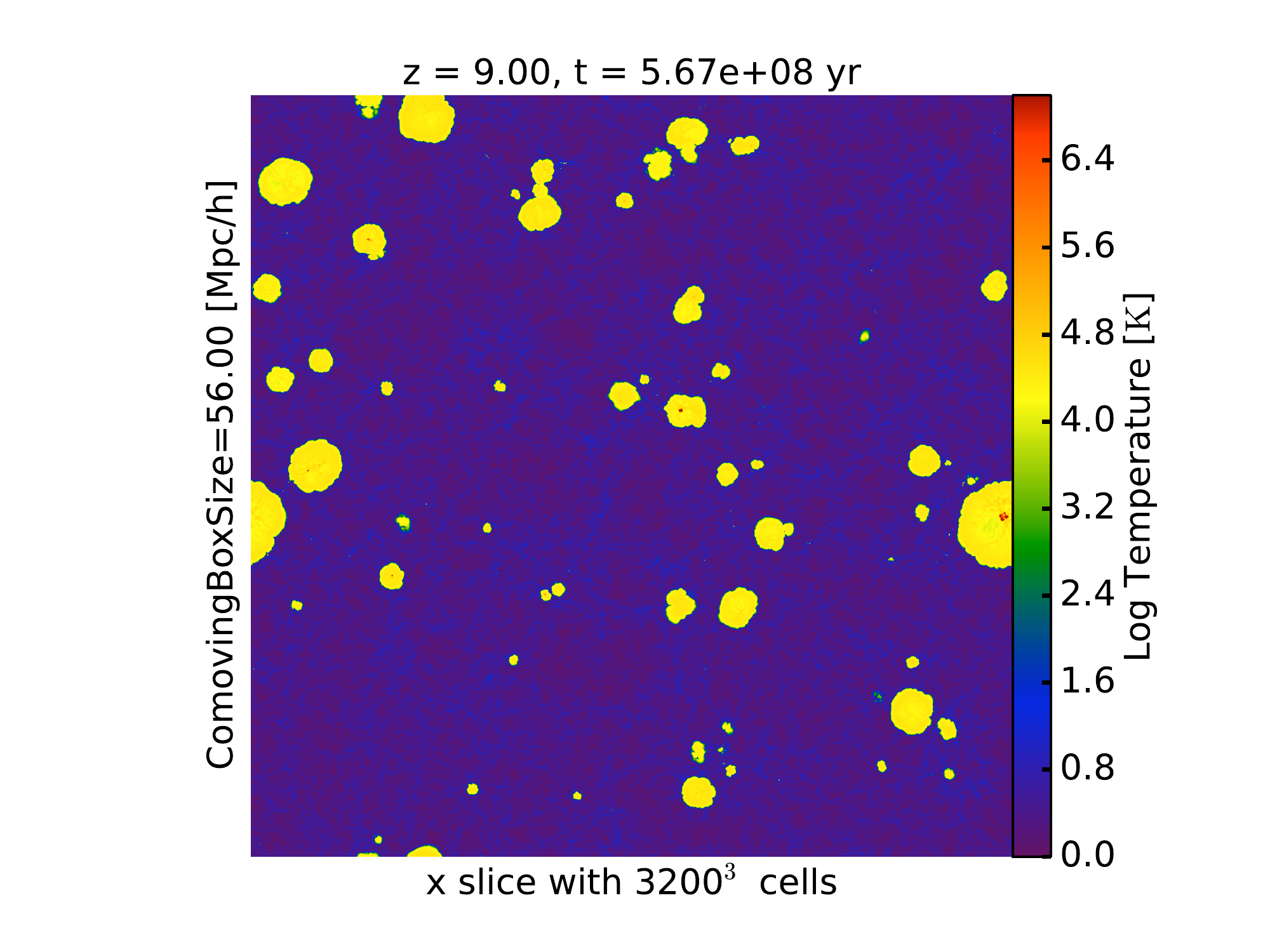}
  \hfill}
\centerline{\hfill
  \includegraphics[scale=0.75, trim=0.5cm 0.5cm 0.5cm 0.5cm]{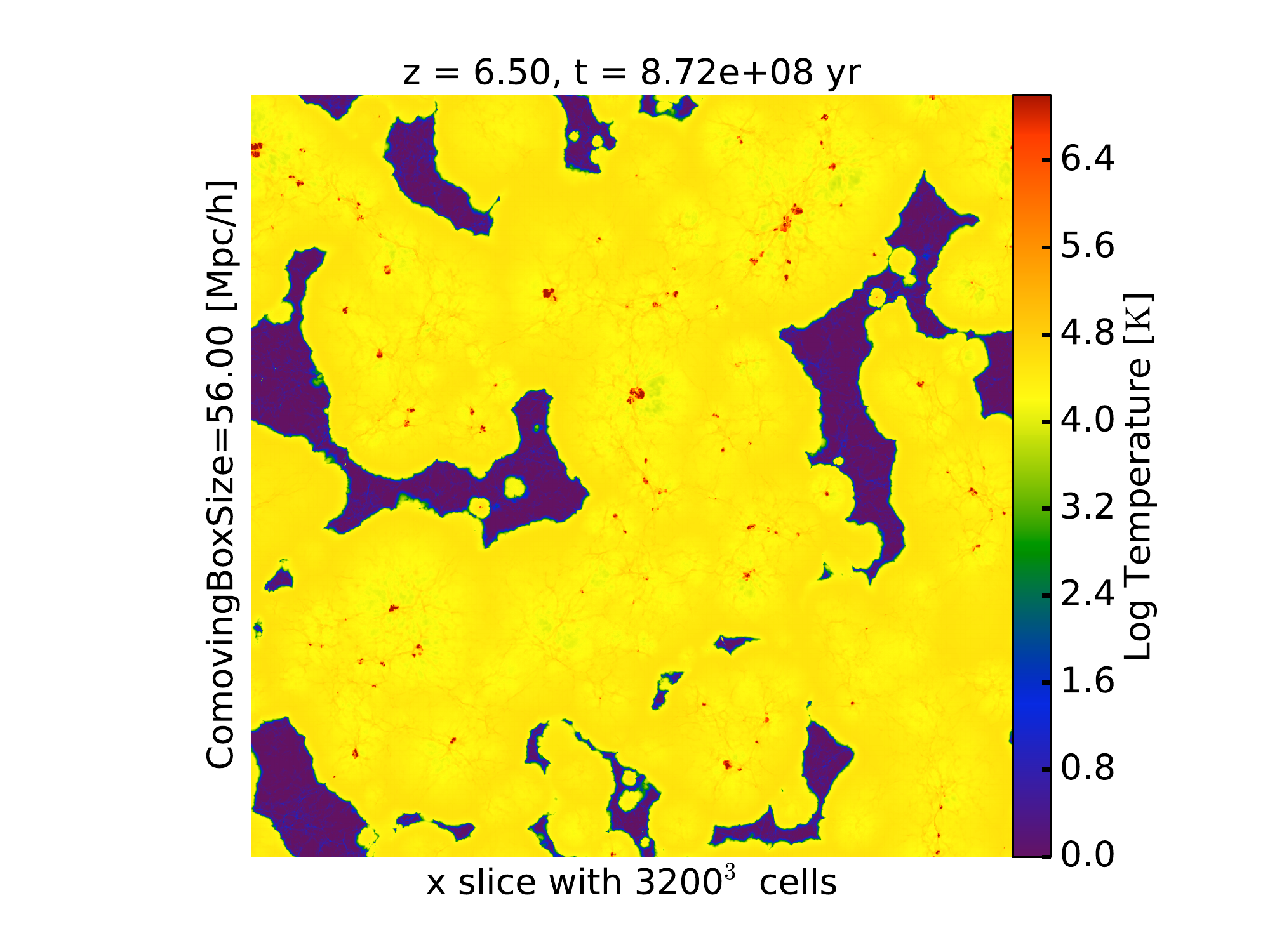}
  \hfill}
  \caption{Application of the numerical methods described in this paper to cosmological hydrogen reionization. Shown are slices of gas temperature at z=9 and 6.5 through a (80 Mpc)$^3$ simulation volume resolved with mesh of $3200^3$ Eulerian cells and $3200^3$ dark matter particles. Ionized regions appear yellow.}
  \label{fig:simulation}
\end{figure}

Fig. \ref{fig:simulation} shows slices through the simulation box of gas temperature at z=9 and 6.5, when the volume is roughly 5\% and 90\% ionized, respectively. Reionization begins at $z \approx 14$ with the first luminous sources inflating isolated HII regions, and completes at 
$z \approx 5.8$ after the HII regions merge and overlap. The \hii regions are roughly spherical until they begin to merge, which is
indicative of the photon budget for reionization being dominated by fewer, more luminous sources, as opposed to numerous low luminosity sources \cite{Zahn07,Iliev2012}. In general appearance they are not dissimilar to the post-processing results of \cite{Iliev06,TracCen2007,Iliev2012} except that the I-fronts somewhat smoother. This is consistent with what we found in our FLD vs. ray tracing comparison in Sec. \ref{subsec:FLDvsRT}, where the ray tracing results produce a more jagged I-front.
An inspection of the temperature projections shows photoionized gas in yellow, with smaller pockets of shock heated gas near the centers of \hii regions, resulting from thermal feedback from supernovae. We also see that the gas immediately behind the I-front is hotter than closer to the center of the \hii region. That is due to radiative transfer effects, as discussed in detail in \cite{AbelHaehnelt99}. This simulation is analyzed is detail in Norman et al. 2014, {\em in preparation}.

\section{Summary and Conclusions}
\label{sec:conclusions}

We have described an extension of the {\em Enzo} code to enable the fully-coupled numerical simulation 
of inhomogeneous cosmological ionization in reasonably large cosmological volumes. By fully-coupled we mean
all dynamical, radiative, and chemical properties are solved self-consistently on the same mesh, 
as opposed to a postprocessing approach which coarse-grains the radiative transfer, as is done
in other works \citep{Iliev06,Zahn07,TracCen2007,TracCenLoeb2008,ShinTracCen2008,Finlator09,Iliev2012}. 
Star formation and feedback are treated through a parameterized
phenomenological model, which is calibrated to observations. The goal of this work is to achieve a higher
level of self-consistency in modeling processes occuring outside the virial radii of luminous sources to  
better understand how recombinations in the clumpy intergalactic medium retard reionization and how
radiative feedback effects star formation in low mass galaxies. 

In its current incarnation, the model has three principal limitiations. First, it is formulated on a fixed
Eulerian grid, which limits the spatial resolution that can be achieved. With a judicious choice of grid
sizes and resolutions, one can sample the dark matter halo mass function over a significant range of scales,
thereby including important sources and sinks of ionizing radiation. One can do a good job resolving the Jeans length 
in the diffuse IGM, which is important for ``Jeans smoothing" \citep{Gnedin00b}. However one cannot resolve the
internal structure of halos, which is important for calculating star formation rates and ionizing escape fractions. 
In this work we do not claim to be modeling these aspects self-consistently, but rather calibrate these unknown
parameters to observations. In a forthcoming paper (Reynolds et al., {\em in prep.})
we present the extention of our method to adaptive
mesh refinement (AMR), which directly addresses the numerical resolution limitation. 

The second model limitation is the use of FLD to model the transport of radiation, as opposed to a higher
order moment method such as OTVET \citep{GnedinAbel2001,Petkova09}. FLD has the well known deficiency of
not casting shadows behind opaque objects. However, as we have shown in Sec. \ref{subsec:FLDvsRT}, casting
shadows is not required to correctly predict the late evolution of the ionized volume fraction in a cosmological reionization
simulation. The early phases of reionization 
proceed somewhat faster with FLD as compared to a ray tracing solution, which can be attributed to a lack of shadowing as follows: at early times, when \hii regions are isolated, gas is ionized by a single dominant source. Any clump of gas will be irradiated from one side. We have shown from the photoevaporating clump test that FLD photoevaporates the cloud more quickly than ray tracing since the radiation envelops the cloud and irradiates it from all sides. At later times, when \hii regions have percolated, neutral gas is irradiated by multiple sources in different directions. FLD becomes a better approximation to this situation as time goes on. In the limit of a fully ionized IGM and once a UV background is established, FLD becomes an even better approximation since the assumption of isotropy is built in. Our {\em a priori} assumption that
small scale features like shadows will have little effect on the later stages of large scale reionization processses are borne out
by these validation tests. For simulating smaller scale processes where shadows may be important, such as the 
effect of halo substructure on the escape fraction of ionizing radiation or the presistence of self-shielded denser gas embedded in \hii regions we note that our implicit
solution methodology is easily 
extended to higher-order moment methods \citep{HayesNorman2003}. 

The third model limitation is our simplified model for the radiation spectrum, which at the moment consists of 
monochromatic and grey with an assumed fixed SED. For simulating hydrogen reionization by soft UV radiation
from stellar sources
this spectral model is quite adequate compared to a multifrequency model (see RT09, Sec. 4.1). However for
harder radiation sources, such as Pop III stars and AGN, our model makes I-fronts that are too sharp, and does
not produce the preheating of gas ahead of the I-front by more penetrating, higher energy photons (``spectral
hardening"). The principal difference between our model and a multifrequency/multigroup model is in the temperature
distribution of the gas. Our model will slightly overpredict the temperature inside an \hii region, and underpredict
the temperature outside of it, because all of the radiation energy is absorbed inside the I-front. Another way to
think about this is that in the multifrequency model in which the highest energy photons leak out of the \hii region,
the characterisitic temperature of the radiation field inside the \hii region is lower than outside of it. The standard
approach for dealing with the limitiations of our spectral model is to move to a multifrequency or multigroup 
discretization of the radiation field \citep{MirochaEtAl2012}. This is straightforward in practice, however the computational cost
increases linearly with the number of frequencies/groups. With the speed and memory of modern supercomputers
is this not a severe limitation, except for the very largest grids. Indeed we have implemented a multigroup FLD 
version of our method which is undergoing testing at the present time (Reynolds et al., {\em in prep.}). 

Despite these limitations, the method is robust and acceptably fast. On verification tests for which analytic 
solutions are known, we have shown the method to be capable of high accuracy; the accuracy being
governed by grid resolution and the error tolerance parameter in the radiation diffusion calculation.
In validation tests, for which no analytic solution exists, we have shown that our method gives results which are 
qualitatively and quantitatively similar with those obtained with ray tracing and Monte Carlo methods \citep{IlievEtAl2006,IlievEtAl2009,WiseAbel11},
with what differences exist understood to be the result of the geometric
simplification of the radiation field inherent in FLD, and the difference in radiation spectrum modeling.
 
Regarding the speed of our method, we have shown by direct comparison that a radiation hydrodynamic
simulation of cosmological reionization costs about $8 \times$ that of a corresponding pure hydrodynamic
model in which the IGM is ionized by a uniform UV background. We have not compared it to a postprocessing
radiative transfer code using ray tracing, although this would be a useful thing to do. Our method, which 
exhibits $\mathcal O(N\log N)$ scaling, should be competitive with, and possibly even beat ray tracing methods for 
very large numbers of sources.

Our method is highly scalable, as discussed in Sec. \ref{subsec:scalability}. Since the 
FLD solver dominates the execution time, the scalability of the entire simulation is largely determined by the scalability of the FLD solver, which was presented in an earlier work \citep{ReynoldsHayesPaschosNorman2009}. There it is shown 
that our multigrid solution procedure exhibits optimal $\log p$ scaling on a non-cosmological weak scaling test. 
That we can simulate very large problems is demostrated in Sec. \ref{sec:example_simulation}. This simulation is 64 times the volume of the simulation discussed in \cite{So2014}, and is executed on 64 times as many cores, making it in effect a weak scaling test. After the $\log p$ scaling of the FLD solver is taken into account, we find that the large simulation is operating at 72\% parallel efficiency relative to the smaller simulation. There is potential for improving this through load balancing the dark matter particles across processors, and using a more scalable 3D FFT solver.

This research was partially supported by National Science Foundation grants AST-0808184 and AST-1109243 to MLN and DRR. JHW acknowledges partial support by NSF grants AST-1211626 and AST-1333360. Simulations were performed on the {\em Kraken}
supercomputer operated for the Extreme Science and Engineering Discovery Environment (XSEDE)
by the National Institute for Computational Science (NICS), ORNL with support from XRAC allocation MCA-TG98N020 to MLN, as well as on the {\em Jaguar} supercomputer operated for the DOE Office of Science at the National Center for Computational Science (NCCS), ORNL with support from INCITE awards AST025 and AST033 to MLN. 
MLN, DRR and GS would like to especially acknowledge the tireless devotion to this project by our co-author and dear colleague Robert Harkness
who passed away shortly before this manuscript was completed.

\bibliography{sources}
\bibliographystyle{apj}
\end{document}